\documentstyle[prl,aps,preprint,epsf,floats,rotate]{revtex}
\tightenlines

\newcommand{\nwc}{\newcommand}
\nwc{\cl}  {\clubsuit}
\nwc{\di}  {\diamondsuit}
\nwc{\sps} {\spadesuit}
\nwc{\hyp} {\hyphenation}
\nwc{\be}  {\begin{equation}}
\nwc{\ee}  {\end{equation}}
\nwc{\ba}  {\begin{array}}
\nwc{\ea}  {\end{array}}
\nwc{\bdm} {\begin{displaymath}}
\nwc{\edm} {\end{displaymath}}
\nwc{\bea} {\be\ba{rcl}}
\nwc{\eea} {\ea\ee}
\nwc{\ben} {\begin{eqnarray}}
\nwc{\een} {\end{eqnarray}}
\nwc{\bda} {\bdm\ba{lcl}}
\nwc{\eda} {\ea\edm}
\nwc{\bc}  {\begin{center}}
\nwc{\ec}  {\end{center}}
\nwc{\ds}  {\displaystyle}
\nwc{\bmat}{\left(\ba}
\nwc{\emat}{\ea\right)}
\nwc{\non} {\nonumber}
\nwc{\bib} {\bibitem}
\nwc{\lra} {\longrightarrow}
\nwc{\Llra}{\Longleftrightarrow}
\nwc{\ra}  {\rightarrow}
\nwc{\Ra}  {\Rightarrow}
\nwc{\lmt} {\longmapsto}
\nwc{\pa} {\partial}
\nwc{\iy}  {\infty}
\nwc{\ovl}  {\overline}
\nwc{\hm}  {\hspace{3mm}}
\nwc{\lf}  {\left}
\nwc{\ri}  {\right}
\nwc{\lm}  {\limits}
\nwc{\lb}  {\lbrack}
\nwc{\rb}  {\rbrack}
\nwc{\ov}  {\over}
\nwc{\pr}  {\prime}
\nwc{\nnn} {\nonumber \vspace{.2cm} \\ }
\nwc{\Sc}  {{\cal S}}
\nwc{\Lc}  {{\cal L}}
\nwc{\Rc}  {{\cal R}}
\nwc{\Dc}  {{\cal D}}
\nwc{\Oc}  {{\cal O}}
\nwc{\Cc}  {{\cal C}}
\nwc{\Pc}  {{\cal P}}
\nwc{\Mc}  {{\cal M}}
\nwc{\Ec}  {{\cal E}}
\nwc{\Fc}  {{\cal F}}
\nwc{\Hc}  {{\cal H}}
\nwc{\Kc}  {{\cal K}}
\nwc{\Xc}  {{\cal X}}
\nwc{\Gc}  {{\cal G}}
\nwc{\Zc}  {{\cal Z}}
\nwc{\Nc}  {{\cal N}}
\nwc{\fca} {{\cal f}}
\nwc{\xc}  {{\cal x}}
\nwc{\Ac}  {{\cal A}}
\nwc{\Bc}  {{\cal B}}
\nwc{\Uc}  {{\cal U}}
\nwc{\Vc}  {{\cal V}}
\nwc{\Th} {\Theta}
\nwc{\th} {\theta}
\nwc{\vth} {\vartheta}
\nwc{\eps}{\epsilon}
\nwc{\si} {\sigma}
\nwc{\Gm} {\Gamma}
\nwc{\gm} {\gamma}
\nwc{\bt} {\beta}
\nwc{\La} {\Lambda}
\nwc{\la} {\lambda}
\nwc{\om} {\omega}
\nwc{\Om} {\Omega}
\nwc{\dt} {\delta}
\nwc{\Si} {\Sigma}
\nwc{\Dt} {\Delta}
\nwc{\al} {\alpha}
\nwc{\vph}{\varphi}
\nwc{\zt} {\zeta}
\def\tr{\mathop{\rm tr}}
\def\Tr{\mathop{\rm Tr}}

\def\VEV#1{\left\langle #1\right\rangle}

\def\abs#1{\left| #1\right|}
\def\pr#1{#1^\prime}
\def\ltap{\raisebox{-.4ex}{\rlap{$\sim$}} \raisebox{.4ex}{$<$}}
\def\gtap{\raisebox{-.4ex}{\rlap{$\sim$}} \raisebox{.4ex}{$>$}}
\nwc{\Id}  {{\bf 1}}
\nwc{\diag} {{\rm diag}}
\nwc{\inv}  {{\rm inv}}
\nwc{\mod}  {{\rm mod}}
\nwc{\hal} {\frac{1}{2}}
\nwc{\tpi}  {2\pi i}
\def\KK{{\rm I\kern -.2em  K}}
\def\NN{{\rm I\kern -.16em N}}
\def\RR{{\rm I\kern -.2em  R}}
\def\ZZ{Z \kern -.43em Z}
\def\QQ{{\rm \kern .25em
             \vrule height1.4ex depth-.12ex width.06em\kern-.31em Q}}
\def\CC{{\rm \kern .25em
             \vrule height1.4ex depth-.12ex width.06em\kern-.31em C}}
\def\ZZZ{Z\kern -0.31em Z}

\def\MeV {\,{\rm  MeV}}
\def\GeV {\,{\rm  GeV}}

\def\npb#1{Nucl. Phys. {\bf B#1}}

\def\plb#1{Phys. Lett. {\bf #1B}}
\def\pr#1{Phys. Rev. {\bf #1}}
\def\pra#1{Phys. Rev. {\bf A#1 }}

\def\zpc#1{Z. Phys. {\bf C#1}}

\def\CNPP#1{Comm. Nucl. Part. Phys.~{\bf #1}}

\def\NP#1{Nucl. Phys.~{\bf #1}}
\def\NPPS#1{Nucl. Phys. Proc. Suppl.~{\bf #1}}
\def\NC#1{Nuovo Cim.~{\bf #1}}

\def\PL#1{Phys. Lett.~{\bf #1}}
\def\PR#1{Phys. Rev.~{\bf #1}}
\def\PRP#1{Phys. Rep.~{\bf #1}}
\def\PRL#1{Phys. Rev. Lett.~{\bf #1}}

\def\RMP#1{Rev. Mod. Phys.~{\bf #1}}

\def\ZP#1{Z. Phys.~{\bf #1}}

\begin{document}
\setcounter{page}{0}
\def\footnoterule{\kern-3pt \hrule width\hsize \kern3pt}
\tighten
\title{QCD in Extreme Conditions\thanks
{This work is supported in part by funds provided by the U.S.
Department of Energy (D.O.E.) under cooperative 
research agreement \# DE-FC02-94ER40818.}\\
  and the\\
  Wilsonian `Exact Renormalization Group'}

\author{J{\"u}rgen Berges\footnote{Email addresses: 
{\tt berges@ctp.mit.edu}}}

\address{Center for Theoretical Physics \\
Laboratory for Nuclear Science \\
and Department of Physics \\
Massachusetts Institute of Technology \\
Cambridge, Massachusetts 02139 \\
{~}}

%\date{MIT-CTP-, hep-ph/}
\date{MIT-CTP-2829}
\maketitle

\thispagestyle{empty}

\begin{abstract}
This is an introduction to the use of nonperturbative flow equations 
in strong interaction physics at nonzero temperature and baryon density.
We investigate the QCD phase diagram as a function of temperature,
chemical potential for baryon number and quark mass 
within the linear quark meson model 
for two flavors. Whereas the renormalization group flow leads to 
spontaneous chiral symmetry breaking in vacuum, the symmetry is restored 
in a second order phase transition at high temperature and vanishing
quark mass. We explicitly 
connect the physics at zero temperature and realistic quark mass with 
the universal behavior near the critical temperature $T_c$ and the chiral 
limit. At high density we find a chiral symmetry restoring first order 
transition. The results imply the presence of a tricritical point with 
long-range correlations in the phase diagram. We end with an outlook to 
densities above the chiral transition, where QCD is expected to behave 
as a color superconductor at low temperature.\\

\noindent
Based on five lectures presented at the 11th Summer School
and Symposium on Nuclear Physics ``Effective Theories of Matter'', 
Seoul National University, June 23--27, 1998.
\end{abstract}

%\clearpage
%\thispagestyle{empty}
\newpage
%\mbox{ }
%\newpage

%\pagestyle{headings}
\setcounter{page}{1}
\vspace*{-2.cm}
\tableofcontents

%\clearpage
%\thispagestyle{empty}
\newpage
%\mbox{ }
%\newpage

\section{Overview}
\label{intro}
These lectures are about some recent developments concerning the physics
of the strong interaction, Quantumchromodynamics (QCD), in extremes
of temperature and baryon density. 
Here ``extremes'' means temperatures
of the order of $10^{12}\,$K or $100 \MeV$ and densities of about a few times
nuclear matter density
$n_0=0.153\, {\rm fm}^{-3}=(105 \MeV)^3$. \footnote{We 
will work in units where $h\!\!\bar{}\,\,=c=1$ such that 
the mass of a particle is equal to its rest energy ($m c^2$)
and also to its inverse Compton wavelength ($m c/h\!\!\bar{}\,\,$). 
It is sometimes convenient to express length in the unit of a 
fermi (${\rm fm}=10^{-13}$cm) because it is the order of the dimension 
of a nucleon. The conversion to units of energy is easily done
through $(1\MeV)^{-1}h\!\!\bar{}\,\, c = 197.33\,$fm .}

The following schematic phase diagram 
gives an idea about what the 
behavior of QCD in thermal and chemical equilibrium may look like
as a function of temperature and chemical potential of quark number density. 
We will use it here to draw the attention to some aspects
which will be explained in more detail in these lectures and to 
point out their experimental relevance.
\vspace*{-0.3in}
\begin{figure}[h]
\begin{center}
\epsfxsize=4.in
\hspace*{0in}
\epsffile{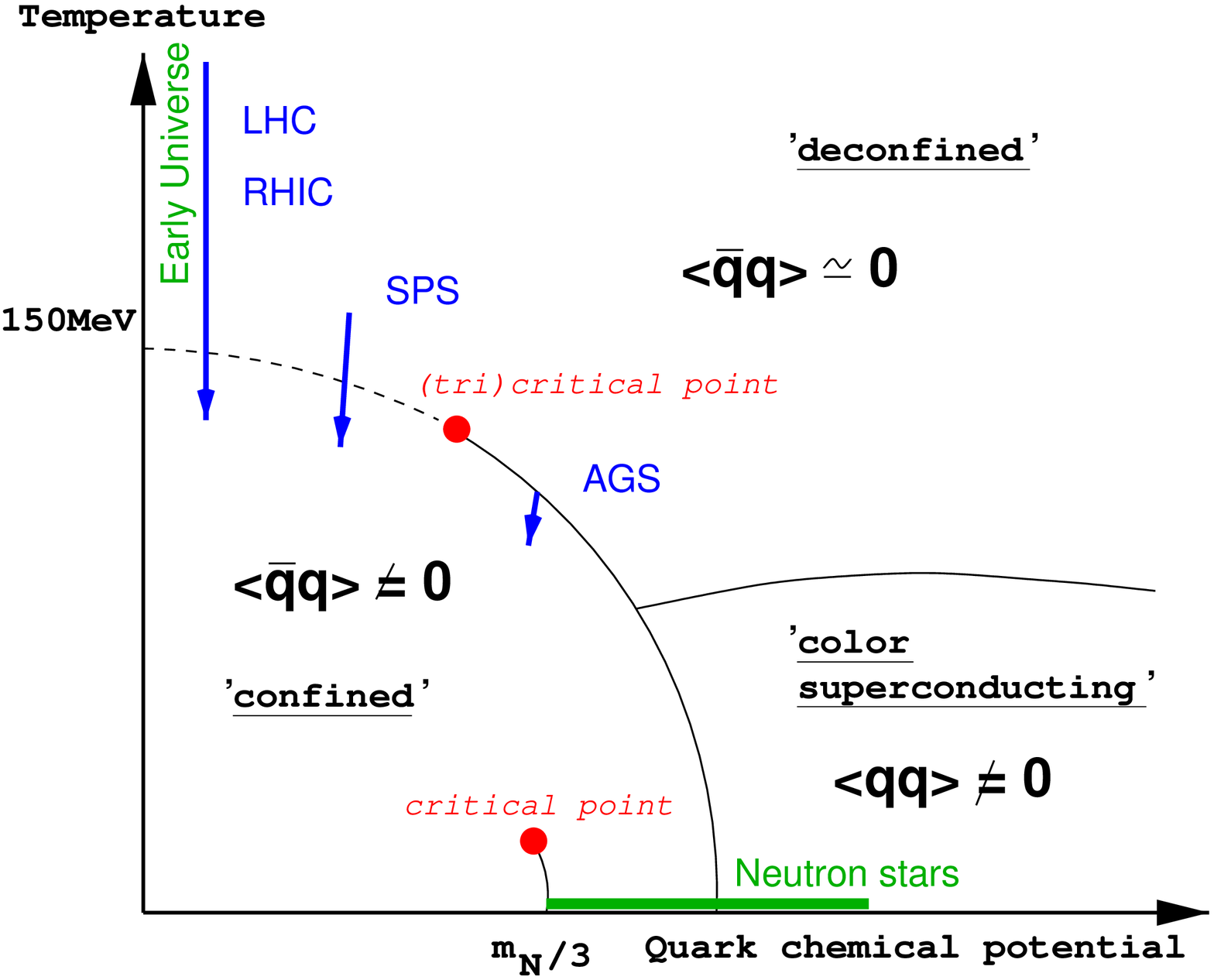}
\end{center} 
%\caption{}
\label{phasediagram}
\end{figure}

\vspace*{-0.8in}
Shortly after the discovery of asymptotic freedom \cite{GWP}, 
it was realized \cite{CP75} that 
at sufficiently high temperature or density QCD may differ in
important aspects from the corresponding zero temperature or
vacuum properties. If one considers the phase structure of QCD
as a function of {\em temperature} one expects around a critical temperature
of $T_c \simeq 150 \MeV$ two qualitative changes. First, the 
quarks and gluons which are confined into hadrons
at zero temperature can be excited independently at sufficiently 
high temperature. The hot state is conventionally called the 
{\bf quark--gluon plasma}. The second aspect has to do with the fact that
the vacuum of QCD contains a condensate of 
quark--antiquark pairs, $\langle \bar{\psi} \psi\rangle \not =0$, which 
spontaneously breaks the (approximate) {\bf chiral symmetry} of QCD and has 
profound implications for the hadron 
spectrum. At high temperature this condensate is expected to
melt, i.e.\ $\langle \bar{\psi} \psi \rangle \simeq 0$, which signals
the {\bf chiral phase transition}. 

QCD at nonzero {\em baryon density} is expected to have a rich phase structure
with different possible phase transitions as the density varies. 
Since nucleons as bound states of quarks have a characteristic size 
$r_{\rm  N}\simeq 1{\rm fm} \simeq(200\MeV)^{-1}$, 
for very high baryon density there is
not enough space to form nucleons and 
one expects a new {\bf quark matter} phase. 
In this phase, similar to the case 
of very high temperature, confinement is not expected to
play an important role and the quark--antiquark condensate 
is absent. However, the high density state is far from trivial. 
It has been realized early \cite{bailin} that at very high 
density, where perturbative QCD can be applied, quark matter
behaves as a color superconductor: Cooper pairs of quarks condense,
opening up an energy gap at the quark Fermi surface.   
Recent investigations \cite{CSC} at
intermediate densities using effective models show the 
formation of quark--quark condensates with phenomenologically significant 
gaps of order $100$ MeV. A first study which takes into account 
the formation of condensates in the conventional 
\mbox{quark--antiquark} channel and in a superconducting quark--quark 
channel reveals a phase diagram~\cite{BR} similar to the one sketched 
in the above figure. New symmetry breaking schemes like
color--flavor locking \cite{CFL} may also be relevant for the study
of {\bf nuclear matter}, if the latter is continuously connected to the
quark matter phase \cite{SW}. The nuclear matter state is
intermediate between a gas of nucleons and quark matter and can 
be associated with a liquid of nucleons. 

Where do we encounter QCD at high temperatures and/or densities?
The astrophysics of {\bf neutron stars} provides a good testing ground 
for the exploration of very dense matter. Neutron stars 
are cold on the nuclear scale with temperatures of about 
$10^5$ to $10^9$ K $(1 \MeV=1.1065 \times 10^{10} {\rm K})$. 
The range of densities is enormous. At the edge, where the pressure is zero,
the density is that of ordinary iron. In contrast, at the center it may be 
a few times nuclear matter density. 

According to the standard hot big bang
cosmology the high temperature transition must have occured during the 
evolution of the {\bf early universe}. For most of its evolution the 
early universe was to a good approximation in thermal equilibrium
and the transition took place about a microsecond after
the big bang, where the temperature dropped to the order of 
$100 \MeV$. 

A promising prospect is to reproduce QCD phase transitions in  
{\bf heavy--ion collisions} in the laboratory. Large efforts in
ultra--relativistic heavy ion collision experiments focus on the observation
of signatures for a high temperature and/or density transition \cite{QGP}.
Intensive searches have been performed at the AGS accelerator (BNL) and 
at the CERN SPS accelerator and soon also at the RHIC collider and
the future LHC. In a relativistic heavy--ion collision, one 
may create a region of the high temperature or density phase in 
approximate local equilibrium. Depending on the initial density 
and temperature, when this region expands and cools it will traverse the 
phase transition at different points in the phase diagram.
A main challenge is to find distinctive, 
qualitative signatures that such a transition has indeed occured
during the course of the collision. 

Certain signatures rely on the observation that near a phase transition
long--range correlations can occur. We are familiar with this phenomenon
from condensed matter systems. An example is a ferromagnet which is 
heated above a certain temperature $T_c$ where its magnetization
disappears. The phase transition is second order and in the vicinity 
of $T_c$ the correlation length between spins grows very large.
Points in the phase diagram which correspond to second order phase 
transitions are called {\bf critical points}. It can be argued 
\cite{PW84-1,RaWi93-1,Raj95-1}
that in the theoretical limit of two massless quark flavors there is a line
of critical points in the phase diagram of QCD (cf.\ the dashed line
in the above phase diagram). 
Indeed, in the real world there are  
two quarks, the $u$ and the $d$, which are particularly light. 
A large correlation length may then be responsible \cite{RaWi93-1,Raj95-1} 
for the creation of large domains, in which the pion field has a 
non--zero expectation value pointing in a fixed direction in
isospin space (``disoriented chiral condensate'').
The observational consequences would be 
strong fluctuations in the number ratio of neutral 
to charged pions \cite{RaWi93-1,Raj95-1}. The light quarks,
however, are not massless and it is a quantitative question if the correlation
length is long enough to allow for distinctive signatures in a heavy--ion
collision. 

Most strikingly, there is the possibility of a truely infinite 
correlation length for a particular temperature and density even
for realistic quark masses. The corresponding critical point in
the phase diagram marks the endpoint of a line of first order phase
transitions. A familiar analogy in
condensed matter physics comes from the
boiling of water. For the liquid--gas system there exists an endpoint
of first order transitions in the phase diagram where the liquid and
the gaseous phase become indistinguishable and which exhibits the 
well-known phenomenon of critical opalescence. It is the precise
analogue to critical opalescence in QCD at nonzero temperature and
density which has drawn much attention recently \cite{BR,SB,SRS}.
In QCD the corresponding critical point marks the endpoint of a line
of first order transitions in which the quark--antiquark condensate
$\langle \bar{\psi} \psi\rangle$ drops discontinuously as the 
temperature or density is increased. In the theoretical limit of vanishing
quark masses this critical endpoint becomes a {\bf tricritical point} 
(cf.\ the above figure).
We note that a critical endpoint is also known for the 
nuclear gas--liquid transition for a temperature of about $10 \MeV$. 
Signatures and critical 
properties of this point have been studied through measurements
of the yields of nuclear fragments in low energy heavy ion collisions
\cite{CK86-1,EMU98-1}.  

Apart from the phenomenological
implications, a large correlation length near a critical point opens the  
possibility that the QCD phase transition is characterized by 
universal properties. The notion of {\bf universality}
for critical phenomena is well--established in statistical physics. 
Universal properties are independent of the details,
like short distance couplings, of the model under investigation.
They only depend on the symmetries, the dimensionality of space
and the field content. 
As a consequence a whole class of models 
is described by the same universal scaling form of the equation of state
in the vicinity of the critical point. The range of 
applicability typically covers very different physical systems
in condensed matter physics and high temperature quantum field
theory. The analogy between critical opalescence
in a liquid--gas system and QCD near a critical endpoint is
indeed quantitatively correct: Both systems are in
the universality class of the well--known 
Ising model.\\

The thermodynamics described above is difficult to tackle analytically.
A major problem is that for the relevant length scales the running
QCD gauge coupling $\alpha_s$ is large making a perturbative approach 
unreliable. The universal QCD properties near critical points may
be computed within a much simpler model in the same universality class. 
However, effective couplings near critical points are typically
not small and the physics is characterized by nonanalytic behavior.
Important information near the phase transition is also
nonuniversal, like the critical temperature $T_c$ or the overall
size of a correlation length. The question how small $m_u$ and
$m_d$ would have to be in order to see a large correlation length 
near $T_c$ at low density, and if it is realized for realistic
quark masses, will require both universal and nonuniversal information.

A very promising approach to treat these questions is the use of 
the Wilsonian `exact renormalization group' applied to an {\bf effective 
field theory}.
More precisely, we employ an exact {\bf nonperturbative flow equation}
for a scale dependent effective action $\Gamma_k$~\cite{Wet93-2}, which is
the generating functional of the $1 PI$ Green functions in the presence of
an infrared momentum cutoff $\sim k$.  In the
language of statistical physics, $\Gamma_k$ is a coarse grained free 
energy with a coarse graining length scale $\sim k^{-1}$.
The renormalization group flow for
the average action $\Gm_k$ interpolates between a given short distance or
classical action $S$ and the standard effective action $\Gm$, which is
obtained by following the flow for $\Gm_k$ to $k=0$.
We will investigate in these lectures the QCD phase diagram within 
the {\bf linear quark meson model} for two quark flavors. 
Truncated {nonperturbative flow equations} are derived at nonzero  
temperature and chemical potential. 
Whereas the renormalization group flow leads to 
spontaneous chiral symmetry breaking in vacuum, the symmetry gets restored 
in a second order phase transition at high temperature for vanishing 
quark mass. The description \cite{BJW} covers both the low temperature chiral 
perturbation theory domain of validity as well as the high temperature 
domain of critical phenomena. In particular, we obtain 
a precise estimate of the universal equation of state in the vicinity 
of critical points \cite{BJW,BTW}. We explicitly 
connect the physics at zero temperature and realistic quark mass with 
the universal behavior near the critical temperature $T_c$ and the chiral 
limit. An important property will be the observation that certain
low energy properties are effectively independent of the details
of the model even away from the phase transition. This behavior 
is caused by a strong attraction of the renormalization group
flow to approximate {partial infrared fixed points} 
({\bf infrared stability}) \cite{JW,BJW}. Within this approach
at high density we find \cite{BJW2} a chiral symmetry restoring first order 
transition. As pointed out above these 
results imply the presence of a tricritical point with 
long--ranged correlations in the phase diagram. The lectures 
end with an outlook to 
densities above the chiral transition, where QCD is expected to behave 
as a color superconductor at low temperature.

\section{Coarse graining and nonperturbative flow equations\label{coarseg}}

\subsection{From short to long distance scales}
\label{shorttolong}

\begin{figure}[h]
\begin{center}
\epsfxsize=3.5in
\hspace*{0in}
\epsffile{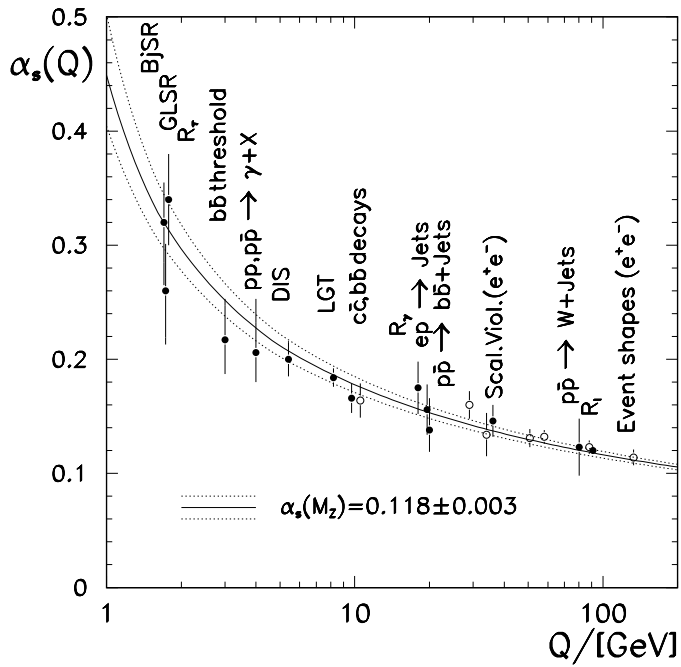}
\end{center}
\caption{Running of the strong gauge coupling 
by various types of measurements compared to theory.}
\label{as}
\end{figure}

Quantum chromodynamics describes qualitatively different physics
at different length scales. The theory is asymptotically free \cite{GWP} 
and at short distances or high energies
the strong interaction dynamics of quarks and gluons can be 
determined from perturbation theory. 
On the other hand, at scales of a few hundred $\MeV$ confinement sets 
in and the spectrum of the theory consists only of color neutral 
states. 
The change of the strong gauge coupling $\alpha_s$ with scale has been
convincingly demonstrated by a comparison of various measurements 
with the QCD prediction as shown in figure~\ref{as}. (Taken
from \cite{Schmelling}.)  
The coupling increases with decreasing momenta $Q$ and
perturbation theory becomes invalid for $Q\, \gtap \, 1.5$GeV. 
Extrapolating QCD from short distance to long distance scales 
is clearly a nonpertur\-ba\-tive problem. 
However, not only effective couplings but also the relevant 
degrees of freedom can change with scale.
Low--energy degrees of freedom in strong interaction physics may
comprise mesons, baryons and glueballs rather than quarks and gluons.
Indeed, at low energies an essential part of strong interaction 
dynamics can be encoded in the masses and interactions of 
mesons. A prominent example of a systematic effective 
description of the low energy behavior of QCD is chiral 
perturbation theory\footnote{See also the lectures of this school from 
C.P.\ Burgess \cite{Burgess} for an introduction and references.} 
\cite{GL82-1}. 
It rather accurately 
describes the dynamics of the lightest hadronic bound states, i.e.\
the Goldstone bosons of spontaneous {chiral symmetry breaking}. 

Spontaneous chiral symmetry breaking is an important building block
for our understanding of the hadron spectrum at low energies. 
The approximate chiral
symmetry, hiding behind this phenomenon, originates in the fact that 
left-- and right--handed components of free massless fermions do not
communicate. In the limit of
vanishing quark masses the classical or short distance QCD 
action does not couple left-- and right--handed quarks. As a consequence 
it exhibits a global chiral invariance 
under $U_L(N_f) \times U_R(N_f) = SU_L(N_f) \times SU_R(N_f) \times 
U_V(1) \times U_A(1)$. Here $N_f$ denotes the number of massless quarks
and the quark fields $\psi$ transform with different unitary 
transformations acting on the left-- and right--handed components
\ben 
\psi_R\equiv\frac{1-\gamma_5}{2}\psi & \longrightarrow &
\Uc_R \psi_R\; ;\;\;\;\Uc_R\in U_R(N)\, , \nnn
\psi_L\equiv\frac{1+\gamma_5}{2}\psi & \longrightarrow & 
\Uc_L \psi_L\; ;\;\;\;\Uc_L\in U_L(N)\; .
\een 
In reality, the quark masses differ from zero and the chiral symmetry 
is only an approximate symmetry. There are two especially light quark
flavors with similar quark masses, $m_u \simeq m_d$. If the masses
of $u$ and $d$ are taken to be the same they form an isospin doublet.
The corresponding approximate isospin symmetry of the QCD action is manifest
at low energies in the observed pattern of bound states which occur in
nearly degenerate multiplets: $(p,n)$, $(\pi^+,\pi^0,\pi^-),\ldots$    
Even though the strange quark $s$ is much heavier than $u$ and $d$
the associated symmetry for three degenerate 
flavors, termed the ``eightfold way'', can be observed as mesonic
and baryonic levels grouped into multiplets of $SU(3)$ --- singlets,
octets, decuplets. The other flavors $c,b,t$ remain singlets. 
If chiral symmetry was realized in the same manner, the energy
levels would appear approximetely as multiplets of   
$SU_L(2) \times SU_R(2)$, or $SU_L(3) \times SU_R(3)$, respectively.
The multiplets would then necessarily contain members of 
opposite parity which is not observed in the hadron spectrum.
Furthermore the pion mass is small compared to the masses of
all other hadrons. Together this indicates
that the approximate chiral symmetry 
$SU_L(N_f) \times SU_R(N_f)$ with $N_f=2$ or $3$ 
is spontaneously broken to the diagonal $SU_V(N_f)$ vector--like
subgroup with the pions as the corresponding Goldstone bosons.\\ 

\vspace*{0.1in}
\setlength{\unitlength}{1cm}
\begin{picture}(18,5)
\put(3.8,0){$SU_V(N_f)\quad\, \,\, \times\quad\,\, \,\, U_V(1)$}
%\put(8.,0){$\,\, \, \times$}
%\put(10.,0){$\, \, U_V(1)$}
\put(3.3,3.){spontaneously}
\put(4.,2.2){broken}
\put(7.,3){baryon}
\put(7.,2.2){number}
\put(9.4,3){broken by}
\put(9.4,2.2){chiral anomaly}
%\put(0.5,5){\tb UV \ts}
\put(1.8,5){$\qquad\! SU_L(N_f)\, \, \times \, \, SU_R(N_f) \, \, \times 
\, \, U_V(1) \, \, \times \, \, U_A(1)$}
\thicklines{
\put(4.6,1.9){\vector(0,-1){1}}
\put(7.7,1.9){\vector(0,-1){1}}
\put(7.7,4.7){\line(0,-1){1}}
\put(10.,4.7){\line(0,-1){1}}
\put(3.6,4.6){\line(1,-1){1}}
\put(5.6,4.6){\line(-1,-1){1}}
}
\end{picture}\\

\noindent
In addition, the axial abelian subgroup $U_A(1)$ of the classical QCD
action is broken in the quantum 
theory by an anomaly of the axial--vector current \cite{Ho}. 
This breaking proceeds without the occurence of a Goldstone boson.
The abelian $U_V(1)$ subgroup corresponds to baryon number 
conservation.

\subsection{Kadanoff--Wilson renormalization group}

A conceptually very appealing approach to bridge the gap between
the short distance and the long distance physics
relies on the general ideas of the Kadanoff--Wilson renormalization
group \cite{Kad66-1,Wil71-1}. The renormalization 
group method consists in systematically 
reducing the number of degrees of freedom by integrating over
short wavelength fluctuations. For an example, one may consider
a spin system on a lattice with lattice spacing $a$.  A possible
strategy for integrating over short wavelength fluctuations is to form
blocks of spins. 
\begin{figure}[h]
\begin{center}
\epsfxsize=3.5in
\hspace*{0in}
\epsffile{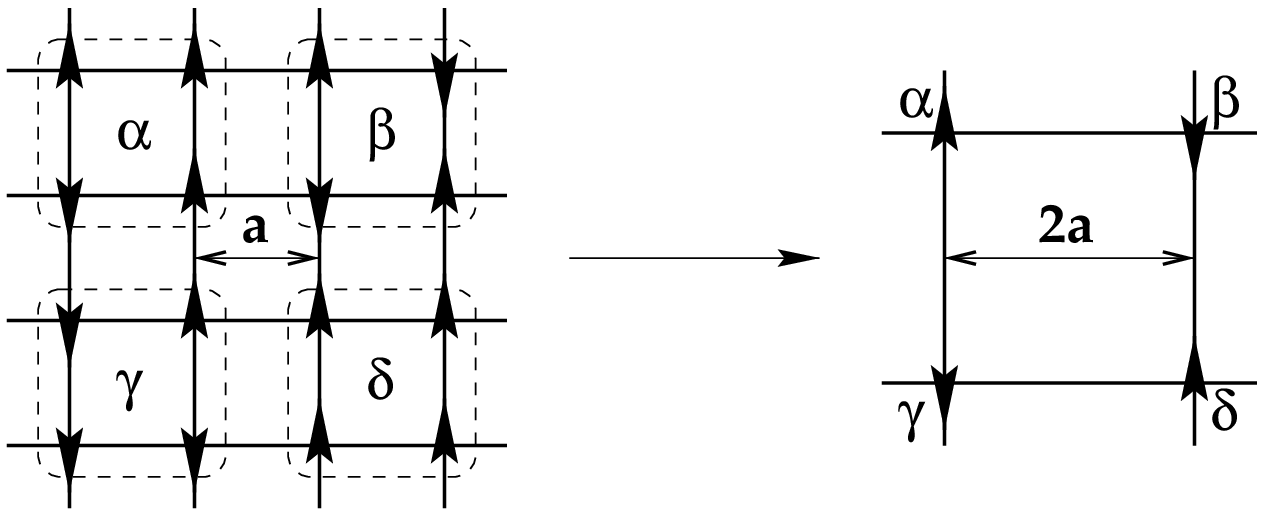}
\caption{}
\end{center}
\end{figure}
In figure 2 the spins are grouped by fours
and to each block one attributes an appropriate ``average'' spin. 
One may visualize such a procedure by imagining the spin system 
observed through a microscope with two different resolutions.
At first one uses a relatively high resolution, the minimum wavelength
for fluctuations is $\simeq a$. Subsequently one uses a longer 
wavelength, permitting the observation of details with dimension
$\simeq 2a$. The resolution becomes less since details having dimension
between $a$ and $2 a$ have been integrated out. As a result one obtains
an effective description for averages of spins on the sites of a coarse 
grain (block) lattice, incorporating the effects of the short
wavelength fluctuations. One should note that for 
the averaging procedure ($a \to 2 a$) fluctuations on scales larger 
than the coarse graining length scale ($2 a$) play no role. 
This observation can be used to obtain a relatively simple 
coarse graining description also in the case of a quantum 
field theory which will be considered in the following.

\subsection{Average action}

I will introduce here the {average action} $\Gamma_k$ \cite{Wet91-1} which 
is based on the quantum field theoretical concept of the effective
action $\Gamma$, i.e.\ the generating functional of 
the $1 PI$ Green functions. 
The field equations derived from the 
{\em effective action} include all quantum effects. 
In thermal and chemical equilibrium
$\Gamma$ includes in addition the thermal 
fluctuations and depends on the temperature and chemical potential. In
statistical physics $\Gamma$ corresponds to the free energy as a functional
of some (space dependent) order parameter. 
\begin{itemize}
\item {The average action $\Gamma_k$ is a simple
generalization of the effective action, with the distinction that only
fluctuations with momenta $q^2\, \gtap\, k^2$ are included.} 
\end{itemize}
This is achieved
by implementing a momentum infrared cutoff $\sim k$ into the functional 
integral which defines the effective action $\Gamma$. In the
language of statistical physics, $\Gamma_k$ is a coarse grained free 
energy with a coarse graining length scale $\sim k^{-1}$. 
Lowering $k$ results in 
a successive inclusion of fluctuations with momenta
$q^2\, \gtap\,\, k^2$ and therefore permits to explore the theory on 
larger and larger length scales. The average action $\Gamma_k$ can be viewed
as the effective action for averages of fields over a volume
with size $k^{-d}$ and is similar in spirit to the action for block--spins
on the sites of a coarse lattice.  
By definition, the average action equals the standard effective action 
for $k=0$, i.e.\ $\Gamma_{0}=\Gamma$, since the infrared cutoff is absent
and therefore all fluctuations are included. On the other hand, 
in a theory with a physical 
ultraviolet cutoff $\Lambda$ we
can associate $\Gamma_\Lambda$ with the microscopic or 
classical action $S$ 
since no fluctuations below $\Lambda$ are effectively included.
Thus the average action has the important property that 
\begin{itemize}
\item {$\Gamma_k$ interpolates between the classical 
action $S$ 
and the effective action $\Gamma$ as $k$ is lowered from the ultraviolet
cutoff $\Lambda$ to zero:} 
$\lim\limits_{k \to \Lambda} \Gamma_k = S$ , 
$\,\, \lim\limits_{k \to 0} \Gamma_k = \Gamma$ .
\end{itemize}
The ability to follow the evolution to $k\ra0$ is equivalent to the
ability to solve the theory. Most importantly, the dependence of the
average action on the scale $k$ is described by an exact
nonperturbative flow equation which is presented in the next section.\\

Let us consider the construction of $\Gamma_k$ for a 
simple model with real scalar fields $\chi_a$, $a=1 \ldots N$, in $d$ 
Euclidean dimensions with classical action $S$ and sources $J_{a}$. 
We start with the path integral representation of the 
generating functional for the connected Green functions 
\be
W_k[J]=\ln \int D \chi \exp\left(-S_{}[\chi]-\Dt S_k[\chi]+
\int d^dx J_a(x)\chi^a(x) \right)\; ,
\label{genfunc}
\ee
where we have added to the classical action an infrared (IR)
cutoff term $\Dt S_k[\chi]$ which is quadratic in the fields and reads
in momentum space
\be
\Dt S_k[\chi]=\hal \int 
\frac{d^dq}{(2\pi)^d} R_k(q)\chi_a(-q)\chi^a(q).
\ee
Here the infrared cutoff function $R_k$ is required to vanish
for $k \to 0$ and to diverge for $k \to \Lambda$
and fixed $q^2$.
For $\Lambda \to \infty$ this can be achieved, for example, by the 
exponential form
\be
 R_k(q) \sim \frac{q^2}{\displaystyle{e^{q^2/k^2} - 1}}
 \label{Rk(q)}.
\ee
For fluctuations with small momenta
$q^2\ll k^2$ this cutoff behaves as $R_k(q^2)\sim k^2$ and allows 
for a simple interpretation. Since 
$\Dt S_k[\chi]$ is quadratic in the fields, this means
that all Fourier modes of $\chi$ with momenta smaller than 
$k$ acquire an effective mass $\sim k$. This additional mass term
acts as an effective IR cutoff for the low momentum modes. 
In contrast, for $q^2\gg k^2$ the function $R_k(q^2)$ vanishes 
such that the functional integration of the high momentum modes
is not disturbed. The term $\Dt S_k[\chi]$ added to the classical
action is the main ingredient for the construction of an effective
action that contains all fluctuations with momenta $q^2\, \gtap\, k^2$
whereas fluctuations with $q^2\, \ltap\, k^2$ are suppressed.

The expectation value of $\chi$, i.e.\ the {\em macroscopic field} $\phi$, 
in the presence of $\Dt S_k[\chi]$ and $J$ reads 
\be
\label{classicfield}
\phi^a(x) \equiv \langle\chi^a(x)\rangle = 
\frac{\dt W_k[J]}{\dt J_a(x)}.
\ee
We note that the relation between $\phi$ and $J$ is
$k$--dependent, $\phi=\phi_k(J)$ and therefore \mbox{$J=J_k(\phi)$}.
In terms of $W_k$ the average action is
defined via a modified Legendre transform
\be
\label{GaDef}
\Gm_k[\phi]=-W_k[J]+\int d^dx J_a(x)\phi^a(x)-\Dt S_k[\phi]
\ee
where we have subtracted the term $\Dt S_k[\phi]$ on the r.h.s.  
This subtraction of the infrared cutoff term as a function of the
macroscopic field $\phi$ is crucial for the definition of a
reasonable coarse grained free energy with the property
$\lim_{k\to \Lambda} \Gamma_k=S$. It guarantees
that the only difference between $\Gamma_k$ and $\Gamma$ is the
effective infrared cutoff in the fluctuations.\\

\vspace*{0.2in}
\hrule
\vspace*{0.2in}

{\sc Exercise}: Check of the properties $\quad$
(i) $\lim\limits_{k\to 0} \Gamma_k = \Gamma$, $\quad$
(ii) $\lim\limits_{k\to \Lambda} \Gamma_k = S$.
\begin{itemize}
\item[(i)] The first property follows immediately  with 
$\lim_{k \to 0} R_k = 0$ from the absence of any IR
cutoff term in the above (standard) definitions for the effective action. 
\item[(ii)]
To establish the property $\Gamma_{\Lambda}=S$ we consider an
integral equation for $\Gamma_k$ which is equivalent to (\ref{GaDef}).
In an obvious matrix notation we use (\ref{genfunc})
\ben
\exp\Big(W_k[J]\Big)&=& \int D \chi \exp\left(-S[\chi]+\int J \chi 
-\frac{1}{2}\int \chi\, R_k \chi \right)\; .
\nonumber
\een
Eliminating in this equation $W_k$ and $J$ with (\ref{GaDef}) 
\ben
-W_k[J]=\Gamma_k[\phi]+\frac{1}{2}\int \phi\, R_k\, \phi - \int J \phi
&\quad \Rightarrow \quad&
J=\frac{\dt \Gamma_k}{\dt \phi}+\phi\, R_k 
\nonumber
\een
we obtain the integral equation
\be
\label{IntegralEquation}
\exp(-\Gamma_k[\phi])=\ds{\int D \chi \exp\left(-S[\chi]
+\int \frac{\dt \Gamma_k}{\dt \phi}
[\chi -\phi]\right)} \ds{\exp\left( -\frac{1}{2}
[\chi -\phi]\, R_k\, [\chi -\phi]  \right)}
\; .
\ee
For $k \to \Lambda$ the cutoff function $R_k$ diverges and  
the term $\exp( -[\chi -\phi] R_k [\chi -\phi]/2) $ behaves 
as a delta functional $\sim \dt[\chi-\phi]$, thus leading to the 
property $\Gamma_k \to S$
in this limit.
\end{itemize}

\vspace*{0.1in}
\hrule
\vspace*{0.3in}

Let us point out a few properties of the average action: 
\begin{enumerate}
\item All symmetries of the model which are respected by the IR cutoff
  term $\Delta S_k$ are also symmetries of $\Gamma_k$. In
  particular, this concerns translation and rotation invariance.
%  In consequence, $\Gamma_k$ may be expanded in terms of invariants
%  with respect to these symmetries with couplings depending on $k$. 
\item The construction of the average action can be easily 
  generalized to fermionic degrees of freedom. In particular, it is possible 
  to incorporate chiral fermions since a
  chirally invariant cutoff $R_k$ can be formulated~\cite{Wet90-1}.
\item Gauge theories can be formulated along similar
  lines~\cite{RW93-1,Bec96-1,BAM94-1,EHW94-1} even
  though $\Delta S_k$ may not be gauge invariant. In this case the
  usual Ward identities receive corrections for which one can derive
  closed expressions~\cite{EHW94-1}. These corrections vanish for
  $k \to 0$.
\item Despite the similar spirit one should note the
  difference in viewpoint to the Kadanoff--Wilson block--spin
  action \cite{Kad66-1,Wil71-1}. The Wilsonian effective 
  action realizes that physics 
  with a given characteristic length scale $l$ can be conveniently 
  described by a
  functional integral with an ultraviolet cutoff $\Lambda$ for the
  momenta where $l^{-1}$ should be smaller than $\Lambda$, but not
  necessarily by a large factor. The Wilsonian effective action
  $S_\Lambda^{\rm W}$ replaces then the classical action in the
  functional integral. It is obtained by integrating out the
  fluctuations with momenta $q^2\gtap\, \Lambda^2$. 
  The $n$--point functions have to be computed from $S_\Lambda^{\rm W}$ by
  further functional integration and are independent of
  $\Lambda$. 

  In contrast, the average action $\Gamma_k$ realizes the concept of a
  coarse grained free energy with a coarse graining length scale 
  $\sim k^{-1}$. For each value of $k$
  the average action is related to the generating functional
  of a theory with a different action $S_k=S+\Delta_k S$. 
  The $n$--point functions derived from $\Gamma_k$ depend on $k$.
  To obtain $k$--independent $n$--point functions their
  characteristic momentum scale $l^{-1}$ should be much larger than
  $k$. The standard effective 
  action $\Gm$ is obtained by following the flow for $\Gm_k$ to $k=0$, 
  thus removing the infrared cutoff in the end. The Wilsonian effective 
  action does not generate the $1PI$ Green functions~\cite{KKS92-1}.\\

\end{enumerate}

\subsection{Exact flow equation}

The dependence of the average action $\Gamma_k$ on the 
coarse graining scale $k$ is
described by an {exact nonperturbative flow equation} ~\cite{Wet93-2}
\begin{equation}
  \frac{\partial}{\partial t}\Gm_k[\phi] =  \hal\Tr\left\{\left[
  \Gm_k^{(2)}[\phi]+R_k\right]^{-1}\frac{\partial}{\partial t}
  R_k\right\} \; . 
  \label{ERGE}
\end{equation}
Here $t=\ln(k/\La)$ denotes the logarithmic scale variable with 
some arbitrary momentum scale $\La$. 
The trace involves only one integration as well as a summation over internal
indices, and in momentum space it reads 
$\Tr=\sum_a \int d^dq/(2\pi)^d$ for the $a=1,\ldots,N$ component 
scalar field theory.
The exact flow equation describes
the scale dependence of $\Gamma_k$ in terms of the inverse
average propagator $\Gm_k^{(2)}$ as given by
the second functional derivative of $\Gm_k$ with respect 
to the field components
\begin{equation}
  \left(\Gamma_k^{(2)}\right)_{a b}(q,q^\prime)=
  \frac{\delta^2\Gamma_k}
  {\delta\phi^a(-q)\delta\phi^b(q^\prime)}\; . 
\end{equation}

\vspace*{0.2in}
\hrule
\vspace*{0.2in}

{\sc Exercise}: Derivation of the exact flow equation (\ref{ERGE}).
\begin{itemize}
\item[] Let us write 
\ben
\Gamma_k[\phi]&=&\tilde{\Gamma}_k[\phi]-\Delta S_k[\phi]
\label{GtG}
\een
where according to (\ref{GaDef})
\ben
\tilde{\Gamma}_k[\phi]=-W_k[J]+\int d^dx J(x)\phi(x) 
\een
and $J=J_k(\phi)$. We consider for simplicity a one--component
field and derive first the scale dependence of $\tilde{\Gamma}$:
\ben
\frac{\partial}{\partial t} \tilde{\Gamma}_k[\phi] &=&
-\left(\frac{\partial W_k}{\partial t}  \right) [J]
-\int d^dx \frac{\delta W_k}{\delta J(x)}
\frac{\partial J(x)}{\partial t} + \int d^dx \phi(x) 
\frac{\partial J(x)}{\partial t} \; .
\label{tildeG}
\een
With $\phi(x)=\delta W_k/\delta J(x)$ the last two terms 
in (\ref{tildeG}) cancel. The $t$--derivative 
of $W_k$ is obtained from its defining functional integral 
(\ref{genfunc}) and yields
\ben
\frac{\partial}{\partial t} \tilde{\Gamma}_k[\phi] &=&
\langle \frac{\partial}{\partial t}\Delta S_k[\chi] \rangle =
\langle \frac{1}{2} \int d^dx d^dy \chi(x) 
\frac{\partial}{\partial t}R_k(x,y) \chi(y) \rangle \; . 
\label{tildeG2}
\een
where $R_k(x,y)\equiv R_k(-\partial^2_x) \delta(x-y)$. Let
$G(x,y)=\delta^2 W_k/\delta J(x)\delta J(y)$
denote the connected $2$--point function and decompose
\ben
\langle \chi(x) \chi(y) \rangle = G(x,y) + \langle \chi(x) \rangle
\langle \chi(y) \rangle \equiv G(x,y) + \phi(x) \phi(y) \; .
\een
Plugging this decomposition into (\ref{tildeG2}) the scale
dependence of $\tilde{\Gamma}_k$ can be expressed as
\ben
\frac{\partial}{\partial t} \tilde{\Gamma}_k[\phi] &=&
\frac{1}{2} \int d^dx d^dy \left\{ G(x,y)  \frac{\partial}{\partial t} 
R_k(x,y) 
+ \phi(x) \frac{\partial}{\partial t} R_k(x,y) \phi(y) \right\} \nnn
&\equiv& 
\frac{1}{2} \Tr\left\{ G \frac{\partial}{\partial t} R_k\right\}
+\frac{\partial}{\partial t} \Delta S_k[\phi] \; . 
\een
The exact flow equation for the average action $\Gamma_k$ 
follows now with (\ref{GtG})
\ben
\frac{\partial}{\partial t} \Gamma_k[\phi] &=&
\frac{1}{2} \Tr\left\{ G \frac{\partial}{\partial t} R_k\right\} \nnn
&=& \frac{1}{2} \Tr\left\{ \left[
  \Gm_k^{(2)}[\phi]+R_k\right]^{-1} \frac{\partial}{\partial t} R_k
\right\} 
\een
where we have used that 
$\tilde{\Gamma}_k^{(2)}(x,y)\equiv 
\delta^2\tilde{\Gamma}_k/\delta\phi(x)\delta\phi(y)=
\delta J(x)/\delta \phi(y)$
is the inverse of $G(x,y)\equiv \delta^2 W_k/\delta J(x)\delta J(y)=
\delta \phi(x)/\delta J(y)$ to obtain the last equation. 
\end{itemize}

\vspace*{0.1in}
\hrule
\vspace*{0.3in}

Let us point out a few properties of the exact flow equation:
\begin{enumerate}
\item The flow
equation (\ref{ERGE}) closely resembles a one--loop equation.
Replacing $\Gamma_k^{(2)}$ by the second functional derivative of the
classical action, $S^{(2)}$, one obtains the corresponding one--loop
result. Indeed, the one--loop formula for
$\Gamma_k$ reads
\begin{equation}
  \label{AAA72}
  \Gamma_k[\phi]=S[\phi]+
  \frac{1}{2}\Tr\ln\left(
  S^{(2)}[\phi]+R_k\right)
\end{equation}
and taking a $t$--derivative of
(\ref{AAA72}) gives a one--loop flow equation very similar to
(\ref{ERGE}). The ``renormalization
group improvement'' $S^{(2)}\ra\Gamma_k^{(2)}$ turns the one--loop
flow equation into an exact nonperturbative flow equation
which includes
the effects from all loops.
Replacing the propagator and vertices appearing in
$\Gamma_k^{(2)}$ by the ones derived from the classical
action, but with running $k$--dependent couplings, and expanding
the result to lowest non--trivial order in the coupling constants one
recovers standard renormalization group improved one--loop
perturbation theory.
%\item
%It also turns the equation into a functional 
%differential equation. Possible methods for its solution include
%standard perturbation theory in the case of a small coupling,
%the $1/N$--expansion \cite{} or the $\epsilon$-expansion \cite{}.
%Particularly suitable for our purposes is the 
%derivative expansion.
\item The additional cutoff function $R_k$ with a form like
the one given in eq.\ (\ref{Rk(q)}) renders the momentum integration 
implied in the trace of (\ref{ERGE}) both 
infrared and ultraviolet finite. In particular, the direct 
implementation of
the additional mass--like term $R_k \sim k^2$ for $q^2 \ll k^2$
into the inverse average propagator makes the formulation suitable
for dealing with theories which are plagued by infrared problems
in perturbation theory. We note that
the derivation of the exact flow equation does not 
depend on the particular choice of the cutoff function. 
Ultraviolet finiteness, however, is related
to a fast decay of $\partial_t R_k$ for $q^2\gg k^2$. 
If for some other
choice of $R_k$ the right hand side of the flow equation would not
remain ultraviolet finite this would indicate 
that the high momentum modes have
not yet been integrated out completely in the computation of
$\Gamma_k$. Unless stated otherwise we will always assume a
sufficiently fast decaying choice of $R_k$ in the following.  

Of course, the particular choice for the infrared cutoff function 
should have no effect on the physical results for $k \to 0$.
Different choices of $R_k$ correspond
to different trajectories in the space of effective actions along
which the unique infrared limit $\Gamma_0$ is reached.
Nevertheless, once approximations are applied not
only the trajectory but also its end point may depend on the precise
definition of the function $R_k$. This dependence may be used
to study the robustness of the approximation.
\item
Flow equations for $n$--point functions can be easily obtained
from (\ref{ERGE}) by differentiation. The flow equation for the
two--point function $\Gamma_k^{(2)}$ involves the three and
four--point functions, $\Gm_k^{(3)}$ and $\Gm_k^{(4)}$, respectively:
\ben
\frac{\partial}{\partial t} \Gamma_k^{(2)}(q,q) &=&
\frac{\partial}{\partial t} \frac{\partial^2\Gamma_k}
  {\partial\phi(-q)\partial\phi(q)}\nnn
&=& \frac{1}{2} \Tr \left\{ \frac{\partial R_k}{\partial t}
\frac{\partial}{\partial \phi(-q)} (-1) 
\left[\Gm_k^{(2)}+R_k\right]^{-1} \Gm_k^{(3)} 
\left[\Gm_k^{(2)}+R_k\right]^{-1}\right\}\nnn
&=& \Tr \left\{ \frac{\partial R_k}{\partial t}
\left[\Gm_k^{(2)}+R_k\right]^{-1} \Gm_k^{(3)} 
\left[\Gm_k^{(2)}+R_k\right]^{-1} \Gm_k^{(3)} 
\left[\Gm_k^{(2)}+R_k\right]^{-1}\right\}\nnn
&-&\frac{1}{2} \Tr \left\{ \frac{\partial R_k}{\partial t}
\left[\Gm_k^{(2)}+R_k\right]^{-1} \Gm_k^{(4)} 
\left[\Gm_k^{(2)}+R_k\right]^{-1}\right\} \; .
\een
In general, the flow equation for $\Gm_k^{(n)}$ involves
$\Gm_k^{(n+1)}$ and $\Gm_k^{(n+2)}$. 
\item 
We emphasize that the flow equation (\ref{ERGE}) is equivalent 
to the Wilsonian exact renormalization group 
equation \cite{Wil71-1,WH73-1,NC77,Wei76-1,Pol84-1,Has86-1}.  
The latter describes how
the Wilsonian effective action $S_\Lambda^{\rm W}$ changes with the 
ultraviolet cutoff $\Lambda$. Polchinski's continuum
version of the Wilsonian flow
equation~\cite{Pol84-1} can be transformed into eq.\ (\ref{ERGE}) 
by means of a Legendre transform and a suitable variable
redefinition~\cite{BAM,BAM93-1}.\footnote{See also the contribution
from C.\ Kim \cite{Kim} this school.}
\item
Extensions of the
flow equations to gauge fields 
\cite{RW93-1,Bec96-1,BAM94-1,EHW94-1} and 
fermions~\cite{Wet90-1,CKM97-1} are
available.  
\end{enumerate}

\subsection{Elements of effective field theory}

A strict derivation of an effective low energy description from QCD is still 
missing. Yet, predictions from effective low energy models often show
convincing agreement with the results of real and numerical experiments.
Let us consider some important aspects for the success of an effective 
field theory description. They find a natural theoretical basis in the 
framework of the average action.

1) {\bf Decoupling of ``heavy'' degrees of freedom} --- 
At any scale $k$ only fluctuations with momenta in a small range around $k$
influence the renormalization group flow of the average action $\Gamma_k$.
This expresses the fact that the momentum integration implied by the
trace on the r.h.s.\ of 
the exact flow equation (\ref{ERGE}) is dominated by momenta 
$q^2 \simeq k^2$, schematically

\vspace*{0.2in}
$\qquad\qquad$ $\displaystyle{\frac{\partial \Gamma_k}
{\partial k}\,\,\,\, =\,\,\,\,\frac{1}{2}\,\,\,\,}$
\parbox{1.in}{
\epsfxsize=.4in
\epsffile{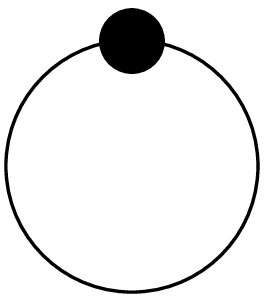}}$\qquad\qquad\qquad\quad$
\parbox{1.in}{
\epsfxsize=1.8in
\epsffile{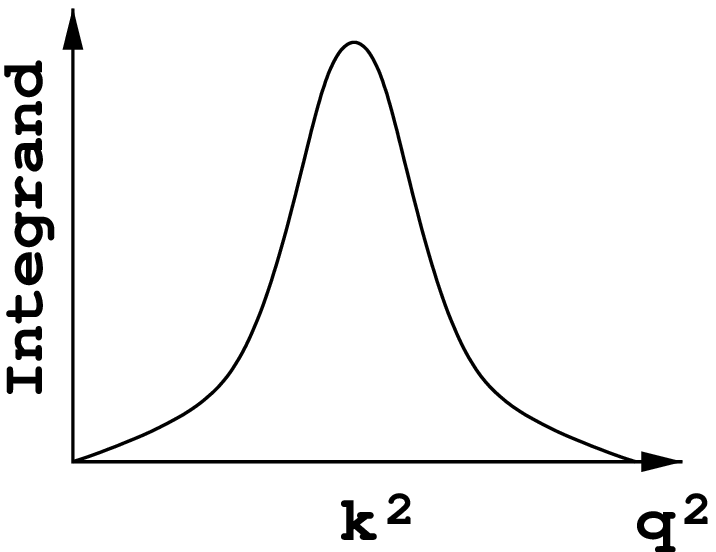}} 
\vspace*{0.2in}

\noindent
with the full $k$--dependent propagator associated to the propagator
line and the dot denotes the insertion $\partial_t R_k$. The effects of
the high momentum modes with $q^2\, \gtap\, k^2$ determine the precise 
form of the average action $\Gamma_k$ at the scale $k$. 
Modes with $q^2\, \gtap\, k^2$ have been ``integrated out''.
In particular, for any given $k=k_0$
their only effect is to determine the initial value $\Gamma_{k_0}$ for 
the solution of the flow 
equation for the low momentum fluctuations with $q^2\, \ltap\, k_0^2$. 
%The effects of the low momentum fluctuations can, therefore, be 
%described in terms of an effective theory where the high momentum modes
%have been integrated out.

Most importantly, for each 
range of $k$ only those degrees of freedom have to be included which are 
relevant in the corresponding momentum range. 
In QCD these may comprise compounds of quarks and gluons. ``Heavy''
degrees of freedom effectively decouple from the flow of
$\Gamma_k$ once $k$ drops below their mass $m$, since $m$ represents
a physical infrared cutoff for fluctuations with momenta 
$q^2 \simeq k^2\, \ltap\, m^2$. We will observe in section 
\ref{FlowEquationsAndInfraredStability} the occurence of mass 
threshold functions, which explicitly describe the decoupling
of heavy modes, as one of the important nonperturbative ingredients 
of the flow equations.

As a prominent example for an effective field theory, 
chiral perturbation theory describes the IR behavior 
of QCD in terms of the lightest mesons, i.e.\ the Goldstone bosons of 
spontaneous chiral symmetry breaking. This yields a very successful 
effective formulation of strong
interactions dynamics for momentum scales up to a few hundred $\MeV$.
For somewhat higher scales additional
degrees of freedom like the sigma meson or the light quark flavors will
become important and should be included explicitly. The linear
quark meson model based on these degrees of freedom will be introduced 
in the next section. 

2) {\bf Infrared stability} --- The predictive power of an effective
low energy description crucially depends on how sensitively the infrared value
$\Gamma=\lim_{k\to 0}\Gamma_k$ depends on the initial value 
$\Gamma_{k_0}$. Here $k_0$ plays the role of an ultraviolet cutoff 
scale of the low energy description. 
Indeed, as the coarse graining 
scale $k$ is lowered from $k_0$ to zero, the ``resolution'' is smeared 
out and the detailed information of the short
distance physics can be lost. (On the other hand, the ``observable
volume'' is increased and long distance aspects such as collective
phenomena become visible.) 

There is a prominent example for insensitivity of long
distance properties to details at short distances: Systems of
statistical mechanics near {\em critical points}, where they 
undergo a second order phase transition. For the example of
a spin system near its critical temperature the equation of state,
which relates for a given temperature the magnetization to an
external magnetic field, is independent of the microscopic details 
up to the short distance value of two parameters. These can be related to the 
deviation from the critical temperature and to the deviation 
from a nonzero magnetic field. Stated differently,
only two parameters of the short distance effective action 
$\Gamma_{k_0}$ have to be
``finetuned'' in order to be at the phase transition.
These few {\em relevant} parameters are typically accompanied by 
many {\em irrelevant} parameters which parametrize 
$\Gamma_k$.\footnote{For a more detailed introduction and a systematic 
classification of operators as relevant, marginal or irrelevant see 
e.g.\ ref.\ \cite{Bel}.}  
As the coarse graining scale $k$ is lowered
from $k_0$ to zero, the running (appropriately rescaled) 
irrelevant parameters are attracted to a {\bf fixed point}, 
whereas relevant parameters are driven away
from this point. Figure 3 shows, schematically, the vicinity
of a fixed point for the case of one irrelevant parameter and 
one relevant parameter.  
\begin{figure}[h]
\begin{center}
\epsfxsize=2.2in
\hspace*{0in}
\epsffile{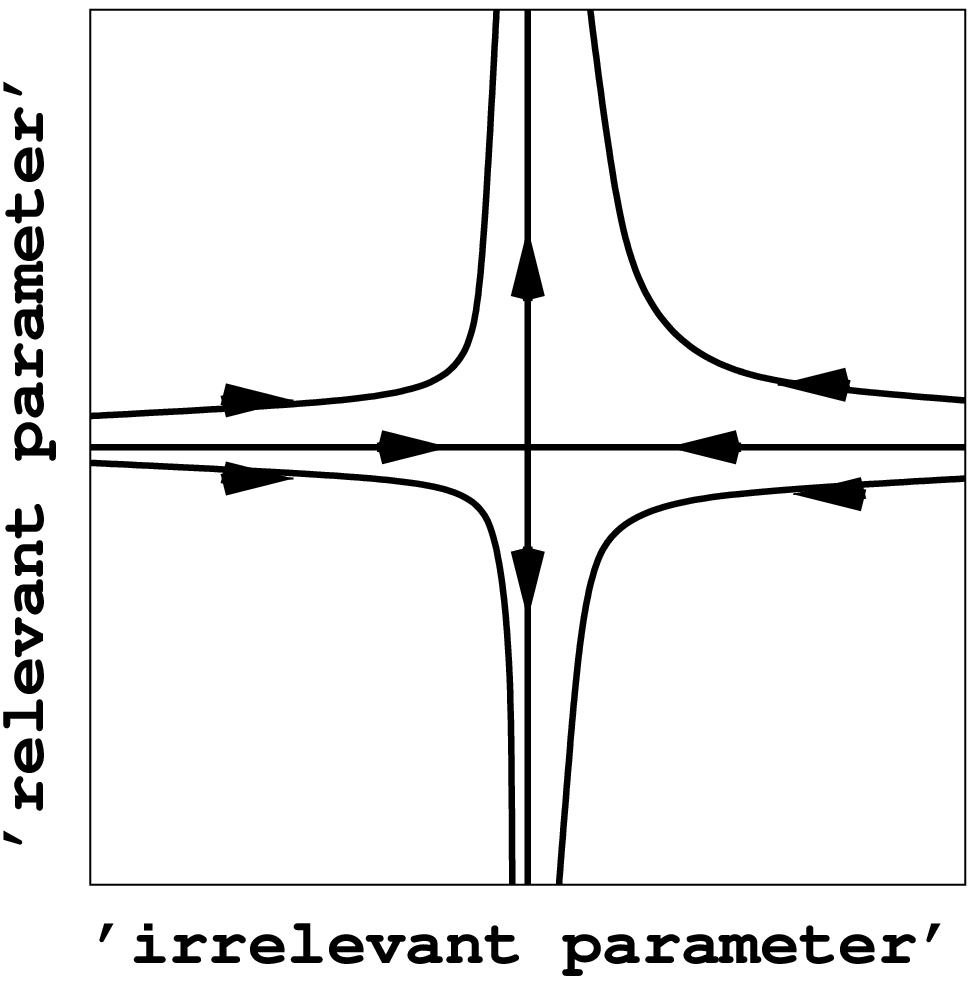}
\end{center}
\caption{}
\end{figure}
\vspace*{-0.1in}
The arrows 
indicate the renormalization group flow. 
Since the irrelevant parameter is driven to the 
same fixed point irrespective of its initial value (once the relevant 
parameter is tuned to criticality), the short distance 
information about it gets lost in the infrared. This behavior
is crucial for the phenomenon of universality for critical phenomena
(cf.\ section \ref{CriticalBehavior}).

We will observe 
within the linear quark meson an important different example
of the insensitivity of physics at long distances to short 
distance details even away from the phase transition.
A crucial observation is the strong attraction of 
the renormalization group flow to approximate fixed points. We will discuss
this remarkable property in section \ref{FlowEquationsAndInfraredStability}.

3) {\bf Symmetries} --- Symmetries constrain the possible form of an 
effective model. 
It is an important property that the coarse
grained free energy $\Gamma_k$
respects all symmetries of the model, in particular rotation and
translation symmetries. In consequence, $\Gamma_k$ can be expanded in terms of
invariants with respect to the symmetries with couplings depending
on $k$. We will frequently employ an expansion of 
$\Gamma_k$ where the invariants are ordered according to the number of 
derivatives of the fields. For the example of the $O(N)$--symmetric scalar 
theory this yields $(a=1,\ldots,N)$
  \begin{equation}
    \label{AAA63}
    \Gamma_k=\int d^d x\left\{
    U_k(\rho)+\frac{1}{2}Z_{k}(\rho)
    \partial^\mu\phi_a\partial_\mu\phi^a+
    \frac{1}{4}Y_{k}(\rho)
    \partial^\mu\rho\partial_\mu\rho + \Oc(\partial^4)\right\}
  \end{equation}
with $\rho\equiv \phi^a\phi_a$. Here $U_k(\rho)$ corresponds to the
most general $O(N)$--symmetric non--derivative, potential term. The 
derivative terms contain a field dependent wave function renormalization
factor $Z_{k}(\rho)$ plus a function $Y_{k}(\rho)$ accounting for a
possible different index structure of the kinetic term for $N \ge 2$. 
Going further would require the consideration of terms with four derivatives
and so on. The (approximate) chiral symmetry of QCD will play an important
role for the construction of the quark meson model which will
be discussed in the following.

\section{The linear quark meson model}
\label{TheQuarkMesonModelAtT=0}

Before discussing the nonzero temperature and density behavior of strong
interaction physics we will review some of its zero temperature
features. This will be done within the framework of a linear quark
meson model as an effective description for QCD for scales below the
mesonic compositeness scale of approximately $k_\Phi\simeq600\MeV$.
Relating this model to QCD in a semi--quantitative way in subsection
\ref{ASemiQuantitativePicture} will allow us to gain some information
on the initial value for the effective average action at the
scale $k_\Phi$.  We emphasize, however, that the
quantitative aspects of the derivation of the effective quark meson
model from QCD will not be relevant for our practical calculations in
the mesonic sector. This is related to the infrared stability  
mentioned in the previous section
and which will be made quantitative in 
\ref{FlowEquationsAndInfraredStability}.

\subsection{A short scale history}
\label{ASemiQuantitativePicture}

We give here a brief
semi--quantitative picture of the relevant scales 
and the physical degrees of freedom which appear 
in relation to the
phenomenon of chiral symmetry breaking in vacuum.
See also the reviews in \cite{JW96-4,BJW}.
Most of this will be explained in more detail below
whereas other parts are rather well established
features of strong interaction physics. 

We will distinguish five qualitatively different
ranges of scales:
\begin{enumerate}
\item At sufficiently high momentum scales, say,
  \begin{displaymath}
    k\, \gtap\, k_p\simeq1.5\GeV
  \end{displaymath}
  the relevant degrees of freedom of strong interactions are quarks
  and gluons and their dynamics is well described by perturbative QCD.
\item For decreasing momentum scales in the range
  \begin{displaymath}
    k_\Phi\simeq600\MeV\, \ltap\,  k\, \ltap\,  k_p\simeq1.5\GeV
  \end{displaymath}
  the dynamical degrees of freedom are still quarks and gluons. Yet,
  as $k$ is lowered part of their dynamics becomes dominated by
  effective non--local quark interactions which cannot be fully
  accessed perturbatively.
\item At still lower scales this situation changes dramatically.
  Quarks and gluons are supplemented by mesonic bound states as
  additional degrees of freedom which are formed at a scale
  $k_\Phi\simeq600\MeV$. We emphasize that $k_\Phi$ is well separated
  from $\Lambda_{\rm QCD}\simeq200\MeV$ where confinement sets in and
  from the constituent masses of the quarks $M_q\simeq(300-350)\MeV$.
  This implies that below the compositeness scale $k_\Phi$ there
  exists a hybrid description in term of quarks {\em and} mesons. It
  is important to note that for scales not too much smaller than
  $k_\Phi$ chiral symmetry remains unbroken. This situation holds down
  to a scale $k_{\chi SB}\simeq400\MeV$ at which the scalar meson
  potential develops a non--trivial minimum thus breaking chiral
  symmetry spontaneously. The meson dynamics within the range
  \begin{displaymath}
    k_{\chi SB}\simeq400\MeV\, \ltap\,  k\, \ltap\,  k_\Phi\simeq600\MeV
  \end{displaymath}
  is dominated by light current quarks with a strong Yukawa coupling
  $h^2_k/(4\pi)\gg\alpha_s(k)$ to mesons. We will thus assume that
  the leading gluon effects are included below $k_\Phi$ already in the
  formation of mesons.  Near $k_{\chi SB}$ also fluctuations of the
  light scalar mesons become important as their initially large
  renormalized mass approaches zero. Other hadronic bound states like
  vector mesons or baryons should have masses larger than those of the
  lightest scalar mesons, in particular near $k_{\chi SB}$, and give
  therefore only subleading contributions to the dynamics. This leads
  us to a simple linear model of quarks and scalar mesons as an
  effective description of QCD for scales below $k_\Phi$.
\item As one evolves to scales below $k_{\chi SB}$ the Yukawa coupling
  decreases whereas $\alpha_s$ increases. Of course, getting closer to
  $\Lambda_{\rm QCD}$ it is no longer justified to neglect in the
  quark sector the QCD effects which go beyond the dynamics of the
  simple effective quark meson model.  On the
  other hand, the final IR value of the Yukawa coupling $h$ is fixed
  by the typical values of constituent quark masses $M_q\simeq300\MeV$
  to be $h^2/(4\pi)\simeq3.4$. One may therefore speculate that the
  domination of the Yukawa interaction persists even for the interval
  \begin{displaymath}
    M_q\simeq300\MeV\, \ltap\,  k\, \ltap\,  k_{\rm \chi SB}\simeq400\MeV
  \end{displaymath}
  below which the quarks decouple from the evolution of the mesonic
  degrees of freedom altogether. Of course, details of the gluonic
  interactions are expected to be crucial for an understanding of
  quark and gluon confinement. Strong interaction effects may
  dramatically change the momentum dependence of the quark propagator
  for $k$ and $q^2$ around $\Lambda_{\rm QCD}$.  Yet, there is no
  coupling of the gluons to the color neutral mesons. As long as one
  is only interested in the dynamics of the mesons one is led to
  expect that confinement effects are quantitatively not too
  important.
\item Because of the effective decoupling of the quarks and therefore
  of the whole colored sector the details of confinement have only
  little influence on the mesonic dynamics for scales
  \begin{displaymath}
    k\, \ltap\,  M_q\simeq300\MeV\; .
  \end{displaymath}
  Here quarks and gluons disappear effectively from the spectrum and
  one is left with the pions. For scales below the
  pion mass the flow of the couplings stops.
\end{enumerate}

\vspace*{0.3cm}
\centerline{\em 1.\ `Integrating out gluons'}\vspace*{0.5cm}

Let us now discuss the above picture with different ranges of scales 
in more detail. 
In order to obtain the effective action at the compositeness scale
$k_\Phi$ from short distance QCD two steps have to be carried out. In
the first step one computes at the scale $k_p\simeq1.5\GeV$ an effective
action involving only quarks. This step integrates out the gluon
degrees of freedom in a ``quenched approximation''. More precisely,
one solves a truncated flow equation for QCD with quark and gluon
degrees of freedom in presence of an effective infrared cutoff
$k_p\simeq1.5\GeV$ in the quark propagators. This procedure is outlined
in \cite{W9604227}. The exact flow equation
to be used for this purpose is obtained by lowering the infrared
cutoff $R_k$ for the gluons while keeping the one for the
quarks fixed. Subsequently, the gluons are eliminated by solving the
field equations for the gluon fields as functionals of the quarks.
This will result in a non--trivial momentum dependence of the quark
propagator and effective non--local four and higher quark
interactions. Because of the infrared cutoff $k_p$ the resulting
effective action for the quarks resembles closely the one for heavy
quarks. The dominant effect is the
appearance of an effective quark potential (similar to the one for the
charm quark) which describes the effective four--quark interactions.
First promising results within this approach
include an estimate of the heavy quark effective potential valid 
for momenta $\sqrt{q^2}\, \gtap\, 300 - 500 \MeV$ \cite{BBW97-1,EHW96-1}.
The inverse quark propagator is found in
this computation to remain very well approximated by the simple
classical momentum dependence $\gamma_{\mu}q^{\mu}$.\\

\centerline{\em 2.\ Formation of mesonic bound states}\vspace*{0.5cm}

In the second step one has to lower the infrared cutoff in the
effective non--local quark model in order to extrapolate from $k_p$ to
$k_\Phi$.  This task can be carried out by means of the flow
equation for quarks only, starting at $k_p$ with an initial value
$\Gamma_{k_p}[\psi]$ as obtained after integrating out the gluons. For
fermions the trace in (\ref{ERGE}) has to be replaced by a supertrace in
order to account for the minus sign related to Grassmann
variables~\cite{Wet90-1}.  A first investigation in this
direction~\cite{EW94-1} has used a chirally invariant
four quark interaction (``dressed'' one--gluon exchange)
whose most general momentum dependence was retained
\begin{eqnarray}
 \ds{\Gm_k} &=& \ds{\int\frac{d^4 p}{(2\pi)^4}
 \ovl{\psi}_a^i(p)Z_{\psi,k}(p)\left[
 \gamma^{\mu}p_{\mu}\delta^{ab}+m^{ab}(p)\gamma_5+
 i\tilde{m}^{ab}(p)\right]\psi_{ib}(p)}\nnn 
  &+& \ds{
 \frac{1}{2}\int\left(\prod_{l=1}^4
 \frac{d^4 p_l}{(2\pi)^4}\right)
 \left(2\pi\right)^4\delta(p_1+p_2-p_3-p_4)} 
 \la_k^{(\psi)}(p_1,p_2,p_3,p_4){\cal M}(p_1,p_2,p_3,p_4)\; ,
\label{momint}
\een
\be
 {\cal M}(p_1,p_2,p_3,p_4) = \ds{
-\left\{\bar\psi^i_a(-p_1)\gamma^\mu(T^z)_i^{\
j}\psi_j^a(-p_3)\right\}\left\{\bar\psi^k_b(p_4)\gamma_\mu(T_z)_k
^{\ \ell}\psi_\ell^b(p_2)\right\}
 }\; .
 \label{QCDFourFermi}
\ee
The curled brackets indicate  contractions over spinor indices, 
$i,j,k,l=1...N_c$ are the colour
indices and $a,b=1...N_f$ the flavour indices of the quarks.
By an appropriate Fierz transformation and using the identity
\be\label{4.12}
(T^z)_i^{\ j}(T_z)_k^{\ \ell}=\frac{1}{2}\delta^\ell_i\delta^j_k-
\frac{1}{2N_c}\delta^j_i\delta^\ell_k
\ee
one can split ${\cal M}$ into three terms 
\ben
{\cal M}&=&{\cal M}_\sigma+{\cal M}_\rho+{\cal M}_p\label{4.13}\\
{\cal
M}_\sigma&=&-\frac{1}{2}\left\{\bar\psi^i_a(-p_1)i\psi_i^b(p_2)\right\}
\left\{\bar\psi^j_b(p_4)i\psi_j^a(-p_3)\right\}\nonumber\\
&&-\frac{1}{2}\left\{\bar\psi^i_a(-p_1)\gamma^5
\psi_i^b(p_2)\right\}\left\{
\bar\psi^j_b(p_4)\gamma^5\psi_j^a(-p_3)\right\}\label{sigmac}\\
{\cal M}_\rho&=&\frac{1}{4}\left\{\bar\psi^i_a(-p_1)i
\gamma_\mu\psi_i^b(p_2)
\right\}\left\{\bar\psi^j_b(p_4)i\gamma^\mu\psi_j^a
(-p_3)\right\}\nonumber\\
&&-\frac{1}{4}\left\{\bar\psi^i_a(-p_1)\gamma_\mu\gamma^5
\psi_i^b(p_2)\right\}
\left\{\bar\psi^j_b(p_4)\gamma^\mu\gamma^5\psi_j^a(-p_3)\right\}
\label{4.15}\\
{\cal M}_p&=&-\frac{1}{2N_c}\left\{\bar\psi^i_a(-p_1)i
\gamma_\mu\psi_i^a(-p_3)
\right\}\left\{\bar\psi^j_b(p_4)i\gamma^\mu\psi_j^b(p_2)\right\} \; .
\label{4.16}
\een
In terms of the Lorentz invariants
\ben\label{4.17}
s&=&(p_1+p_2)^2=(p_3+p_4)^2\nonumber\\
t&=&(p_1-p_3)^2=(p_2-p_4)^2
\een
we recognize that the quantum numbers of the fermion bilinears in ${\cal
M}_\sigma$ correspond to colour singlet, flavour non-singlet scalars in the
$s$-channel and similarly for spin-one mesons for ${\cal M}_\rho$. 
Following \cite{EW94-1} we
associate these terms with the scalar mesons of
the linear $\sigma$-model and with the $\rho$-mesons. The bilinears in the
last term ${\cal M}_p$ correspond to a colour and flavour singlet spin-one
boson in the $t$-channel. These are the quantum numbers of the pomeron.
In the following we neglect interactions in
the vector meson and pomeron channels and only retain the
contribution ${\cal M}_\sigma$.  We will discuss this 
approximation in more detail in section \ref{AdditionalDegreesOfFreedom}.
The matrices $m$ and $\tilde{m}$ are hermitian and
$m+i\tilde{m}\gamma_5$ forms therefore the most general quark mass
matrix. (Our chiral conventions~\cite{Wet90-1} where the hermitean part
of the mass matrix is multiplied by $\gamma_5$ may be somewhat unusual
but they are quite convenient for Euclidean calculations.)  With
$V(q^2)$ the heavy quark potential in a Fourier representation, the
initial value at $k_p=1.5\GeV$ was taken as
($\hat{Z}_{\psi,k}=Z_{\psi,k}(p^2=-k_p^2)$)
\begin{equation}
  \label{FFCBC}
  \la_{k_p}^{(\psi)}(p_1,p_2,p_3,p_4)
  \hat{Z}_{\psi,k_p}^{-2}=
  \frac{1}{2}V((p_1-p_3)^2)=
  \frac{2\pi\alpha_s}{(p_1-p_3)^2}+
  \frac{8\pi\la}{\left((p_1-p_3)^2\right)^2}\; .
\end{equation}
This corresponds to an approximation by a one gluon exchange term
$\sim\alpha_s(k_p)$ and a string tension $\la\simeq0.18\GeV^2$ and is
in reasonable agreement with the form computed recently~\cite{BBW97-1}
from the solution of flow equations.  In the
simplified ansatz (\ref{FFCBC}) the string tension introduces a second
scale in addition to $k_p$ and indeed the incorporation
of gluon fluctuations is a crucial ingredient for the emergence of
mesonic bound states. For a more precise treatment~\cite{BBW97-1} of
the four--quark interaction at the scale $k_\Phi$ this second scale is
set by the running of $\alpha_s$ or $\Lambda_{\rm QCD}$.

The evolution equation for the function $\la_k^{(\psi)}$ for $k<k_p$
can be derived from the fermionic version of (\ref{ERGE}) and the
truncation (\ref{QCDFourFermi}). Since $\la_k^{(\psi)}$ depends on six
independent momentum invariants it is a partial differential equation
for a function depending on seven variables and has to be solved
numerically~\cite{EW94-1}.  The ``initial value'' (\ref{FFCBC})
corresponds to the $t$--channel exchange of a ``dressed'' colored
gluonic state and it is exciting to realize that the evolution of
$\la_k^{(\psi)}$ leads at lower scales to a momentum dependence
representing the exchange of colorless mesonic bound states. At
the compositeness scale
\begin{equation}
\label{kphi}
  k_\Phi\simeq630\MeV
\end{equation}
one finds \cite{EW94-1} an approximate factorization
\begin{equation}
\label{BSFact}
  \la_{k_\Phi}^{(\psi)}(p_1,p_2,p_3,p_4)=
  g(p_1,p_2)\tilde{G}(s)g(p_3,p_4)+\ldots
\end{equation}
which indicates the formation of mesonic bound states.  Here
$g(p_1,p_2)$ denotes the amputated Bethe--Salpeter wave function and
$\tilde{G}(s)$ is the mesonic bound state propagator displaying a
pole--like structure in the $s$--channel if it is continued to
negative $s=(p_1+p_2)^2$. The dots indicate the part of
$\la_k^{(\psi)}$ which does not factorize and which will be
neglected in the following. In the limit where the momentum dependence
of $g$ and $\tilde{G}$ is neglected we recover the four--quark
interaction of the Nambu--Jona-Lasinio 
model~\cite{NJL61-1,Bij95-1,klevansky}.\\

\centerline{\em 3.\ Quark meson model}\vspace*{0.5cm}

For scales below the mesonic compositeness scale
$k_\Phi$ a description of strong
interaction physics in terms of quark fields alone would be rather
inefficient. Finding physically reasonable truncations of the
effective average action should be much easier once composite fields
for the mesons are introduced.  The exact renormalization group
equation can indeed be supplemented by an exact formalism for the
introduction of composite field variables or, more generally, a change
of variables~\cite{EW94-1}. For our purpose, this amounts in practice
to inserting at the scale $k_\Phi$ the identities
\begin{eqnarray}
 1 &=& \displaystyle{
 {\rm const}\;
 \int{\cal D}\si_A}\nnn
 && \ds{\hspace{-1.2cm}\times
 \exp\left\{ -\tr
 \left(\si_A^\dagger -
 K_A^\dagger \tilde{G}-m_A^\dagger-
 {\cal O}^\dagger \tilde{G}\right)
 \frac{1}{2\tilde{G}}
 \left(\si_A -\tilde{G}K_A -m_A-
 \tilde{G}{\cal O} \right)\right\} }\nnn
 \label{identity}
  1 &=& \displaystyle{
 {\rm const}\;
 \int{\cal D}\si_H}\\[2mm]
 && \displaystyle{\hspace{-1.2cm}\times
 \exp\left\{-\tr
 \left(\si_H^\dagger -
 K_H^\dagger \tilde{G}-m_H^\dagger-
 {\cal O}^{(5)\dagger} \tilde{G}\right)
 \frac{1}{2\tilde{G}}\left(\si_H -\tilde{G}K_H -m_H-
 \tilde{G}{\cal O}^{(5)} \right)\right\} }\nonumber
\end{eqnarray}
into the functional integral which formally defines the quark
effective average action. Here we have used the shorthand notation
$A^\dagger G B\equiv\int\frac{d^d q}{(2\pi)^d}A_a^*(q)G^{ab}(q)
B_b(q)$, and $K_{A,H}$ are sources for the collective fields
$\si_{A,H}$ which correspond in turn to the anti-hermitian and
hermitian parts of the meson field $\Phi$. They are associated to the
fermion bilinear operators ${\cal O} [\psi]$, ${\cal O}^{(5)}[\psi ]$
whose Fourier components read
\begin{eqnarray}
 \ds{{\cal O}_{\;\; b}^a (q)} &=& \ds{ 
 -i\int\frac{d^4 p}{(2\pi)^4} g(-p,p+q)
 \ovl{\psi}^a (p)\psi_b (p+q) }\nnn
 \ds{{\cal O}_{\;\;\;\;\;\; b}^{(5)a} (q)} &=& \ds{ 
 -\int\frac{d^4 p}{(2\pi)^4} g(-p,p+q)
 \ovl{\psi}^a (p)\gamma_5\psi_b (p+q) }\; .
\end{eqnarray}
The choice of $g(-p,p+q)$ as the bound state wave function
renormalization and of $\tilde{G}(q)$ as its propagator guarantees
that the four--quark interaction contained in (\ref{identity}) cancels
the dominant factorizing part of the QCD--induced non--local
four--quark interaction Eqs.(\ref{QCDFourFermi}), (\ref{BSFact}). In
addition, one may choose
\begin{eqnarray}
 \ds{m_{Hab}^T} &=& \ds{
 m_{ab}(0)g^{-1}(0,0)Z_{\psi,k_\Phi}(0)}\nnn
 \ds{m_{Aab}^T} &=& \ds{
 \tilde{m}_{ab}(0)g^{-1}(0,0)Z_{\psi,k_\Phi}(0)}
\end{eqnarray}
such that the explicit quark mass term cancels out for $q=0$. The
remaining quark bilinear is $\sim m(q)-m(0)Z_{\psi,k_\Phi}(0)g(-q,q)/
[Z_{\psi,k_\Phi}(q)g(0,0)]$. It vanishes for zero momentum and will be
neglected in the following. Without loss of generality we can take $m$
real and diagonal and $\tilde{m}=0$. 

In consequence, we have replaced at the scale $k_\Phi$ the effective
quark action (\ref{QCDFourFermi}) with (\ref{BSFact}) by an effective
quark meson action given by
\begin{eqnarray}
 \ds{\hat{\Gamma}_k} &=& \ds{
 \Gamma_k-\frac{1}{2}\int d^4 x\tr
 \left(\Phi^\dagger\jmath+\jmath^\dagger\Phi\right)}\nnn
 \ds{\Gamma_{k}} &=& \displaystyle{ \int d^4 x
 U_k(\Phi,\Phi^\dagger)
  \label{EffActAnsatz}
 }\\[2mm]
 &+& \displaystyle{ 
 \int\frac{d^4 q}{(2\pi)^d}\Bigg\{
 Z_{\Phi,k}(q) q^2 \tr\left[
 \Phi^\dagger (q)\Phi (q)\right] +
 Z_{\psi,k}(q)\ovl{\psi}_a(q)
 \gamma^\mu q_\mu \psi^a (q)
 }\nonumber\vspace{.2cm}\\
 &+& \displaystyle{
 \int\frac{d^4 p}{(2\pi)^d}\ovl{h} _k (-q,q-p)}\nnn
 &\times& \ds{
 \ovl{\psi}^a(q) \left(
 \frac{1+\gamma_5}{2}\Phi _{ab}(p)-
 \frac{1-\gamma_5}{2}\Phi_{ab}^\dagger (-p) \right)
 \psi^b (q-p) \Bigg\}\nonumber \; .}
\end{eqnarray}
At the scale $k_\Phi$ the inverse scalar propagator is related to
$\tilde{G}(q)$ in (\ref{BSFact}) by
\begin{equation}
 \tilde{G}^{-1}(q^2) = 2\ovl{m}^2_{k_\Phi} +
 2Z_{\Phi,k_\Phi}(q)q^2\; .
\end{equation}
This fixes the term in $U_{k_\Phi}$ which is quadratic in $\Phi$ to be
positive, $U_{k_\Phi}=\ovl{m}^2_{k_\Phi}\tr\Phi^\dagger\Phi+\ldots$.
The higher order terms in $U_{k_\Phi}$ cannot be determined in the
approximation (\ref{QCDFourFermi}) since they correspond to terms
involving six or more quark fields.  The initial value of the Yukawa
coupling corresponds to the ``quark wave function in the meson'' in
(\ref{BSFact}), i.e.
\begin{equation}
 \ovl{h} _{k_\Phi}(-q,q-p) = g(-q,q-p)
\end{equation}
which can be normalized with $\ovl{h}_{k_\Phi}(0,0)=g(0,0)=1$.
We observe that the explicit chiral symmetry breaking from
non--vanishing current quark masses appears now in the form of a meson
source term with
\begin{equation}
  \label{AAA22}
  \jmath=2\ovl{m}^2_{k_\Phi} Z_{\psi,k_\Phi}(0)
  g^{-1}(0,0)\left(
  m_{ab}+i\tilde{m}_{ab}\right)=
  2Z_{\psi,k_\Phi}\ovl{m}^2_{k_\Phi}
  {\rm diag}(m_u,m_d,m_s)\; .
\end{equation}
This induces a non--vanishing $\VEV{\Phi}$ and an effective quark mass
$M_q$ through the Yukawa coupling. We note that the current quark mass
$m_q$ and the constituent quark mass $M_q\sim\ovl{h}_k\VEV{\Phi}$ are
identical at the scale $k_\Phi$. (By solving the field equation for
$\Phi$ as a functional of $\ovl{\psi}$, $\psi$ (with
$U_k=\ovl{m}^2_k\tr\Phi^\dagger\Phi$) one recovers from
(\ref{EffActAnsatz}) the effective quark action (\ref{QCDFourFermi}).  For
a generalization beyond the approximation of a four--quark interaction
or a quadratic potential see ref~\cite{BJW}.)  Spontaneous chiral
symmetry breaking can be described in this language by a
non--vanishing $\VEV{\Phi}$ in the limit $\jmath\ra0$. 

The effective potential
$U_k(\Phi)$ must be invariant under the chiral $SU_L(N) \times
SU_R(N)$ flavor symmetry. In fact, the axial anomaly of QCD breaks the
Abelian $U_A(1)$ symmetry. The resulting $U_A(1)$ violating
multi--quark interactions\footnote{A first investigation for the computation
  of the anomaly term in the fermionic average action can be
  found in~\cite{Paw96-1}.} lead to corresponding $U_A(1)$
violating terms in $U_k(\Phi)$.  Accordingly, the most general
effective potential $U_k$ is a function of the $N+1$ independent $C$
and $P$ conserving $SU_L(N)\times SU_R(N)$ invariants
\begin{equation}
  \begin{array}{rcl}
    \label{Invariants}
    \ds{\rho} &=&
    \ds{\tr\Phi^\dagger\Phi}\; ,\nnn
    \ds{\tau_i} &\sim& \ds{
      \tr\left(\Phi^\dagger\Phi- \frac{1}{N}\rho\right)^i\; ,\;\;\;
      i=2,\ldots,N}\; ,\nnn
    \ds{\xi} &=&
    \ds{\det\Phi+\det\Phi^\dagger}\; .
  \end{array}
\end{equation}
We will concentrate in this work on the two flavor case ($N=2$) and
comment on the effects of including the strange quark in
section~\ref{AdditionalDegreesOfFreedom}. Furthermore we will neglect
isospin violation and therefore consider a singlet source term
$\jmath$ proportional to the average light current quark mass
$\hat{m}\equiv\frac{1}{2}(m_u+m_d)$.  Due to the $U_A(1)$--anomaly
there is a mass split for the mesons described by $\Phi$
\begin{equation}
 \Phi=\hal\left(\si-i\eta^\prime\right)+
 \hal\left( a^k+i\pi^k\right)\tau_k \; .
\end{equation}
The scalar
triplet $(a_0)$ and the pseudoscalar singlet $(\eta^\prime)$ receive a
large mass whereas the pseudoscalar triplet $(\pi)$ and the scalar
singlet $(\sigma)$ remains light. From the measured values
$m_{\eta^\prime},m_{a_0}\simeq1\GeV$ it is evident that a decoupling
of these mesons is presumably a very realistic limit. (In
  thermal equilibrium at high temperature this decoupling is not
  obvious. We will comment on this point in section
  \ref{AdditionalDegreesOfFreedom}.)  It can be achieved in a
chirally invariant way and leads to the well known $O(4)$ symmetric
Gell-Mann--Levy linear sigma model~\cite{GML60-1} which is, however,
coupled to quarks now. This is the two flavor linear quark meson model
which we will study in the next sections.  For this model the
effective potential $U_k$ is a function of $\rho$ only.

The quantities which are directly connected to chiral symmetry
breaking depend on the $k$--dependent expectation value
$\VEV{\Phi}_k=\ovl{\sigma}_{0,k}$ as given by the minimum of the
effective potential 
\begin{equation}
  \label{AAA101}
  \frac{\partial U_k}{\partial\rho}(\rho=2\ovl{\sigma}_{0,k}^2)=
  \frac{\jmath}{2\ovl{\sigma}_{0,k}}.
\end{equation}
In terms of the renormalized expectation value
\begin{equation}
  \label{AAA100}
  \sigma_{0,k}=Z_{\Phi,k}^{1/2}\ovl{\sigma}_{0,k}\; 
\end{equation}
we obtain the following expressions for quantities
as the pion decay constant $f_{\pi}$, chiral condensate
$\VEV{\ovl{\psi}\psi}$, constituent quark mass $M_{q}$ and
pion and sigma mass, $m_{\pi}$ and $m_{\sigma}$, respectively 
($d=4$) \cite{BJW}
\begin{equation}
  \label{AAA65}
  \begin{array}{rcl}
    \ds{f_{\pi,k}} &=& \ds{2\sigma_{0,k}}\; ,\nnn
    \ds{\VEV{\ovl{\psi}\psi}_k} &=& \ds{
      -2\ovl{m}^2_{k_\Phi}\left[Z_{\Phi,k}^{-1/2}
      \sigma_{0,k}-\hat{m}\right]}\; ,\nnn
      \ds{M_{q,k}} &=& \ds{
        h_k\sigma_{0,k}}\; ,\nnn
      \ds{m^2_{\pi,k}} &=& \ds{
        Z_{\Phi,k}^{-1/2}
        \frac{\ovl{m}^2_{k_\Phi}\hat{m}}{\sigma_{0,k}}=
        Z_{\Phi,k}^{-1/2}\frac{\jmath}{2\sigma_{0,k}}}\; ,\nnn
      \ds{m_{\sigma,k}^2} &=& \ds{
        Z_{\Phi,k}^{-1/2}
        \frac{\ovl{m}^2_{k_\Phi}\hat{m}}{\sigma_{0,k}}+
        4\lambda_k\sigma_{0,k}^2}\; .
  \end{array}
\end{equation}
Here we have defined the dimensionless, renormalized couplings
\begin{equation}
  \label{AAA102}
  \begin{array}{rcl}
    \ds{\lambda}_k &=& \ds{Z_{\Phi,k}^{-2}
      \frac{\partial^2U_k}{\partial\rho^2}(\rho=2\ovl{\sigma}_{0,k}^2)}\; ,\nnn
    \ds{h_k} &=& \ds{
      Z_{\Phi,k}^{-1/2}Z_{\psi,k}^{-1}\ovl{h}_k}\; .
  \end{array}
\end{equation}
We are interested in the ``physical values'' of the
quantities (\ref{AAA65}) in the limit $k\ra0$ where the infrared
cutoff is removed, i.e.\ $f_{\pi}=f_{\pi,k=0}$,
$m_{\pi}^2=m_{\pi,k=0}^2$, etc.\\

\centerline{\em 4. Initial conditions}\vspace*{0.5cm}

At the scale $k_\Phi$ the propagator $\tilde{G}$ and the wave function
$g(-q,q-p)$ should be optimized for a most complete elimination of
terms quartic in the quark fields. In the present context we will,
however, neglect the momentum dependence of $Z_{\psi,k}$, $Z_{\Phi,k}$
and $\ovl{h}_k$.  
We will choose a normalization of $\psi,\Phi$ such that
$Z_{\psi,k_\Phi}=\ovl{h}_{k_\Phi}=1$. We therefore need as initial
values at the scale $k_\Phi$ the scalar wave function renormalization
$Z_{\Phi,k_\Phi}$ and the shape of the potential $U_{k_\Phi}$. We will
make here the important assumption that $Z_{\Phi,k}$ is small at the
compositeness scale $k_\Phi$ (similarly to what is usually assumed in
Nambu--Jona-Lasinio--like models) 
\begin{equation}
Z_{\Phi,k_\Phi} \ll 1 \; . \label{compcon}
\end{equation}
This results in a large value of
the renormalized Yukawa coupling
$h_k=Z_{\Phi,k}^{-1/2}Z_{\psi,k}^{-1}\ovl{h}_k$. A large value of
$h_{k_\Phi}$ is phenomenologically suggested by the comparably large
value of the constituent quark mass $M_q$. The latter is related to
the value of the Yukawa coupling for $k\ra0$ and the pion decay
constant $f_\pi=92.4\MeV$ by $M_q=h f_\pi/2$ (with $h=h_{k=0}$), and
$M_q\simeq300\MeV$ implies $h^2/4\pi\simeq3.4$. For increasing $k$ the
value of the Yukawa coupling grows rapidly for $k\gtap M_q$.  Our
assumption of a large initial value for $h_{k_\Phi}$ is therefore
equivalent to the assumption that the truncation (\ref{EffActAnsatz}) can be
used up to the vicinity of the Landau pole of $h_k$. The existence of
a strong Yukawa coupling enhances the predictive power of our approach
considerably.  It implies a fast approach of the running couplings to
partial infrared fixed points as is shown in section 
\ref{FlowEquationsAndInfraredStability} \cite{JW,BJW}. 
In consequence, the
detailed form of $U_{k_\Phi}$ becomes unimportant, except for the
value of one relevant parameter corresponding to the scalar mass term
$\ovl{m}^2_{k_\Phi}$. In this work we fix $\ovl{m}^2_{k_\Phi}$ such
that $f_\pi=92.4\MeV$ for $m_\pi=135\MeV$. The value$f_\pi=92.4\MeV$ 
(for $m_\pi=135\MeV$) sets our unit
of mass for two flavor QCD which is, of course, not directly
accessible by observation. Besides $\ovl{m}^2_{k_\Phi}$ (or $f_\pi$)
the other input parameter used in this work is the constituent quark
mass $M_q$ which determines the scale $k_\Phi$ at which $h_{k_\Phi}$
becomes very large. We consider a range $300\MeV\ltap M_q\ltap350\MeV$
and find a rather weak dependence of our results on the precise value
of $M_q$.

All quantities in our truncation of
$\Gamma_k$ are now fixed and we may follow the flow of
$\Gamma_k$ to $k \to 0$.  In this context it is important that the
formalism for composite fields~\cite{EW94-1} also provides an infrared
cutoff in the meson propagator. The flow equations are therefore
exactly of the form (\ref{ERGE}), with quarks and mesons treated on an
equal footing. At the compositeness scale the quadratic term of
$U_{k_\Phi}=\ovl{m}^2_{k_\Phi}\Tr\Phi^\dagger\Phi+\ldots$ is positive
and the minimum of $U_{k_{\Phi}}$ therefore occurs for $\Phi=0$.
Spontaneous chiral symmetry breaking is described by a non--vanishing
expectation value $\VEV{\Phi}$ in absence of quark masses. This
follows from the change of the shape of the effective potential $U_k$
as $k$ flows from $k_\Phi$ to zero.  The large renormalized Yukawa
coupling rapidly drives the scalar mass term to negative values and
leads to a potential minimum away from the origin at some scale
$k_{\rm \chi SB}<k_\Phi$ such that finally
$\VEV{\Phi}=\ovl{\sigma}_0\neq0$ for $k \to 0$ \cite{EW94-1,JW,BJW}.
This concludes our overview of the general features of chiral symmetry
breaking in the context of flow equations.

\subsection{Flow equations and infrared stability}
\label{FlowEquationsAndInfraredStability}

\centerline{\em 1. Flow equation for the effective potential}\vspace*{0.5cm}

The dependence of the effective action
$\Gamma_k$ on the infrared cutoff scale $k$ 
is given by an exact flow equation (\ref{ERGE}), which
for fermionic fields $\psi$ (quarks) and bosonic fields $\Phi$ (mesons)
reads \cite{Wet93-2,Wet90-1} $(t=\ln (k/k_{\Phi}))$
\begin{equation}
  \label{frame}
  \frac{\pa}{\pa t}\Gm_k[\psi,\Phi] = \frac{1}{2}{\rm Tr} 
  \left\{ \frac{\pa R_{kB}}
    {\pa t} \left(\Gm^{(2)}_k[\psi,\Phi]+R_k\right)^{-1}  \right\} 
    -{\rm Tr} \left\{ \frac{\pa R_{kF}}
    {\pa t} \left(\Gm^{(2)}_k[\psi,\Phi]+R_k\right)^{-1} 
  \right\} \, . \label{bferge}
\end{equation}
Here $\Gamma_k^{(2)}$ is the matrix of second functional derivatives of
$\Gamma_k$ with respect to both fermionic and bosonic field components. The
first trace on the right hand side of~(\ref{frame}) effectively runs only
over the bosonic degrees of freedom. It implies a momentum integration and
a summation over flavor indices. The second trace runs over the
fermionic degrees of freedom and contains in addition a summation over
Dirac and color indices. The infrared cutoff function $R_k$ has a 
block substructure
with entries $R_{kB}$ and $R_{kF}$ for the bosonic and the fermionic fields,
respectively.

We compute the flow equation for the effective potential $U_k$ from
equation (\ref{frame}) using the ansatz (\ref{truncation}) for $\Gamma_k$.
The flow equation has a bosonic and fermionic contribution
\begin{equation}
  \frac{\pa }{\pa t}U_k(\rho)=\frac{\pa }{\pa t}U_{kB}(\rho)
  +\frac{\pa }{\pa t}U_{kF}(\rho) \label{dtu} \, .
\end{equation}
Let us first concentrate on the bosonic part by neglecting for a 
moment the quarks and compute $\partial U_{kB}/\partial t$ . 
The effective potential $U_k$ is obtained
from $\Gamma_k$ evaluated for constant fields. 
The bosonic contribution follows as
\ben
  \displaystyle{\frac{\pa }{\pa t}U_{kB}(\rho)} &=& 
  \displaystyle{\frac{1}{2} \int
    \frac{d^4 q}{(2 \pi)^4} 
    \frac{\pa R_{kB}(q^2)}{\pa t} \left\{
    \frac{3}{Z_{\Phi,k} P_{kB}(q^2) + U_k'(\rho)}\right. 
    }\displaystyle{
    +\left. \frac{1}{Z_{\Phi,k} P_{kB}(q^2) + U_k'(\rho)
      + 2 \rho U_k''(\rho)}\right\}} \; . \label{ubk}
\een
Here primes denote derivatives with respect to $\rho$ 
and one observes the appearance of the (massless) inverse average
propagator
\begin{equation}
P_{kB}(q^2)=q^2+Z_{\Phi,k}^{-1}R_k(q^2)=\frac{q^2}{1-e^{-q^2/k^2}} \; .
\label{pk}
\end{equation}
For $\rho$ different from zero one recognizes the first term 
in (\ref{ubk}) as the contribution from the pions, the Goldstone
bosons of chiral symmetry breaking. (The mass term $U_k'$ vanishes
at the minimum of the potential.)  The second contribution is
then related to the radial or $\sigma$--mode.\\

\vspace*{0.2in}
\hrule
\vspace*{0.2in}

{\sc Exercise}: To obtain $\partial U_{kB}/\partial t$
we calculate the trace in (\ref{bferge}) for small field fluctuations
around a constant background configuration, which we take without
loss of generality to be $\sim \phi \delta_{a1}$ and $\rho=\phi^2/2$. 
The inverse propagator $\Gamma^{(2)}_k$ can be computed from
our ansatz (\ref{EffActAnsatz}) by expanding $U_k$ 
to second order in small fluctuations~$\chi$,
$\Phi_a(x)=\phi \delta_{a1} + \chi_a(x)$ 
\ben
U_k &=& U_k(\rho) + U_k'(\rho) 
\left(\frac{1}{2} \Phi_a(x) \Phi^a(x) - \rho\right)
+ \frac{1}{2} U_k''(\rho) 
\left(\frac{1}{2} \Phi_a(x) \Phi^a(x) - \rho\right)^2
+\ldots \nnn
&=&  U_k(\rho) + U_k'(\rho) 
\left(\frac{1}{2} \chi_a \chi^a + \phi \chi_1\right)
+\frac{1}{2} U_k''(\rho) \phi^2 \chi_1^2 + \ldots \; 
\een
and we obtain 
\ben
\left(\Gamma^{(2)}_k\right)_{11}(q,q') &=& \left(Z_k q^2 + U_k'(\rho) 
+ 2 \rho U_k''(\rho)\right) (2\pi)^d \delta(q-q') \nnn
\left(\Gamma^{(2)}_k\right)_{aa|a\not = 1}(q,q') &=& 
\left(Z_k q^2 + U_k'(\rho)\right) (2\pi)^d \delta(q-q') \; .
\een
which yields (\ref{ubk}). 

\vspace*{0.1in}
\hrule
\vspace*{0.3in}

For the study of phase transitions it
is convenient to work with rescaled, dimensionless and renormalized variables. 
We introduce 
\begin{equation}
  \label{AAA190}
  u(t,\tilde{\rho})\equiv k^{-d}U_k(\rho)\; ,\;\;\;
  \tilde{\rho}\equiv Z_{\Phi,k} k^{2-d}\rho\; ,\;\;\;
  h_k=Z_{\Phi,k}^{-1/2}Z_{\psi,k}^{-1} k^{d-4} \ovl{h}_k\; .
\end{equation}
With
\ben
\displaystyle{\frac{\partial}{\partial t} 
u(t,\tilde{\rho})_{|\tilde{\rho}}} &=& 
\displaystyle{-d u(t,\tilde{\rho}) +
(d-2+\eta_{\Phi}) \tilde{\rho} u'(t,\tilde{\rho})} 
+ \displaystyle{k^{-d} \frac{\partial}{\partial t} 
U\left(\rho(\tilde{\rho})\right)_{|\rho}}
\label{dlesstrans}    
\een
one obtains from (\ref{ubk})
the evolution equation for the 
dimensionless potential. Here the anomalous dimension 
$\eta_{\Phi}$ arises from the $t$-derivative acting on $Z_k$ and 
is given by 
\begin{equation}
\eta_\Phi=\frac{d}{d t}(\ln Z_{\Phi,k}) \; .
\end{equation}
The fermionic contribution to the evolution equation for the
effective potential can be computed without additional effort 
from the ansatz (\ref{EffActAnsatz}) since the fermionic fields appear only
quadratically. The respective flow equations is obtained by taking
the second functional derivative evaluated at $\psi=\bar{\psi}=0$.
Combining the bosonic and the fermionic contributions
one obtains the flow equation \cite{BJW} 
\begin{equation}
  \begin{array}{rcl}
    \ds{\frac{\partial}{\partial t}u} &=& \ds{
      -d u+\left(d-2+\eta_\Phi\right)
      \tilde{\rho}u^\prime}\\[2mm]
    &+& \ds{
      2v_d\left\{
      3l_0^d(u^\prime;\eta_\Phi)+
      l_0^d(u^\prime+2\tilde{\rho}u^{\prime\prime};\eta_\Phi)-
      2^{\frac{d}{2}+1}N_c
      l_0^{(F)d}(\frac{1}{2}\tilde{\rho}h^2;\eta_\psi)
      \right\} }\; . \label{udl}
  \end{array}
\end{equation}
Here $v_d^{-1}\equiv2^{d+1}\pi^{d/2}\Gamma(d/2)$ is a prefactor 
depending on the dimension $d$ and primes now denote
derivatives with respect to $\tilde{\rho}$. The first two terms of
the second line in (\ref{udl}) denote the contributions from
the pions and the $\sigma$--resonance
(cf.\ (\ref{ubk})) and the last 
term corresponds to the fermionic contribution from the $u,d$ quarks.
The number of quark colors
will always be $N_c=3$ in the following.

The symbols $l_n^d$, $l_n^{(F)d}$ in (\ref{udl})
denote bosonic and fermionic mass threshold functions which
contain the momentum integral implied in the traces on the r.h.s.\
of the exact flow equation (\ref{bferge}).  
The threshold functions describe
the decoupling of massive modes and provide an important
non-perturbative ingredient. For instance, the bosonic threshold
functions read
\begin{equation}
 \label{AAA85}
 l_n^d(w;\eta_\Phi)=\frac{n+\delta_{n,0}}{4}v_d^{-1}
 k^{2n-d}\int\frac{d^d q}{(2\pi)^d}
 \frac{1}{Z_{\Phi,k}}\frac{\partial R_k}{\partial t}
 \frac{1}{\left[ P_{kB}(q^2)+k^2w\right]^{n+1}} \;.
\end{equation}
These functions decrease $\sim w^{-(n+1)}$ for $w\gg1$.
Since typically $w=M^2/k^2$ with $M$ a mass, the main
effect of the threshold functions is to cut off fluctuations of
particles with masses $M^2\gg k^2$. Once the scale $k$ is changed
below a certain mass threshold, the corresponding particle no longer
contributes to the evolution and decouples smoothly.

Eq.~(\ref{udl}) is a partial differential equation for the effective
potential $u(t,\tilde{\rho})$ which has to be supplemented by the flow
equation for the Yukawa coupling $h_k$ and expressions for the anomalous
dimensions, where
\begin{equation}  
\eta_\psi =\frac{d}{d t}(\ln Z_{\psi,k}) \; .
\end{equation}
The corresponding flow equations and
further details can be found in ref.\ \cite{BJW,JW}. We will consider
them in the next section in a limit where they can be solved analytically.
We note that the running dimensionless renormalized expectation value
$\kappa\equiv2k^{2-d}Z_{\Phi,k}\ovl{\sigma}_{0,k}^2$, with
$\ovl{\sigma}_{0,k}$ the $k$--dependent expectation value of $\Phi$, 
may be computed for each $k$ directly from the condition (\ref{AAA101})
\begin{equation}
  \label{AAA90}
  u^\prime(t,\kappa)=\frac{\jmath}{\sqrt{2\kappa}}
  k^{-\frac{d+2}{2}}Z_{\Phi,k}^{-1/2}\; .
\end{equation}
\pagebreak

%\vspace*{0.3cm}
\centerline{\em 2. Infrared stability}\vspace*{0.5cm}

Most importantly, one finds that the system of flow equations 
for the effective potential $U_k(\rho)$, the Yukawa
coupling $\bar{h}_k$ and the wave function renormalizations $Z_{\Phi,k}$,
$Z_{\psi,k}$ exhibits
an approximate partial fixed point~\cite{JW,BJW}. 
The small initial
value (\ref{compcon}) of the scalar wave function renormalization
$Z_{\Phi,k_\Phi}$ at the scale $k_\Phi$ results in a large renormalized
meson mass term $Z_{\Phi,k_\Phi}^{-2} U_{k_\Phi}'$ and a large 
renormalized Yukawa coupling 
$h_{k_\Phi}=Z_{\Phi,k_\Phi}^{-1/2}\ovl{h}_{k_\Phi}$ ($Z_{\psi,k_\Phi}=1$). 
We can therefore neglect in the flow
equations all scalar contributions with threshold functions involving
the large meson masses.  This yields the simplified equations \cite{BJW,JW}
for the rescaled quantities ($d=4,v_4^{-1}=32\pi^2$)
\begin{equation}
  \label{AAA110}
  \begin{array}{rcl}
    \ds{\frac{\partial}{\partial t}u} &=& \displaystyle{
      -4u+\left(2+\eta_\Phi\right)
      \tilde{\rho}u^\prime
      -\frac{N_c}{2\pi^2}
      l_0^{(F)4}(\frac{1}{2}\tilde{\rho}h^2)\; ,
      }\nnn
    \displaystyle{\frac{d}{d t}h^2} &=& \displaystyle{
      \eta_\Phi h^2 \; ,
      }\nnn
      \displaystyle{\eta_\Phi} &=& \displaystyle{
        \frac{N_c}{8\pi^2} h^2\; ,
        }\nnn
      \ds{\eta_\psi} &=& \ds{0}\; .
    \end{array}
\end{equation}
Of course, this approximation is only valid
for the initial range of running below $k_\Phi$ before the
(dimensionless) renormalized scalar mass squared
$u^\prime(t,\tilde{\rho}=0)$ approaches zero near the chiral symmetry
breaking scale.  The system (\ref{AAA110}) is exactly soluble 
and we find
\begin{equation}
  \label{AAA113}
  \begin{array}{rcl}
    \ds{h^2(t)} &=& \ds{
      Z_\Phi^{-1}(t)=
      \frac{h_I^2}{1-\frac{N_c}{8\pi^2}h_I^2 t}\; ,
      }\nnn
    \ds{u(t,\tilde{\rho})} &=& \ds{
      e^{-4t}u_I(e^{2t}\tilde{\rho}\frac{h^2(t)}{h_I^2})-
      \frac{N_c}{2\pi^2}\int_0^t d r e^{-4r}
      l_0^{(F)4}(\frac{1}{2}h^2(t)\tilde{\rho}e^{2r}) }\; .
  \end{array}
\end{equation}
Here $u_I(\tilde{\rho})\equiv u(0,\tilde{\rho})$ denotes the effective
average potential at the compositeness scale and $h_I^2$ is the
initial value of $h^2$ at $k_\Phi$, i.e. for $t=0$. To make the behavior
more transparent we
consider an expansion of the initial value effective potential
$u_I(\tilde{\rho})$ in powers of $\tilde{\rho}$ around
$\tilde{\rho}=0$
\begin{equation}
  \label{AAA140}
  u_I(\tilde{\rho})=
  \sum_{n=0}^\infty
  \frac{u_I^{(n)}(0)}{n!}\tilde{\rho}^n \; .
\end{equation}
Expanding also $l_0^{(F)4}$ in eq.~(\ref{AAA113}) in powers of its
argument one finds for $n>2$
\begin{equation}
  \label{LLL00}
  \ds{\frac{u^{(n)}(t,0)}{h^{2n}(t)}} = \ds{
    e^{2(n-2)t}\frac{u_I^{(n)}(0)}{h_I^{2n}}+
    \frac{N_c}{\pi^2}
    \frac{(-1)^n (n-1)!}{2^{n+2}(n-2)}
    l_n^{(F)4}(0)
    \left[1-e^{2(n-2)t}\right]}\; .
\end{equation}
For decreasing $t\ra-\infty$ the initial values $u_I^{(n)}$ become
rapidly unimportant and $u^{(n)}/h^{2n}$ approaches a fixed point.
For $n=2$, i.e., for the quartic coupling, one finds
\begin{equation}
  \label{LLL01}
  \frac{u^{(2)}(t,0)}{h^2(t)}=
  1-\frac{1-\frac{u_I^{(2)}(0)}{h_I^2}}
  {1-\frac{N_c}{8\pi^2}h_I^2 t}
\end{equation}
leading to a fixed point value $(u^{(2)}/h^2)_*=1$. As a consequence
of this fixed point behavior the system looses all its ``memory'' on
the initial values $u_I^{(n\ge2)}$ at the compositeness scale
$k_\Phi$! 
Furthermore, the attraction to partial infrared fixed points continues
also for the range of $k$ where the scalar fluctuations cannot be
neglected anymore.  However, the initial value for the bare dimensionless mass
parameter
\begin{equation}
  \label{AAA142}
  \frac{u_I^\prime(0)}{h_I^2}=
  \frac{\ovl{m}^2_{k_\Phi}}{k_\Phi^2}
\end{equation}
is never negligible.   
In other words, for $h_I\ra\infty$ the IR behavior of the linear quark
meson model will depend (in addition to the value of the compositeness
scale $k_\Phi$ and the quark mass $\hat{m}$) only on one parameter,
$\ovl{m}^2_{k_\Phi}$.  We have numerically verified this feature by
starting with different values for $u_I^{(2)}(0)$.  Indeed, the
differences in the physical observables were found to be small.  This
IR stability of the flow equations leads to a large
degree of predictive power!  For definiteness we will perform our
numerical analysis of the full system of flow equations \cite{BJW} 
with the idealized initial value
$u_I(\tilde{\rho})=u_I^\prime(0)\tilde{\rho}$ in the limit
$h_I^2\ra\infty$. It should be stressed, though, that deviations from
this idealization will lead only to small numerical deviations in the
IR behavior of the linear quark meson model as long as say 
$h_I\gtap 15$.

With this knowledge at hand we may now fix the remaining three
parameters of our model, $k_\Phi$, $\ovl{m}^2_{k_\Phi}$ and $\hat{m}$ by
using $f_\pi=92.4\MeV$, the pion mass $M_\pi=135\MeV$ and the
constituent quark mass $M_q$ as phenomenological input.  Because of
the uncertainty regarding the precise value of $M_q$ we give in table
\ref{tab1} the results for several values of $M_q$.
\begin{table}
\begin{center}
\begin{tabular}{|c|c||c|c|c||c|c|c|c|} \hline
  $\frac{M_q}{\MeV}$ &
  $\frac{\lambda_I}{h_I^2}$ &
  $\frac{k_\Phi}{\MeV}$ &
  $\frac{\ovl{m}^2_{k_\Phi}}{k_\Phi^2}$ &
  $\frac{\jmath^{1/3}}{\MeV}$ &
  $\frac{\hat{m}(k_\Phi)}{\MeV}$ &
  $\frac{\hat{m}(1\GeV)}{\MeV}$ &
  $\frac{\VEV{\ovl{\psi}\psi}(1\GeV)}{\MeV^3}$ &
  $\frac{f_\pi^{(0)}}{\MeV}$
  \\[0.5mm] \hline\hline
  $303$ &
  $1$ &
  $618$ &
  $0.0265$ &
  $66.8$ &
  $14.7$ &
  $11.4$ &
  $-(186)^3$ &
  $80.8$
  \\ \hline
  $300$ &
  $0$ &
  $602$ &
  $0.026$ &
  $66.8$ &
  $15.8$ &
  $12.0$ &
  $-(183)^3$ &
  $80.2$
  \\ \hline
  $310$ &
  $0$ &
  $585$ &
  $0.025$ &
  $66.1$ &
  $16.9$ &
  $12.5$ &
  $-(180)^3$ &
  $80.5$
  \\ \hline
  $339$ &
  $0$ &
  $552$ &
  $0.0225$ &
  $64.4$ &
  $19.5$ &
  $13.7$ &
  $-(174)^3$ &
  $81.4$
  \\ \hline
\end{tabular}
\caption{The table shows the dependence on the
  constituent quark mass $M_q$ of the input parameters $k_\Phi$,
  $\ovl{m}^2_{k_\Phi}/k_\Phi^2$ and $\jmath$ as well as some of our
  ``predictions''. The phenomenological input used here besides $M_q$
  is $f_\pi=92.4\MeV$, $m_\pi=135\MeV$.The first line corresponds to
  the values for $M_q$ and $\lambda_I$ used in the remainder of this
  work. The other three lines demonstrate the insensitivity of our
  results with respect to the precise values of these
  parameters.}
\label{tab1}
\end{center}
\end{table}
The first line of table~\ref{tab1} corresponds to the choice of $M_q$
and $\lambda_I\equiv u_I^{\prime\prime}(\kappa)$ which we will use for
the forthcoming analysis of the model at finite temperature.  As
argued analytically above the dependence on the value of $\lambda_I$
is weak for large enough $h_I$ as demonstrated numerically by the
second line. Moreover, we notice that our results, and in particular
the value of $\jmath$, are rather insensitive with respect to the
precise value of $M_q$. It is remarkable that the values for $k_\Phi$
and $\ovl{m}_{k_\Phi}$ are not very different from those computed in
ref.~\cite{EW94-1}. 

\section{Hot QCD and the chiral phase transition\label{hotqcd}}

\subsection{Thermal equilibrium and dimensional reduction}
\label{FiniteTemperatureFormalism}

Recall that from the partition function
\begin{equation}
Z={\rm Tr}\left\{ \displaystyle{e^{-\beta H}}\right\} 
= \int_{\rm periodic} D\chi\, e^{-S[\chi]}
\label{parfun}
\end{equation}
all standard thermodynamic properties may be determined.
The trace over the statistical density matrix 
$\hat{\rho}=\exp(-\beta H)$ with Hamiltonian $H$ and temperature
$T=1/\beta$ can be expressed as a functional integral \cite{Kap},
as indicated for a scalar field
$\chi$ with action $S$ by the second equation.
We supplement $Z$ by a source term $J$ with the substitution
$S[\chi]~\to~S[\chi]+\int J\chi$ in (\ref{parfun}).
The temperature dependent effective
potential $U=\Gamma \, T/V$ for a constant field 
$\phi \equiv \langle \chi \rangle$, or the Helmholtz free energy,
is then 
\begin{equation}
U(\phi;T)=-\frac{T}{V} \ln Z_J + J \phi \; .
\end{equation}
At its minima the effective potential is related to the
energy density $E/V$, the entropy density $S/V$ and
the pressure $p$ by
\begin{equation}
U_{| \rm min}=\frac{E}{V}-T\frac{S}{V}=-p \; . 
\end{equation}

The term ``periodic'' in (\ref{parfun}) means that the integration
over the field is constrained so that
\begin{equation}
\Phi(x_0+\beta,\vec{x})=\Phi(x_0,\vec{x})=
\sum\limits_{j\in\ZZZ} \Phi_j(\vec{x}) e^{-i \omega_j^B} \qquad, \quad
\omega_j^B=2j \pi T \; .
\end{equation}
This is a consequence of the trace operation, which means that the
system returns to its original state after Euclidean ``time'' $\beta$.
The field can be expanded and we note that the lowest Fourier
mode for bosonic fields vanishes $\omega_0^B=0$. For fermionic 
fields $\psi$ the procedure is analogous with the important
distinction that they are required to be antiperiodic
\begin{equation}
\psi(x_0+\beta,\vec{x})=-\psi(x_0,\vec{x})=
\sum\limits_{j\in\ZZZ} \psi_j(\vec{x}) e^{-i \omega_j^F}\qquad, \quad
\omega_j^F=(2j+1) \pi T
\end{equation}
as a consequence of the anticommutation property of the 
Grassmann fields \cite{Kap}. In contrast to the bosonic case 
the antiperiodicity results in a nonvanishing lowest
Fourier mode $\omega_0^F=\pi T$.

The extension of flow equations to
non--vanishing temperature $T$ is now straightforward~\cite{TetWet}. The 
(anti--)periodic boundary conditions for (fermionic) bosonic fields in
the Euclidean time direction leads to the replacement
\begin{equation}
  \label{AAA120}
  \int\frac{d^d q}{(2\pi)^d}f(q^2)\ra
  T\sum_{j\in\ZZZ}\int\frac{d^{d-1}\vec{q}}{(2\pi)^{d-1}}
  f(q_0^2(j)+\vec{q}^{\,2})
\end{equation}
in the trace of the flow equation (\ref{bferge}) 
when represented as a momentum
integration, with a discrete spectrum for the zero component
$q_0(j)=\omega_j^B$ for bosons and $q_0(j)=\omega_j^F$ for fermions.
Hence, for $T>0$ a four--dimensional QFT can be interpreted as a
three--dimensional model with each bosonic or fermionic degree of
freedom now coming in an infinite number of copies labeled by
$j\in\ZZ$ (Matsubara modes). Each mode acquires an additional
temperature dependent effective mass term $q_0^2(j)$ except
for the bosonic zero mode which vanishes identically. At high
temperature all massive Matsubara modes decouple from
the dynamics of the system. In this case, one therefore expects to observe an
effective three--dimensional theory with the bosonic zero mode as the
only relevant degree of freedom. 
One may visualize this behavior by noting that for a given characteristic
length scale $l$ much larger than the inverse temperature $\beta$ 
the compact 
Euclidean ``time'' dimension cannot be resolved anymore, as is
shown in the figure below. This phenomenon is
known as ``dimensional reduction''.
\begin{figure}[h]
\begin{center}
\epsfxsize=2.5in
\hspace*{0in}
\epsffile{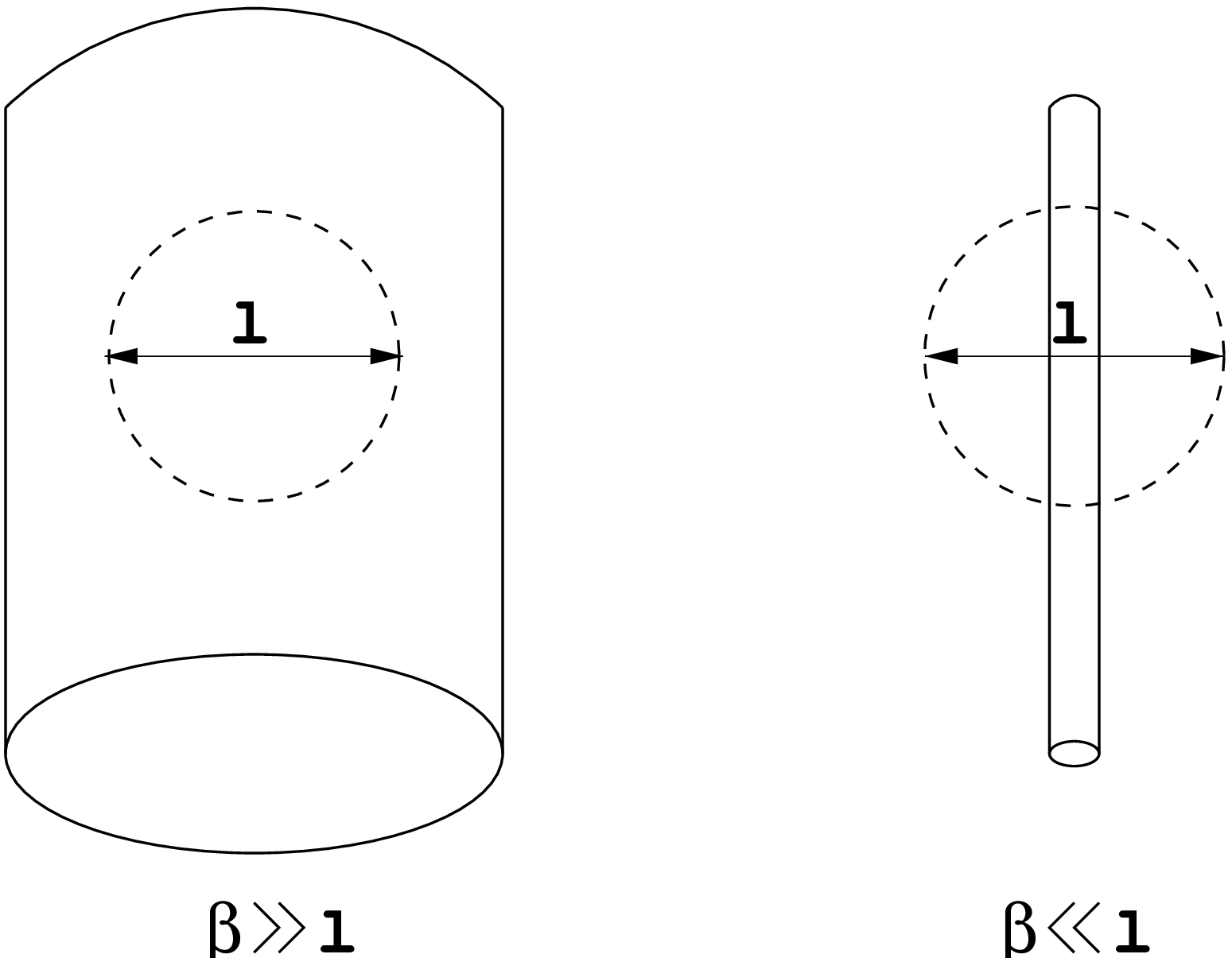}
\end{center}
\end{figure}

The phenomenon of dimensional reduction can be observed
directly from the nonperturbative flow equations. 
The replacement
(\ref{AAA120}) in (\ref{bferge}) manifests itself in the flow equations
only through a change to
$T$--dependent threshold functions.  For instance, the dimensionless
threshold functions $l_n^d(w;\eta_\Phi)$ defined in eq.~(\ref{AAA85}) are
replaced by
\begin{equation}
  \label{AAA200}
  l_n^d(w,\frac{T}{k};\eta_\Phi)\equiv
  \frac{n+\delta_{n,0}}{4}v_d^{-1}k^{2n-d}
  T\sum_{j\in\ZZZ}\int
  \frac{d^{d-1}\vec{q}}{(2\pi)^{d-1}}
  \left(\frac{1}{Z_{\Phi,k}}\frac{\partial R_k(q^2)}{\partial t}\right)
  \frac{1}{\left[P(q^2)+k^2 w\right]^{n+1}}
\end{equation}
where $q^2=q_0^2+\vec{q}^{\,2}$ and $q_0=2\pi j T$. In the
limit $k\gg T$ the sum over Matsubara modes approaches the integration
over a continuous range of $q_0$ and we recover the zero temperature
threshold function $l_n^d(w;\eta_\Phi)$.  In the opposite limit $k\ll
T$ the massive Matsubara modes ($l\neq0$) are suppressed and we expect
to find a $d-1$ dimensional behavior of $l_n^d$. In fact, one obtains
from~(\ref{AAA200})
\begin{equation}
  \label{AAA201}
  \begin{array}{rclcrcl}
    \ds{l_n^d(w,T/k;\eta_\Phi)} &\simeq& \ds{
      l_n^{d}(w;\eta_\Phi)}
    &{\rm for}& \ds{T\ll k}\; ,\nnn
    \ds{l_n^d(w,T/k;\eta_\Phi)} &\simeq& \ds{
      \frac{T}{k}\frac{v_{d-1}}{v_d}
      l_n^{d-1}(w;\eta_\Phi)}
    &{\rm for}& \ds{T\gg k}\; .
  \end{array}
\end{equation}
For our choice of the infrared cutoff function $R_k$,
eq.~(\ref{pk}), the temperature dependent Matsubara modes in
$l_n^d(w,T/k;\eta_\Phi)$ are exponentially suppressed for $T\ll k$.  
Nevertheless, all bosonic threshold
functions are proportional to $T/k$ for $T\gg k$ whereas those with
fermionic contributions vanish in this limit. 
This behavior is demonstrated in figure
\ref{Thresh} where we have plotted the quotients
$l_1^4(w,T/k)/l_1^4(w)$ and $l_1^{(F)4}(w,T/k)/l_1^{(F)4}(w)$ of
bosonic and fermionic threshold functions, respectively.
\begin{figure}
\unitlength1.0cm
\begin{center}
\begin{picture}(13.,18.0)

\put(0.0,9.5){
\epsfysize=11.cm
\rotate[r]{\epsffile{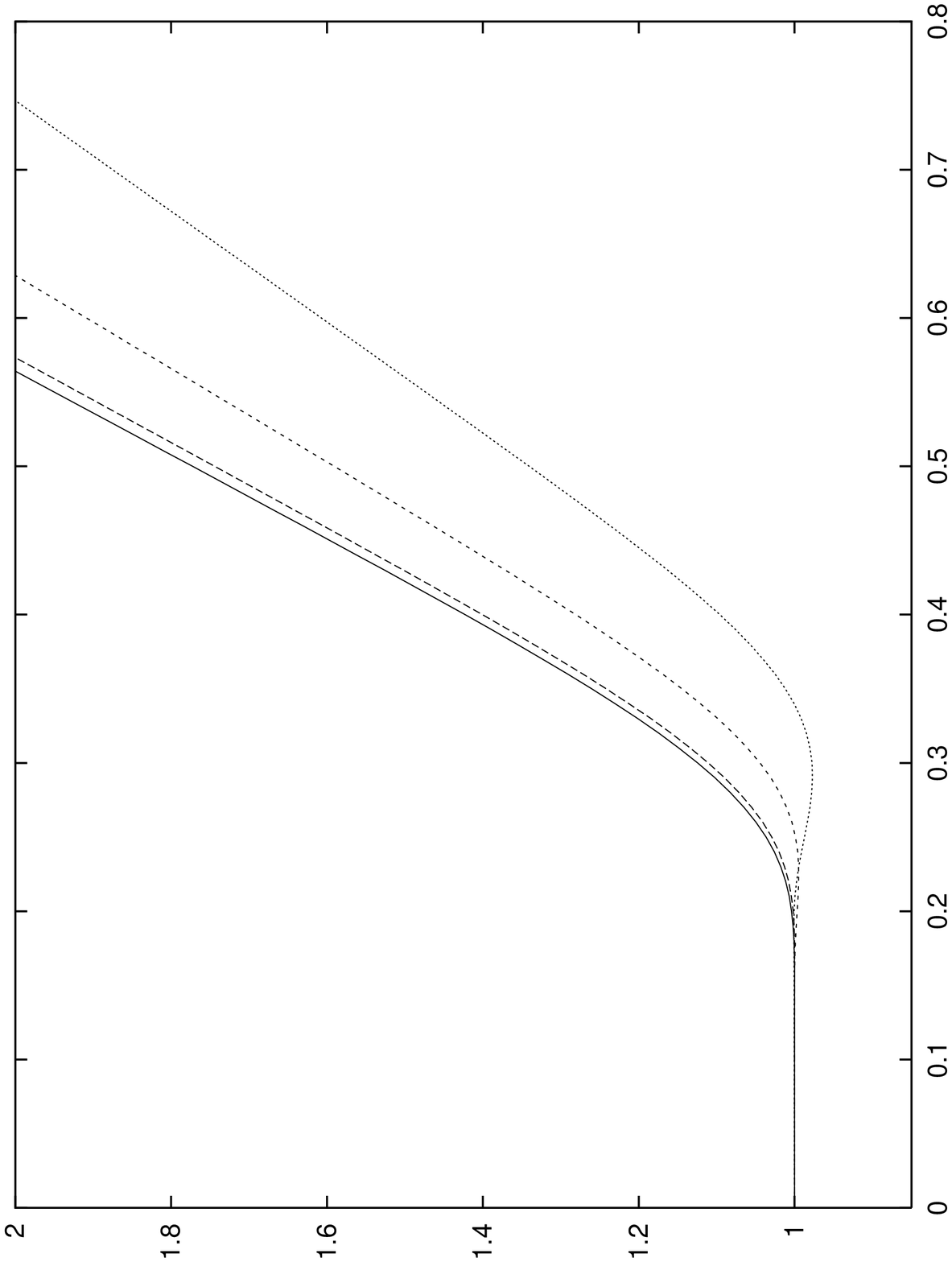}}
}
\put(-1.0,16.5){\bf $\ds{\frac{l_1^4\left(w,\ds{\frac{T}{k}}\right)}
    {l_1^4(w)}}$}
\put(1.5,16.5){\bf $\ds{(a)}$}

\put(0.0,0.5){
\epsfysize=11.cm
\rotate[r]{\epsffile{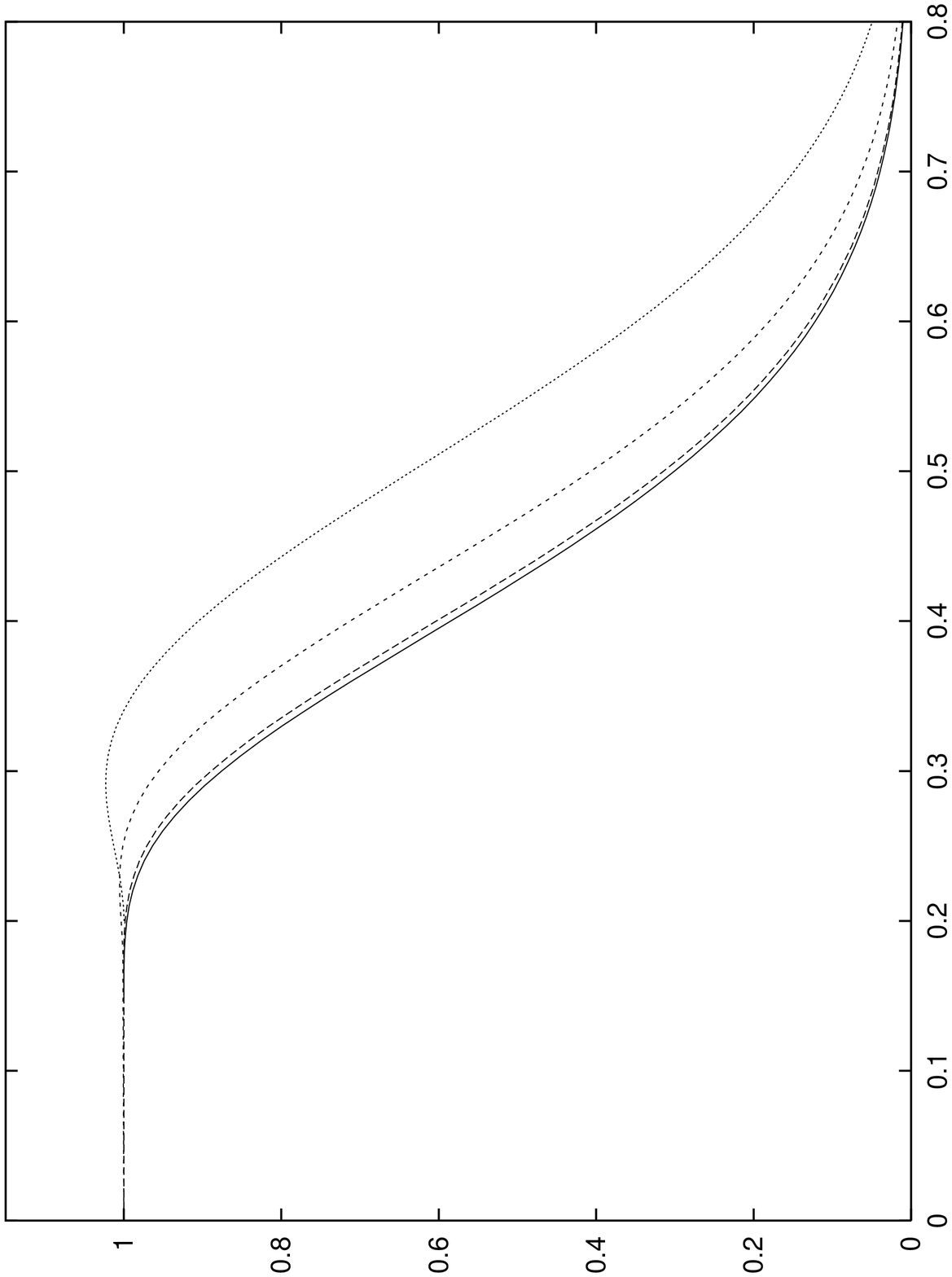}}
}
\put(-1.2,7.5){\bf
  $\ds{\frac{l_1^{(F)4}\left(w,\ds{\frac{T}{k}}\right)}
    {l_1^{(F)4}(w)}}$}
\put(6.1,-0.2){\bf $\ds{T/k}$}
\put(1.5,7.5){\bf $\ds{(b)}$}
\end{picture}
\end{center}
\caption{The plot shows the temperature 
  dependence of the bosonic (a) and the fermionic (b) threshold
  functions $l_1^4(w,T/k)$ and $l_1^{(F)4}(w,T/k)$, respectively, for
  different values of the dimensionless mass term $w$.  The solid line
  corresponds to $w=0$ whereas the dotted ones correspond to $w=0.1$,
  $w=1$ and $w=10$ with decreasing size of the dots.  For $T \gg k$
  the bosonic threshold function becomes proportional to $T/k$ whereas
  the fermionic one tends to zero.  In this range the theory with
  properly rescaled variables behaves as a classical
  three--dimensional theory.}
\label{Thresh}
\end{figure}
One observes that for $k\gg T$ both threshold functions essentially
behave as for zero temperature. For growing $T$ or decreasing $k$ this
changes as more and more Matsubara modes decouple until finally all
massive modes are suppressed. The bosonic threshold function $l^4_1$
shows for $k \ll T$ the linear dependence on $T/k$ derived in
eq.~(\ref{AAA201}).  In particular, for the bosonic excitations the
threshold function for $w\ll1$ can be approximated with reasonable
accuracy by $l_n^4(w;\eta_\Phi)$ for $T/k<0.25$ and by
$(4T/k)l_n^3(w;\eta_\Phi)$ for $T/k>0.25$. The fermionic threshold
function $l_1^{(F)4}$ tends to zero for $k\ll T$ since there is no
massless fermionic zero mode, i.e.~in this limit all fermionic
contributions to the flow equations are suppressed.  On the other
hand, the fermions remain quantitatively relevant up to $T/k\simeq0.6$
because of the relatively long tail in figure~\ref{Thresh}b.  
The formalism of the average action automatically provides
the tools for a smooth decoupling of the massive Matsubara modes as
the momentum scale $k$ is lowered from $k\gg T$ to $k\ll T$.  It therefore
allows us to directly link the low--$T$, four--dimensional QFT to the
effective three--dimensional high--$T$ theory. 
Whereas for $k\gg T$ the model is most efficiently
described in terms of standard four--dimensional fields $\Phi$ a
choice of rescaled three--dimensional variables
$\Phi_{3}=\Phi/\sqrt{T}$ becomes better adapted for $k\ll T$.
Accordingly, for high temperatures one will use the rescaled 
dimensionless potential
\begin{equation}
  \label{CCC01}
  u_{3}(t,\tilde{\rho}_{3})=\frac{k}{T}
  u(t,\tilde{\rho})\; ;\;\;\;
  \tilde{\rho}_{3}=\frac{k}{T}\tilde{\rho}\; .
\end{equation}

For our numerical calculations at non--vanishing temperature we
exploit the discussed behavior of the threshold functions by using the
zero temperature flow equations in the range $k\ge10T$. For smaller
values of $k$ we approximate the infinite Matsubara sums
(cf.~eq.~(\ref{AAA200})) by a finite series such that the numerical
uncertainty at $k=10T$ is better than $10^{-4}$. This approximation
becomes exact in the limit $k\ll10T$.

In section \ref{ASemiQuantitativePicture} we have considered the
relevant fluctuations that contribute to the flow of $\Gamma_k$ in
dependence on the scale $k$. In thermal equilibrium 
$\Gamma_k$ also depends on the temperature $T$ and one may ask for the
relevance of thermal fluctuations at a given scale $k$.  In
particular, for not too high values of $T$ the ``initial condition''
$\Gamma_{k_\Phi}$ for the solution of the flow equations should
essentially be independent of temperature.  This will allow us to fix
$\Gamma_{k_\Phi}$ from phenomenological input at $T=0$ and to compute
the temperature dependent quantities in the infrared ($k \to 0$).  We
note that the thermal fluctuations which contribute to the r.h.s.\ of
the flow equation for the meson potential (\ref{udl}) are
effectively suppressed for $T \ltap k/4$.  
Clearly for $T \gtap k_{\Phi}/3$
temperature effects become important at the compositeness scale. We
expect the linear quark meson model with a compositeness scale
$k_{\Phi} \simeq 600 \MeV$ to be a valid description for two flavor
QCD below a temperature of about\footnote{There will be an effective
  temperature dependence of $\Gamma_{k_{\Phi}}$ induced by the
  fluctuations of other degrees of freedom besides the quarks, the
  pions and the sigma which are taken into account here.  We will
  comment on this issue in section \ref{AdditionalDegreesOfFreedom}.
  For realistic three flavor QCD the thermal kaon fluctuations will
  become important for $T\gtap 170\MeV$.} $170 \MeV$.
We compute the quantities of interest for temperatures $T\ltap 170\MeV$
by solving numerically the $T$--dependent version of the flow
equations by lowering $k$ from $k_\Phi$ to
zero. For this range of temperatures we use the initial values as
given in the first line of table \ref{tab1}. We observe
only a minor dependence of our results on the constituent quark mass 
for the considered
range of values $M_q \simeq 300 - 350 \MeV$. 
In particular, the value for the critical temperature
$T_c$ of the model remains almost unaffected by this variation.

\subsection{High temperature chiral phase transition}
\label{TheQuarkMesonModelAtTNeq0}

We have pointed out in section \ref{intro} that
strong interactions in thermal equilibrium at high temperature $T$ 
differ in important aspects from the well tested vacuum or zero
temperature properties. A phase transition at some critical
temperature $T_c$ or a relatively sharp crossover may separate the
high and low temperature physics~\cite{MO96-1}. 
It was realized early that the transition should be
closely related to a qualitative change in the chiral condensate
according to the general observation that spontaneous symmetry
breaking tends to be absent in a high temperature situation. A series
of stimulating contributions~\cite{PW84-1,RaWi93-1,Raj95-1} pointed
out that for sufficiently small up and down quark masses, $m_u$ and
$m_d$, and for a sufficiently large mass of the strange quark, $m_s$,
the chiral transition is expected to belong to the universality class
of the $O(4)$ Heisenberg model. It
was suggested~\cite{RaWi93-1,Raj95-1} that a large correlation length
may be responsible for important fluctuations or lead to a disoriented
chiral condensate. One main question we are going
to answer using nonperturbative flow equations is: How
small $m_u$ and $m_d$ would have to be in order to see a large
correlation length near $T_c$ and if this scenario could be realized
for realistic values of the current quark masses.
  
Figure \ref{ccc_T} shows our
results \cite{BJW} for the chiral condensate
$\VEV{\ovl{\psi}\psi}$ as a function of the temperature
$T$ for various values of the average quark mass 
$\hat{m}=(m_u+m_d)/2$.
\begin{figure}
\unitlength1.0cm

\begin{center}
\begin{picture}(13.,7.0)

\put(0.0,0.0){
\epsfysize=11.cm
\rotate[r]{\epsffile{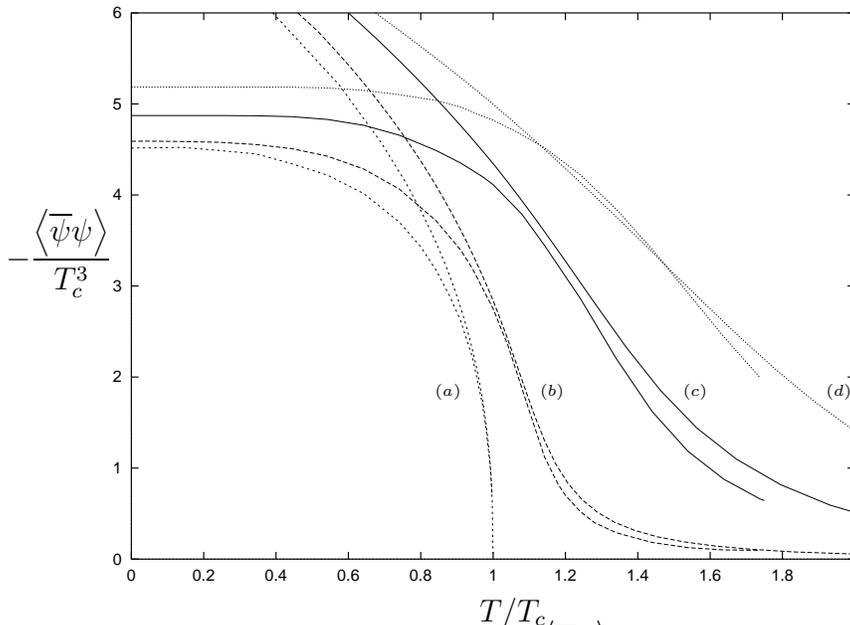}}
}
\put(-0.5,4.2){\bf $\ds{-\frac{\VEV{\ovl{\psi}\psi}}{T_{c}^3}}$}
\put(5.8,-0.5){\bf $\ds{T/T_{c}}$}
\put(5.2,2.5){\tiny $(a)$}
\put(6.6,2.5){\tiny $(b)$}
\put(8.5,2.5){\tiny $(c)$}
\put(10.4,2.5){\tiny $(d)$}

\end{picture}
\end{center}
\caption{The plot shows the chiral condensate
  $\VEV{\ovl{\psi}\psi}$ as a function of temperature $T$.  Lines
  $(a)$, $(b)$, $(c)$, $(d)$ correspond at zero temperature to
  $m_\pi=0,45\MeV,135\MeV,230\MeV$, respectively. For each pair of
  curves the lower one represents the full $T$--dependence of
  $\VEV{\ovl{\psi}\psi}$ whereas the upper one shows for comparison the
  universal scaling form of the equation of state for the $O(4)$
  Heisenberg model. The critical temperature for zero quark mass is
  $T_c=100.7\MeV$. The chiral condensate is normalized at a scale
  $k_{\Phi}\simeq 620\MeV$.}
\label{ccc_T}
\end{figure}
Curve $(a)$ gives the temperature dependence of $\VEV{\ovl{\psi}\psi}$
in the chiral limit $\hat{m}=0$. We first consider only the lower
curve which corresponds to the full result.
One observes that the order parameter $\VEV{\ovl{\psi}\psi}$
goes continuously (but non--analytically)
to zero as $T$ approaches the critical temperature in the massless
limit $T_c=100.7 \MeV$. The transition from the phase with
spontaneous chiral symmetry breaking to the symmetric phase
is second order. The curves $(b)$, $(c)$
and $(d)$ are for non--vanishing values of the average current quark
mass $\hat{m}$. The transition turns into a smooth crossover. 
Curve $(c)$ corresponds to $\hat{m}_{\rm phys}$ or,
equivalently, $m_\pi(T=0)=135\MeV$. The
transition turns out to be much less dramatic than for $\hat{m}=0$. We
have also plotted in curve $(b)$ the results for comparably small
quark masses $\simeq1\MeV$, i.e.~$\hat{m}=\hat{m}_{\rm phys}/10$, for
which the $T=0$ value of $m_\pi$ equals $45\MeV$. The crossover is
considerably sharper but a substantial deviation from the chiral limit
remains even for such small values of $\hat{m}$.

For comparison, the upper curves in figure
\ref{ccc_T} use the universal scaling form of the equation of state
of the {\em three dimensional} $O(4)$--symmetric Heisenberg model which
will be computed explicitly in section \ref{CriticalBehavior}. 
We see perfect agreement 
of both curves in the chiral limit for $T$
sufficiently close to $T_c$ which is a manifestation
of universality and the phenomenon of dimensional reduction. 
This demonstrates the
capability of our method to cover the nonanalytic critical behavior and, in
particular, to reproduce the critical exponents of the $O(4)$--model
(cf.\ section \ref{CriticalBehavior}).
Away from the chiral limit we find that the $O(4)$ universal equation 
of state provides a reasonable approximation for $\VEV{\ovl{\psi}\psi}$
in the crossover region $T=(1.2-1.5)T_c$.  

In order to facilitate comparison with lattice simulations which are typically
performed for larger values of $m_\pi$ we also present results for
$m_\pi(T=0)=230\MeV$ in curve $(d)$. One may define a ``pseudocritical
temperature'' $T_{pc}$ associated to the smooth crossover as the
inflection point of $\VEV{\ovl{\psi}\psi}(T)$ as usually done in
lattice simulations. Our results for $T_{pc}$ are presented in 
table \ref{tab11} for the four different values of $\hat{m}$ or, 
equivalently, $m_\pi(T=0)$.
\begin{table}
\begin{center}
\begin{tabular}{|c||c|c|c|c|} \hline
  $\qquad{{m_\pi}/{\MeV}}\quad$ &
  $0\qquad$ &
  $45\qquad$ &
  $135\qquad$ &
  $230\qquad$
  \\[1.0mm] \hline
  $\qquad{{T_{pc}}/{\MeV}}\quad$ &
  $100.7\qquad$ &
  $\simeq110\qquad$ &
  $\simeq130\qquad$ &
  $\simeq150\qquad$
  \\[1mm] \hline
\end{tabular}
\caption{The table shows the critical and
  ``pseudocritical'' temperatures for various values of the zero
  temperature pion mass. Here $T_{pc}$ is defined as the
  inflection point of $\VEV{\ovl{\psi}\psi}(T)$.}
\label{tab11}
\end{center}
\end{table}
The value for the pseudocritical temperature for $m_{\pi}=230 \MeV$
compares well with the lattice results for two flavor QCD
(cf.~section~\ref{CriticalBehavior}). One should mention, though, that
a determination of $T_{pc}$ according to this definition is subject to
sizeable numerical uncertainties for large pion masses as the curve in
figure \ref{ccc_T} is almost linear around the inflection point for
quite a large temperature range.  A problematic point in lattice
simulations is the extrapolation to realistic values of $m_\pi$ or
even to the chiral limit. Our results may serve here as an analytic
guide. The overall picture shows the approximate validity of the
$O(4)$ scaling behavior over a large temperature interval in the
vicinity of and above $T_c$ once the (non--universal) amplitudes are
properly computed. We point out that the link between the universal
behavior near $T_c$ and zero current quark mass on the one hand and the
known physical properties at $T=0$ for realistic quark masses on the
other hand is crucial to obtain all non--universal information
near $T_c$. 

A second important result is  the temperature
dependence of the space--like pion correlation length
$m_\pi^{-1}(T)$. (We will often call $m_\pi(T)$ the temperature
dependent pion mass since it coincides with the physical pion mass for
$T=0$.) Figure \ref{mpi_T} shows $m_\pi(T)$ and one again observes the
second order phase transition in the chiral limit $\hat{m}=0$. 
\begin{figure}
\unitlength1.0cm
\begin{center}
\begin{picture}(13.,7.0)

\put(0.0,0.0){
\epsfysize=11.cm
\rotate[r]{\epsffile{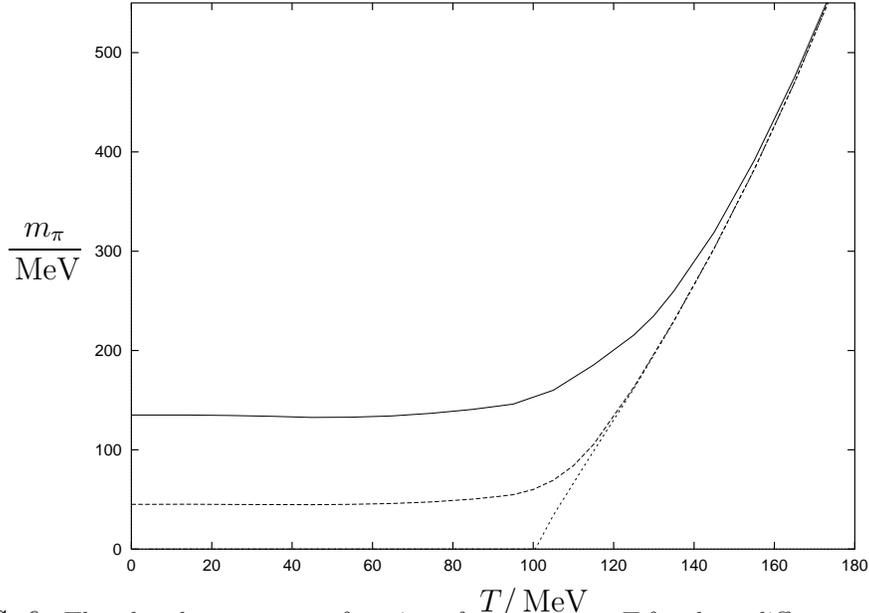}}
}
\put(-0.5,4.2){\bf $\ds{\frac{m_\pi}{\MeV}}$}
\put(5.8,-0.5){\bf $\ds{T/\MeV}$}

\end{picture}
\end{center}
\caption{\footnotesize The plot shows $m_\pi$ as a function of
  temperature $T$ for three different values of the average light
  current quark mass $\hat{m}$. The solid line corresponds to the
  realistic value $\hat{m}=\hat{m}_{\rm phys}$ whereas the dotted line
  represents the situation without explicit chiral symmetry breaking,
  i.e., $\hat{m}=0$. The intermediate, dashed line assumes 
  $\hat{m}=\hat{m}_{\rm phys}/10$.}
\label{mpi_T}
\end{figure}
For $T<T_c$ the pions are massless Goldstone bosons whereas for
$T>T_c$ they form with the sigma a degenerate vector of
$O(4)$ with mass increasing as a function of temperature.  For
$\hat{m}=0$ the behavior for small positive $T-T_c$ is characterized
by the critical exponent $\nu$, i.e.
$m_\pi(T)=\left(\xi^+\right)^{-1}T_c \left( (T-T_c)/T_c\right)^\nu$
and we obtain $\nu=0.787$, $\xi^+=0.270$. For $\hat{m}>0$ we find that
$m_\pi(T)$ remains almost constant for $T\ltap T_c$ with only a very
slight dip for $T$ near $T_c/2$. For $T>T_c$ the correlation length
decreases rapidly and for $T\gg T_c$ the precise value of $\hat{m}$
becomes irrelevant. We see that the universal critical behavior near
$T_c$ is quite smoothly connected to $T=0$.  The full functional
dependence of $m_\pi(T,\hat{m})$ allows us to compute the overall size
of the pion correlation length near the critical temperature and we
find $ m_\pi(T_{pc})\simeq 1.7 m_\pi(0)$ for the realistic value
$\hat{m}_{\rm phys}$. This correlation length is even smaller than the
vacuum ($T=0$) one and gives no indication for strong fluctuations of
pions with long wavelength.\footnote{For a QCD phase transition
far from equilibrium long wavelength modes of the pion field 
can be amplified \cite{RaWi93-1,Raj95-1}.}  We will discuss the possibility of
a tricritical point \cite{fir,BR,SB} with a massless excitation in the 
two--flavor case at non--zero
baryon number density or for three flavors \cite{PW84-1,RaWi93-1,Raj95-1,GGP} 
even at vanishing density
in section \ref{tricriticalpoint}.    
We also point out that the present
investigation for the two flavor case does not take into account a
speculative ``effective restoration'' of the axial $U_A(1)$ symmetry
at high temperature \cite{PW84-1,Shu94-1}. We will comment on these
issues in section~\ref{AdditionalDegreesOfFreedom}.

In figure \ref{Usig} we
display the derivative of the potential with respect to the
renormalized field $\phi_R=(Z_\Phi\rho/2)^{1/2}$, for different values
of $T$.
\begin{figure}
\unitlength1.0cm
\begin{center}
\begin{picture}(13.,7.0)
\put(0.0,0.0){
\epsfysize=11.cm
\rotate[r]{\epsffile{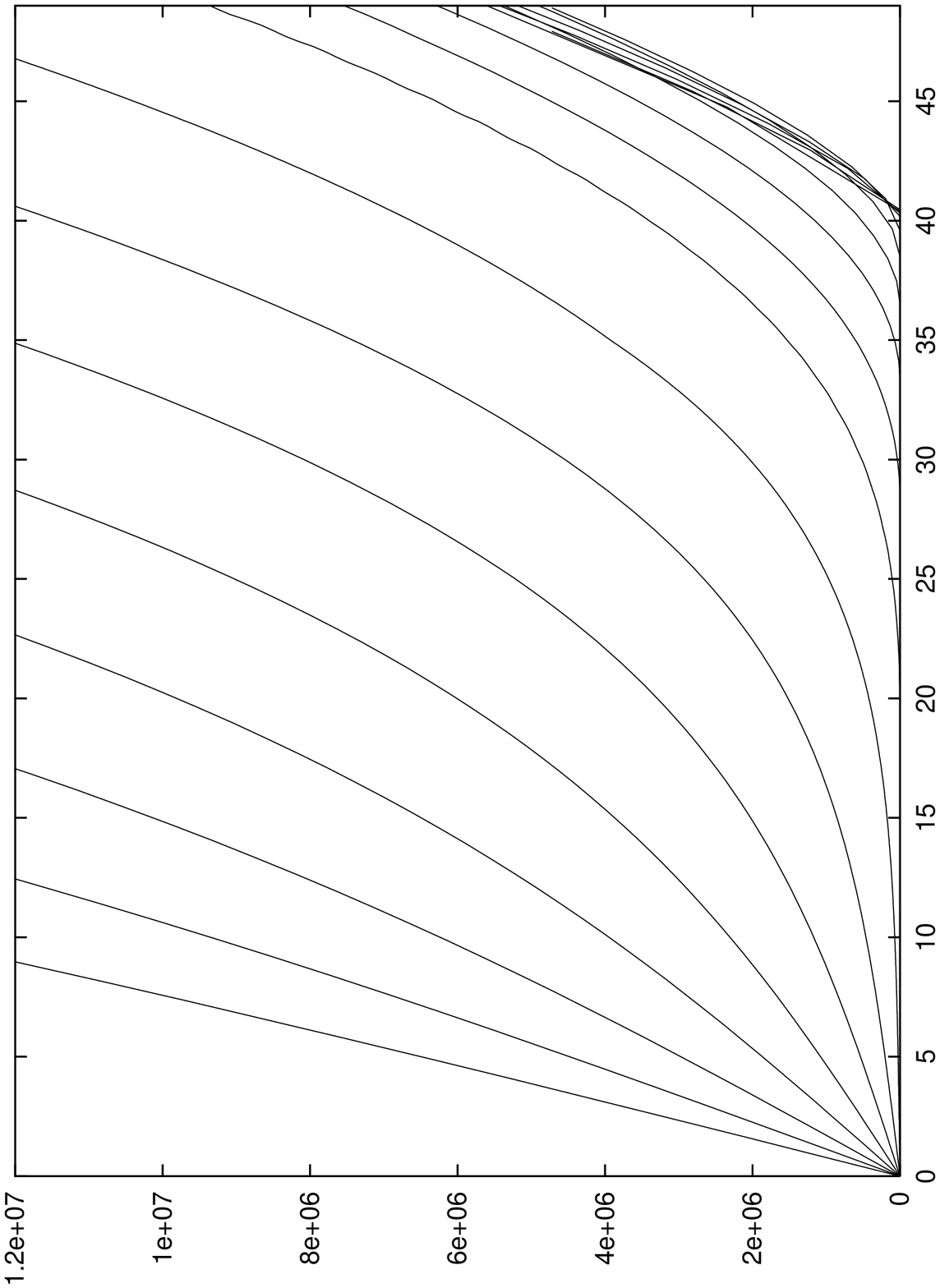}}
}
\put(-1.3,4.2){\bf $\ds{\frac{\partial U(T)/\partial \phi_R}
    {\MeV^{3}}}$}
\put(5.8,-0.5){\bf $\ds{\phi_R/\MeV}$}
\end{picture}
\end{center}
\caption{The plot shows the derivative of the
  meson potential $U(T)$ with respect to the renormalized field
  $\phi_R=(Z_\Phi\rho/2)^{1/2}$ for different values of $T$.  The
  first curve on the left corresponds to $T=175 \MeV$. The successive
  curves to the right differ in temperature by $\Delta T=10 \MeV$ down
  to $T=5 \MeV$. }
\label{Usig}
\end{figure}
The curves cover a temperature range $T = (5 - 175) \MeV$.  The first
one to the left corresponds to $T=175 \MeV$ and neighboring curves
differ in temperature by $\Delta T = 10 \MeV$. One observes only a
weak dependence of $\partial U(T)/\partial\phi_R$ on the temperature
for $T\ltap60\MeV$.  
Evaluated at the minimum of the effective potential, $\phi_R=\sigma_{0}$, 
this function connects the renormalized field expectation value with
$m_{\pi}(T)$, the source $\jmath$ and the mesonic wave function
renormalization $Z_{\Phi}(T)$ according to
\begin{equation}
  \label{Usigeq}
  \ds{\frac{\partial U(T)}{\partial\phi_R}}
  (\phi_R=\sigma_{0})=
  \ds{\frac{2\jmath}{Z_{\Phi}^{1/2}(T)}}=4 \sigma_{0}(T) 
  m_{\pi}^2(T) \; .
\end{equation} 
We point out that we have concentrated here only on the
meson field dependent part of the effective action which is related to
chiral symmetry breaking. The meson field independent part of the free
energy also depends on $T$ and only part of this temperature
dependence is induced by the scalar and quark fluctuations considered
in the present work. Most likely, the gluon degrees of freedom cannot
be neglected for this purpose. This is the reason why we do not give
results for ``overall quantities'' like energy density or pressure as
a function of $T$.

We close this section with a short assessment of the validity of our
effective quark meson model as an effective description of two flavor
QCD at non--vanishing temperature.  The identification of
qualitatively different scale intervals which appear in the context of
chiral symmetry breaking, as presented in section
\ref{ASemiQuantitativePicture} for the zero temperature case, can be
generalized to $T \neq 0$: For scales below $k_{\Phi}$ there exists a
hybrid description in terms of quarks and mesons. For $k_{\chi SB}
\leq k \ltap 600 \MeV$ chiral symmetry remains unbroken where the
symmetry breaking scale $k_{\chi SB}(T)$ decreases with increasing
temperature. Also the constituent quark mass decreases with $T$. 
The running Yukawa coupling depends only
mildly on temperature for $T\ltap 120\MeV$. \cite{BJW} (Only
near the critical temperature and for $\hat{m}=0$ the running is
extended because of massless pion fluctuations.) On the other hand,
for $k\ltap 4T$ the effective three--dimensional gauge coupling
increases faster than at $T=0$ leading to an increase of $\Lambda_{\rm
  QCD}(T)$ with $T$~\cite{RW1}. As $k$ gets closer to the scale
$\Lambda_{\rm QCD}(T)$ it is no longer justified to neglect in the
quark sector confinement effects which go beyond the dynamics of our
present quark meson model.  Here it is important to note that the
quarks remain quantitatively relevant for the evolution of the meson
degrees of freedom only for scales $k \gtap T/0.6$
(cf.~figure~\ref{Thresh}, section~\ref{FiniteTemperatureFormalism}).  In
the limit $k \ll T/0.6$ all fermionic Matsubara modes decouple from
the evolution of the meson potential. Possible sizeable confinement
corrections to the meson physics may occur if $\Lambda_{\rm QCD}(T)$
becomes larger than the maximum of $M_q(T)$ and $T/0.6$. 
This is particularly dangerous for small
$\hat{m}$ in a temperature interval around $T_c$. Nevertheless, the
situation is not dramatically different from the zero temperature case
since only a relatively small range of $k$ is concerned. We do not
expect that the neglected QCD non--localities lead to qualitative
changes.  Quantitative modifications, especially for small $\hat{m}$
and $\abs{T-T_c}$ remain possible. This would only effect the
non--universal amplitudes (see sect.~\ref{CriticalBehavior}). The size
of these corrections depends on the strength of (non--local)
deviations of the quark propagator and the Yukawa coupling from the
values computed in the quark meson model.

\subsection{Universal scaling equation of state}
\label{CriticalBehavior}

In this section \cite{BJW,BTW} 
we study the linear quark meson model in the vicinity
of the critical temperature $T_c$ close to the chiral limit
$\hat{m}=0$. In this region we find that the sigma mass
$m_\sigma^{-1}$ is much larger than the inverse temperature $T^{-1}$,
and one observes an effectively three--dimensional behavior of the
high temperature quantum field theory.  We also note that the fermions
are no longer present in the dimensionally reduced system as has been
discussed in section \ref{FiniteTemperatureFormalism}. We therefore
have to deal with a purely bosonic $O(4)$--symmetric linear sigma
model.  At the phase transition the correlation length becomes
infinite and the effective three--dimensional theory is dominated by
classical statistical fluctuations. In particular, the critical
exponents which describe the singular behavior of various quantities
near the second order phase transition are those of the corresponding
classical system.

Many properties of this system are universal, i.e.~they only depend
on its symmetry ($O(4)$), the dimensionality of space (three) and its
degrees of freedom (four real scalar components). Universality means
that the long--range properties of the system do not depend on the
details of the specific model like its short distance
interactions. Nevertheless, important properties as the value of the
critical temperature are non--universal. We emphasize that although we
have to deal with an effectively three--dimensional bosonic theory,
the non--universal properties of the system crucially depend on the
details of the four--dimensional theory and, in particular, on the
fermions. 

Our aim is a computation of the scaling form of the equation of state which
relates for arbitrary $T$ near $T_c$ 
the derivative of the free energy or effective potential $U$
to the average current quark mass $\hat{m}$. 
At the critical temperature and in the chiral limit there is no scale
present in the theory. In the vicinity of $T_c$ and for small enough
$\hat{m}$ one therefore expects a scaling behavior of the dimensionless
average potential $u_k=k^{-d}U_k$ as a function of the
rescaled field 
$\tilde{\rho}= Z_{\Phi,k} k^{2-d}\rho$~\cite{TW94-1,ABBTW95}.
(See also the review in \cite{JB}.) 

There are only two independent
scales close to the transition point which can be related to the
deviation from the critical temperature, $T-T_c$, and to the explicit
symmetry breaking by a nonvanishing 
quark mass $\hat{m}$.  As a consequence,
the properly rescaled potential can only depend on one scaling
variable.  A possible choice is the Widom scaling
variable~\cite{Widom}
\begin{equation}
  \label{XXX20}
  x=\frac{\left( T-T_c\right)/T_c}
  {\left(2\ovl{\sigma}_0/T_c\right)^{1/\beta}}\; .
\end{equation}
Here $\beta$ is the critical exponent of the order parameter
$\ovl{\sigma}_0$ in the chiral limit $\hat{m}=0$ (see equation
(\ref{NNN21})).  With
$U^\prime(\rho=2\ovl{\sigma}_0^2)=\jmath/(2\ovl{\sigma}_0)$ the Widom
scaling form of the equation of state reads~\cite{Widom}
\begin{equation}
  \label{XXX21}
  \frac{\jmath}{T_c^3}=
  \left(\frac{2\ovl{\sigma}_0}{T_c}\right)^\delta f(x)
\end{equation}
where the exponent $\delta$ is related to the behavior of the order
parameter according to (\ref{NNN21b}).  The equation of state
(\ref{XXX21}) is written for convenience directly in terms of
four--dimensional quantities.  They are related to the corresponding
effective variables of the three--dimensional theory by appropriate
powers of $T_c$.  The source $\jmath$ is determined by the average
current quark mass $\hat{m}$ as $\jmath=2\ovl{m}^2_{k_\Phi}\hat{m}$.
The mass term at the compositeness scale, $\ovl{m}^2_{k_\Phi}$, also
relates the chiral condensate to the order parameter according to
$\VEV{\ovl{\psi}\psi}=-2\ovl{m}^2_{k_\Phi}(\ovl{\sigma}_0-\hat{m})$.  The
critical temperature of the linear quark meson model was found in
section \ref{TheQuarkMesonModelAtTNeq0} to be \mbox{$T_c=100.7\MeV$}.

The scaling function $f$ is universal up to the model specific
normalization of $x$ and itself. Accordingly, all models in the same
universality class can be related by a rescaling of $\ovl{\sigma}_0$
and $T-T_c$. The non--universal normalizations for the quark meson
model discussed here are defined according to
\begin{equation}
  \label{norm}
  f(0)=D\quad, \qquad f(-B^{-1/\beta})=0\; .
\end{equation}
We find $D=1.82\cdot10^{-4}$, $B=7.41$ and our result for $\beta$ is
given in table~\ref{tab2}. Apart from the immediate vicinity of the
zero of $f(x)$ we find the following two parameter fit
for the scaling function,
\begin{equation}
  \label{ffit}
  \begin{array}{rcl}
    \ds{f_{\rm fit}(x)}&=&\ds{1.816 \cdot 10^{-4} (1+136.1\, x)^2 \,
      (1+160.9\, \theta\,
      x)^{\Delta}}\nnn
    &&
    \ds{(1+160.9\, (0.9446\, \theta^{\Delta})^{-1/(\gamma-2-\Delta)} 
      \, x)^{\gamma-2-\Delta}}
  \end{array}
\end{equation}
to reproduce the numerical results for $f$ and $df/dx$ at the $1-2\%$
level with $\theta=0.625$ $(0.656)$, $\Delta=-0.490$ $(-0.550)$ for $x
> 0$ $(x < 0)$ and $\gamma$ as given in table \ref{tab2}.  The
universal properties of the scaling function can be compared with
results obtained by other methods for the three--dimensional $O(4)$
Heisenberg model.  In figure \ref{scalfunc} we display our results
along with those obtained from lattice Monte Carlo simulation
\cite{Tou}, second order epsilon expansion \cite{BWW73-1} and mean
field theory.
\begin{figure}
\unitlength1.0cm
\begin{center}
\begin{picture}(17.,12.)
\put(0.3,5.5){$\ds{\frac{2\ovl{\sigma}_0/T_c}
{(\jmath/T_c^3 D)^{1/\delta}}}$}
\put(8.5,-0.2){$\ds{\frac{(T-T_c)/T_c}
{(\jmath/T_c^3 B^{\delta} D)^{1/\beta \delta}}}$}
\put(8.,2.5){\footnotesize $\mbox{average action}$}
\put(4.19,11.2){\footnotesize $\mbox{average action}$}
\put(13.8,2.2){\footnotesize $\epsilon$}
\put(3.53,11.19){\footnotesize $\epsilon$}
\put(11.3,2.05){\footnotesize $\mbox{MC}$}
\put(3.9,10.){\footnotesize $\mbox{MC}$}
\put(13.2,1.8){\footnotesize $\mbox{mf}$}
\put(6.5,9.2){\footnotesize $\mbox{mf}$}
\put(-1.,-5.9){
\epsfysize=21.cm
\epsfxsize=18.cm
\epsffile{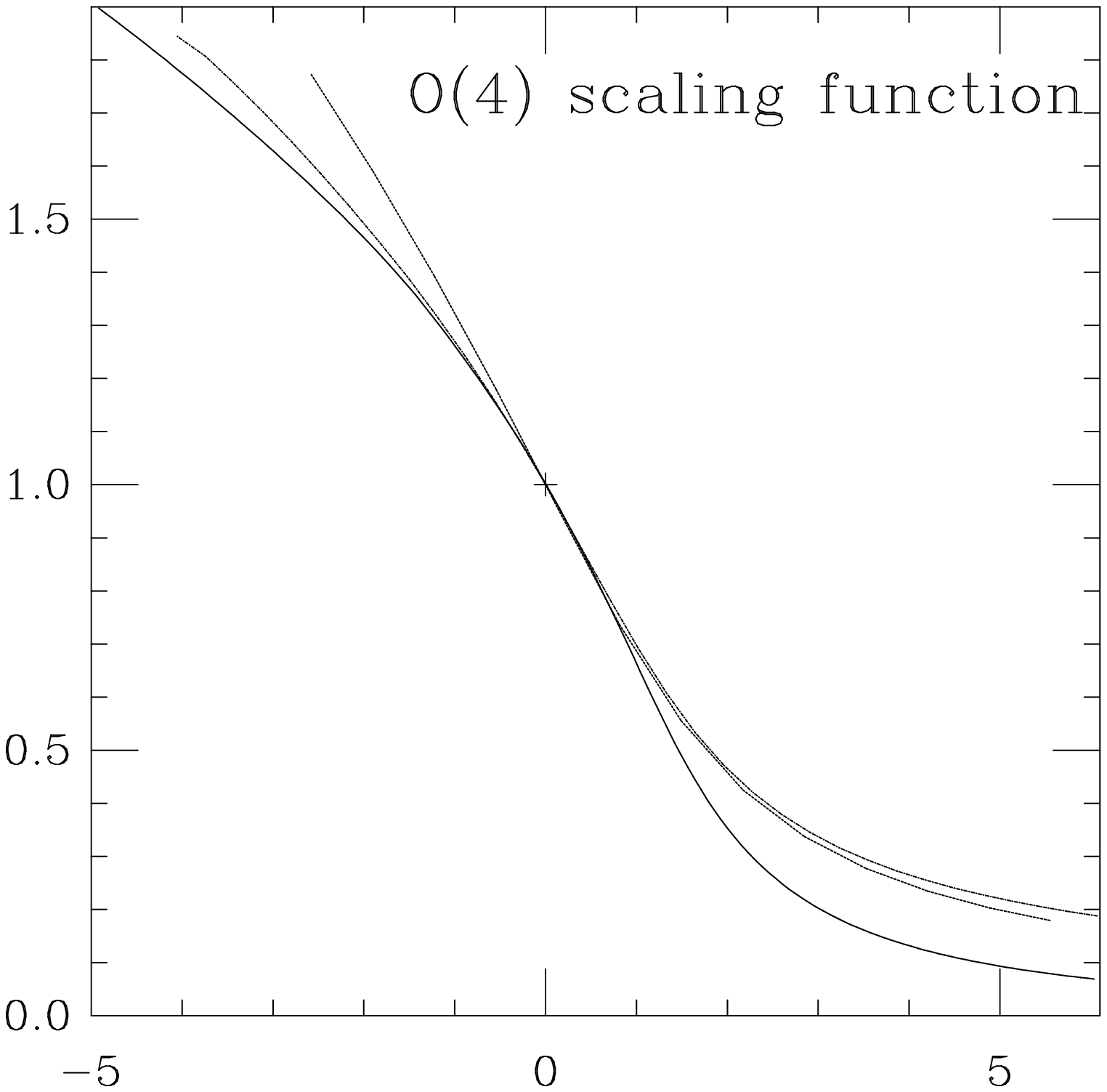}}
\put(1.25,0.483){
\epsfysize=13.45cm
\epsfxsize=11.22cm
\rotate[r]{\epsffile{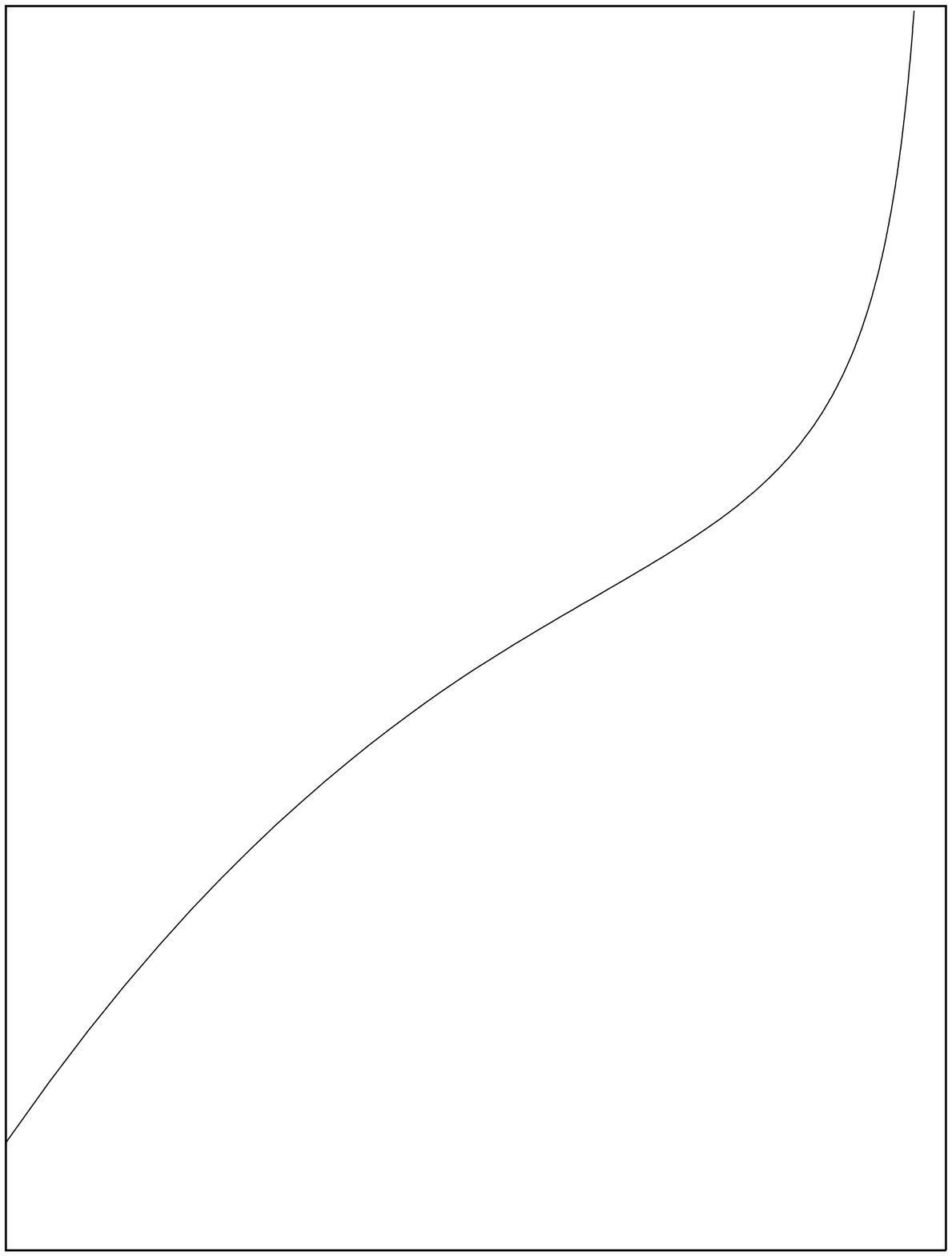}}}
%\put(0.,0.){\framebox(10.,10.)}
\end{picture}
\end{center}
\caption[]{The figure shows a comparison of our results, denoted by ``average
  action'', with results of other methods for the scaling function of
  the three--dimensional $O(4)$ Heisenberg model. We have labeled the
  axes for convenience in terms of the expectation value
  $\ovl{\sigma}_0$ and the source $\jmath$ of the corresponding
  four--dimensional theory.  The constants $B$ and $D$ specify the
  non--universal amplitudes of the model (cf.~eq.~\ref{norm}).  The
  curve labeled by ``MC'' represents a fit to lattice Monte Carlo
  data. The second order epsilon expansion \cite{BWW73-1} and mean
  field results are denoted by ``$\epsilon$'' and ``mf'',
  respectively.  Apart from our results the curves are taken
  from ref.~\cite{Tou}.
  \label{scalfunc}
  }
\end{figure}
We observe a good agreement of average action, lattice and epsilon
expansion results within a few per cent for $T < T_c$. Above $T_c$ the
average action and the lattice curve go quite close to each other with
a substantial deviation from the epsilon expansion and mean field
scaling function. (We note that the question of a better
  agreement of the curves for $T < T_c$ or $T > T_c$ depends on the
  chosen non--universal normalization conditions for $x$ and $f$,
  eq.\ (\ref{norm}).)

Before we use the scaling function $f(x)$ to discuss the general
temperature and quark mass dependent case, we consider the limits
$T=T_c$ and $\hat{m}=0$, respectively.  In these limits the behavior
of the various quantities is determined solely by critical amplitudes
and exponents. In the spontaneously broken phase ($T<T_c$) and in the
chiral limit we observe that the renormalized and unrenormalized order
parameters scale according to
\begin{equation}
  \label{NNN21}
  \begin{array}{rcl}
    \ds{\frac{2\sigma_0(T)}{T_c}} &=& \ds{
      \left(2E\right)^{1/2}
        \left(\frac{T_c-T}{T_c}\right)^{\nu/2}
      }\; ,\nnn
    \ds{\frac{2\ovl{\sigma}_0(T)}{T_c}} &=& \ds{
      B \left(\frac{T_c-T}{T_c}\right)^{\beta}
      }\; ,
  \end{array}
\end{equation}
respectively, with $E=0.814$ and the value of $B$ given above.  In the
symmetric phase the renormalized mass $m=m_\pi=m_\si$ and the
unrenormalized mass $\ovl{m}=Z_\Phi^{1/2}m$ behave as
\begin{equation}
  \label{NNN21a}
  \begin{array}{rcl}
    \ds{\frac{m(T)}{T_c}} &=& \ds{
      \left(\xi^+\right)^{-1}
      \left(\frac{T-T_c}{T_c}\right)^\nu
      }\; ,\nnn
    \ds{\frac{\ovl{m}(T)}{T_c}} &=& \ds{
      \left( C^+\right)^{-1/2}
      \left(\frac{T-T_c}{T_c}\right)^{\gamma/2}
       \; , }
  \end{array}
\end{equation}
where $\xi^+=0.270$, $C^+=2.79$. For $T=T_c$ and
non--vanishing current quark mass we have
\begin{equation}
  \label{NNN21b}
  \begin{array}{rcl}
    \ds{\frac{2\ovl{\sigma}_0}{T_c}} &=& \ds{
      D^{-1/\delta}
        \left(\frac{\jmath}{T_c^3}\right)^{1/\delta}
      }
  \end{array}
\end{equation}
with the value of $D$ given above. 

Though the five amplitudes $E$, $B$, $\xi^+$, $C^+$ and $D$ are not
universal there are ratios of amplitudes which are invariant under a
rescaling of $\ovl{\sigma}_0$ and $T-T_c$. Our results for the
universal amplitude ratios are
\begin{equation}
  \label{ABC01}
  \begin{array}{rcl}
    \ds{R_\chi} &=& \ds{C^+ D B^{\delta-1}=1.02}\; ,\nnn
    \ds{\tilde{R}_\xi} &=& \ds{
      (\xi^+)^{\beta/\nu}D^{1/(\delta+1)}B=0.852}\; ,\nnn
    \ds{\xi^+ E} &=& \ds{0.220}\; .
  \end{array}
\end{equation}
Those for the critical exponents are given in table \ref{tab2}.
\begin{table}
\begin{center}
\begin{tabular}{|c||l|l|l|l|l|} \hline
   &
  $\nu$ &
  $\gamma$ &
  $\delta$ &
  $\beta$ &
  $\eta$
  \\[0.5mm] \hline\hline
  average action &
  $0.787$ &
  $1.548$ &
  $4.80$ &
  $0.407$ &
  $0.0344$
  \\ \hline
  FD &
  $0.73(2)$ &
  $1.44(4)$ &
  $4.82(5)$ &
  $0.38(1)$ &
  $0.03(1)$
  \\ \hline
  MC &
  $0.7479(90)$ &
  $1.477(18)$ &
  $4.851(22)$ &
  $0.3836(46)$ &
  $0.0254(38)$
  \\ \hline
\end{tabular}
\caption[]{The table shows the critical exponents 
  corresponding to the three--dimensional $O(4)$--Heisenberg model.
  Our results are denoted by ``average action'' whereas ``FD''
  labels the exponents obtained from perturbation series at fixed
  dimension to seven loops \cite{BMN78-1}. 
  The bottom line contains lattice Monte Carlo
  results \cite{KK95-1}.
  \label{tab2}}
\end{center}
\end{table}
For comparison table~\ref{tab2} also
gives the results from perturbation series at fixed dimension to 
seven--loop order~\cite{BMN78-1,ZJ} as well as lattice Monte Carlo
results~\cite{KK95-1} which have been used for the lattice form of the
scaling function in figure~\ref{scalfunc}.\footnote{See also
  ref.~\cite{MT97-1} and references therein for a calculation
  of critical exponents using similar methods as in this work.} 
There are only two independent
amplitudes and critical exponents, respectively. They are related by
the usual scaling relations of the three--dimensional scalar
$O(N)$--model~\cite{ZJ} which we have explicitly verified by the
independent calculation of our exponents.

We turn to the discussion of the scaling behavior of the chiral
condensate $\VEV{\ovl{\psi}\psi}$ for the general case of a temperature
and quark mass dependence.  In figure~\ref{ccc_T} 
we have displayed our results for the scaling equation of state in
terms of the chiral condensate
\begin{equation}
  \label{XXX30}
  \VEV{\ovl{\psi}\psi}=
  -\ovl{m}^2_{k_\Phi}T_c
  \left(\frac{\jmath/T_c^3}{f(x)}\right)^{1/\delta}+
    \jmath
\end{equation}
as a function of $T/T_c=1+x(\jmath/T_c^3 f(x))^{1/\beta\delta}$ for
different quark masses or, equivalently, different values of $\jmath$.
The curves shown in figure \ref{ccc_T} correspond to quark masses
$\hat{m}=0$, $\hat{m}=\hat{m}_{\rm phys}/10$, $\hat{m}=\hat{m}_{\rm
  phys}$ and $\hat{m}=3.5\hat{m}_{\rm phys}$ or, equivalently, to zero
temperature pion masses $m_\pi=0$, $m_\pi=45\MeV$, $m_\pi=135\MeV$ and
$m_\pi=230\MeV$, respectively. 
The scaling form (\ref{XXX30}) for the chiral condensate is exact only
in the limit $T\to T_c$, $\jmath\ra0$.  It is interesting to find the
range of temperatures and quark masses for which $\VEV{\ovl{\psi}\psi}$
approximately shows the scaling behavior (\ref{XXX30}).  This can be
infered from a comparison (see figure\ \ref{ccc_T}) with our full
non--universal solution for the $T$ and $\jmath$ dependence of
$\VEV{\ovl{\psi}\psi}$ as described in
section~\ref{TheQuarkMesonModelAtTNeq0}. For $m_\pi=0$ one observes
approximate scaling behavior for temperatures $T\, \gtap\, 90\MeV$. This
situation persists up to a pion mass of $m_\pi=45\MeV$. Even for the
realistic case, $m_\pi=135\MeV$, and to a somewhat lesser extent for
$m_\pi=230\MeV$ the scaling curve reasonably reflects the physical
behavior for $T\gtap T_c$. For temperatures below $T_c$, however, the
zero temperature mass scales become important and the scaling
arguments leading to universality break down.

The above comparison may help to shed some light on the use of
universality arguments away from the critical temperature and the
chiral limit. One observes that for temperatures above $T_c$ the
scaling assumption leads to quantitatively reasonable results even for
a pion mass almost twice as large as the physical value. This in turn
has been used for two flavor lattice QCD as theoretical input to guide
extrapolation of results to light current quark masses.  From
simulations based on a range of pion masses $0.3\ltap
m_\pi/m_\rho\ltap0.7$ and temperatures $0<T\ltap250\MeV$ a
``pseudocritical temperature'' of approximately $140\MeV$ with a weak
quark mass dependence is reported~\cite{MILC97-1}. Here the
``pseudocritical temperature'' $T_{pc}$ is defined as the inflection
point of $\VEV{\ovl{\psi}\psi}$ as a function of temperature.   
For comparison with lattice data we have displayed in
figure \ref{ccc_T} the temperature dependence of the chiral condensate
for a pion mass $m_\pi=230\MeV$.  From the free energy of the linear
quark meson model we obtain in this case a pseudocritical temperature
of about $150\MeV$ in reasonable agreement with lattice results 
\cite{MILC97-1,Laermann}. In contrast, for the critical temperature in
the chiral limit we obtain $T_c=100.7\MeV$.  This value is
considerably smaller than the lattice results of about $(140 - 150)
\MeV$ obtained by extrapolating to zero quark mass in
ref.~\cite{MILC97-1}.  We point out that for pion masses as large as
$230\MeV$ the condensate $\VEV{\ovl{\psi}\psi}(T)$ is almost linear
around the inflection point for quite a large range of temperature.
This makes a precise determination of $T_c$ somewhat difficult.
Furthermore, figure \ref{ccc_T} shows that the scaling form of
$\VEV{\ovl{\psi}\psi}(T)$ underestimates the slope of the physical
curve. Used as a fit with $T_c$ as a parameter this can lead to an
overestimate of the pseudocritical temperature in the chiral limit.

The linear quark meson model exhibits a second order phase transition
for two quark flavors in the chiral limit. As a consequence the model
predicts a scaling behavior near the critical temperature and the
chiral limit which can, in principle, be tested in lattice
simulations. For the quark masses used in the present lattice studies
the order and universality class of the transition in two flavor QCD
remain a partially open question. Though there are results from the
lattice giving support for critical scaling
there are also simulations with two flavors that reveal
significant finite size effects and problems with
$O(4)$ scaling~\cite{Laermann}.

\subsection{Additional degrees of freedom}
\label{AdditionalDegreesOfFreedom}

So far we have investigated the chiral phase transition of QCD as
described by the linear $O(4)$--model containing the three pions and
the sigma resonance as well as the up and down quarks as degrees of
freedom. Of course, it is clear that the spectrum of QCD is much
richer than the states incorporated in our model. It is therefore
important to ask to what extent the neglected degrees of freedom like
the strange quark, strange (pseudo)scalar mesons, (axial)vector
mesons, baryons, etc., might be important for the chiral dynamics of
QCD.  Before doing so it is instructive to first look into the
opposite direction and investigate the difference between the linear
quark meson model described here and chiral perturbation theory based
on the non--linear sigma model~\cite{GL82-1}. In some sense, chiral
perturbation theory is the minimal model of chiral symmetry breaking
containing only the Goldstone degrees of freedom\footnote{
For vanishing temperature it has
been demonstrated~\cite{QuMa,JW97-1} that the results of
chiral perturbation theory can be reproduced within the linear meson
model once certain higher dimensional operators in its effective
action are taken into account for the three flavor case.}. 
By construction it
is therefore only valid in the spontaneously broken phase. 
For small temperatures (and
momentum scales) chiral perturbation theory is expected to give a reliable
description whereas for high temperatures or close to $T_c$ one
expects sizeable corrections.  

From~\cite{GL87-1} we infer the three--loop result for the
temperature dependence of the chiral condensate in the chiral limit
for $N$ light flavors
\begin{equation}
  \label{BBB100}
  \begin{array}{rcl}
    \ds{\VEV{\ovl{\psi}\psi}(T)_{\chi PT}} &=& \ds{
      \VEV{\ovl{\psi}\psi}_{\chi PT}(0)
      \Bigg\{1-\frac{N^2-1}{N}\frac{T^2}{12F_0^2}-
        \frac{N^2-1}{2N^2}
        \left(\frac{T^2}{12F_0^2}\right)^2}\nnn
    &+& \ds{
      N(N^2-1)\left(\frac{T^2}{12F_0^2}\right)^3
      \ln\frac{T}{\Gamma_1}
      \Bigg\} +\Oc(T^8)}\; .
  \end{array}
\end{equation}
The scale $\Gamma_1$ can be determined from the $D$--wave isospin zero
$\pi\pi$ scattering length and is given by $\Gamma_1=(470\pm100)\MeV$.
The constant $F_0$ is (in the chiral limit) identical to the pion
decay constant $F_0=f_\pi^{(0)}=80.8\MeV$ (cf.\ table \ref{tab1}). 
In figure \ref{cc_T} we
have plotted the chiral condensate as a function of $T/F_0$ for both,
chiral perturbation theory according to (\ref{BBB100}) and for the
linear quark meson model.
\begin{figure}
\unitlength1.0cm
\begin{center}
\begin{picture}(13.,7.0)

\put(0.0,0.0){
\epsfysize=11.cm
\rotate[r]{\epsffile{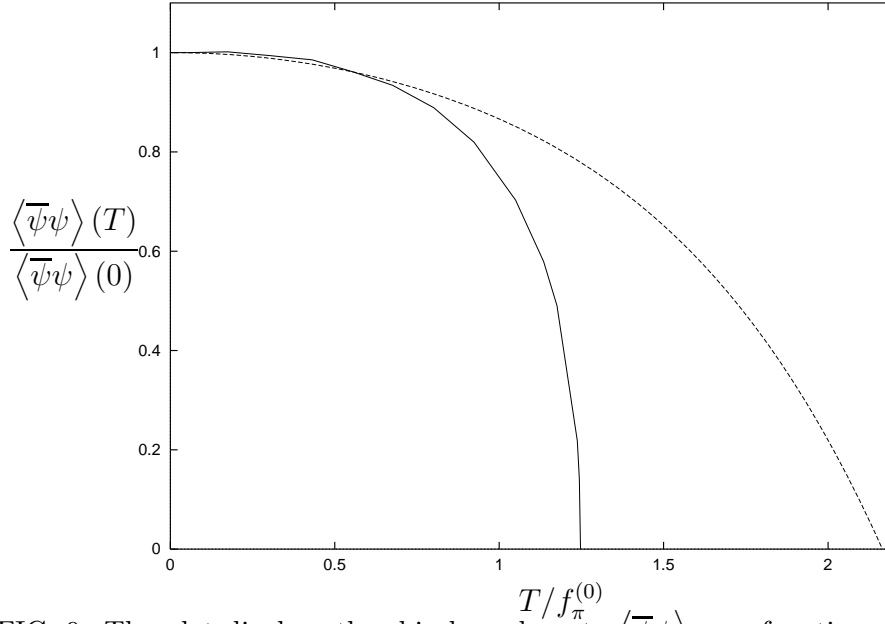}}
}
\put(-1.0,4.2){\bf 
  $\ds{\frac{\VEV{\ovl{\psi}\psi}(T)}{\VEV{\ovl{\psi}\psi}(0)} }$}
\put(5.8,-0.5){\bf $\ds{T/f_\pi^{(0)}}$}

\end{picture}
\end{center}
\caption{The plot displays the chiral condensate
  $\VEV{\ovl{\psi}\psi}$ as a function of $T/f_\pi^{(0)}$. The solid
  line corresponds to our results for vanishing average current quark
  mass $\hat{m}=0$ whereas the dashed line shows the corresponding
  three--loop chiral perturbation theory result for
  $\Gamma_1=470\MeV$.}
\label{cc_T}
\end{figure}
As expected the agreement for small $T$ is very good. Nevertheless,
small numerical deviations are manifest even for $T\ll T_c$
due to quark and sigma meson loop contributions. We observe
for larger values of $T$, say for $T\, \gtap\, 0.8 f_\pi^{(0)}$, 
that the deviations
become significant. Chiral
perturbation theory is not expected to correctly 
reproduce the critical behavior of
the system near its second order phase transition.

Within the language of chiral perturbation theory the neglected
effects of thermal quark fluctuations may be described by an effective
temperature dependence of the parameter $F_0(T)$. We notice that the
temperature at which these corrections become important equals
approximately one third of the constituent quark mass $M_q(T)$ or the
sigma mass $m_\sigma(T)$, respectively, in perfect agreement with
figure~\ref{Thresh}. As suggested by this figure the onset of the
effects from thermal fluctuations of heavy particles with a
$T$--dependent mass $m_H(T)$ is rather sudden for $T\gtap m_H(T)/3$.
These considerations also apply to our two flavor quark meson model.
Within full QCD we expect temperature dependent initial values at
$k_\Phi$.

The dominant contribution to the temperature dependence of the initial
values presumably arises from the influence of the mesons containing
strange quarks as well as the strange quark itself.  Here the quantity
$\ovl{m}^2_{k_\Phi}$ seems to be the most important one.  (The
temperature dependence of higher couplings like $\lambda(T)$ is not
very relevant if the IR attractive behavior remains valid, i.e.~if
$Z_{\Phi,k_\Phi}$ remains small for the range of temperatures
considered. We neglect a possible $T$--dependence of the current quark
mass $\hat{m}$.) In particular, for three flavors the potential
$U_{k_\Phi}$ contains a term
\begin{equation}
  \label{LLL12}
  -\frac{1}{2}\ovl{\nu}_{k_\Phi}
  \left(\det\Phi+\det\Phi^\dagger\right)=
  -\ovl{\nu}_{k_\Phi}\vph_s\Phi_{uu}\Phi_{dd}+\ldots
\end{equation}
which reflects the axial $U_A(1)$ anomaly. It yields a contribution to
the effective mass term proportional to the expectation value
$\VEV{\Phi_{ss}}\equiv\vph_s$, i.e.
\begin{equation}
  \label{LLL13}
  \Delta\ovl{m}^2_{k_\Phi}=
  -\frac{1}{2}\ovl{\nu}_{k_\Phi}\vph_s\; .
\end{equation}
Both, $\ovl{\nu}_{k_\Phi}$ and $\vph_s$, depend on $T$.  We expect
these corrections to become relevant only for temperatures exceeding
$m_K(T)/3$ or $M_s(T)/3$. We note that the temperature dependent kaon
and strange quark masses, $m_K(T)$ and $M_s(T)$, respectively, may be
somewhat different from their zero temperature values but we do not
expect them to be much smaller. A typical value for these scales is
around $500\MeV$. Correspondingly, the thermal fluctuations neglected
in our model should become important for $T\gtap170\MeV$. It is even
conceivable that a discontinuity appears in $\vph_s(T)$ for
sufficiently high $T$ (say $T\simeq170\MeV$). This would be reflected
by a discontinuity in the initial values of the $O(4)$--model leading
to a first order transition within this model.  Obviously, these
questions should be addressed in the framework of the three flavor
$SU_L(3)\times SU_R(3)$ quark meson model. Work in this direction is
in progress.

We note that the temperature dependence of $\ovl{\nu}(T)\vph_s(T)$ is
closely related to the question of an effective high temperature
restoration of the axial $U_A(1)$ symmetry~\cite{PW84-1,Shu94-1}.  The
$\eta^\prime$ mass term is directly proportional to this combination,
$m_{\eta^\prime}^2(T)-m_\pi^2(T)\simeq\frac{3}{2}\ovl{\nu}(T)
\vph_s(T)$ \cite{JuWet96}. Approximate $U_A(1)$ restoration would occur
if $\vph_s(T)$ or $\ovl{\nu}(T)$ would decrease sizeable for large $T$.
For realistic QCD this question should be addressed by a three flavor
study. Within two flavor QCD the combination $\ovl{\nu}_k\vph_s$ is
replaced by an effective anomalous mass term $\ovl{\nu}_k^{(2)}$.   
We add that this question has also been
studied within full two flavor QCD in lattice
simulations~\cite{Laermann} but no final conclusion can be drawn yet.

To summarize, we have found that the effective two flavor quark meson
model presumably gives a good description of the temperature effects
in two flavor QCD for a temperature range $T\ltap170\MeV$. Its
reliability should be best for low temperature where our results agree
with chiral perturbation theory.  However, the range of validity is
considerably extended as compared to chiral perturbation theory and
includes, in particular, the critical behavior of the second order
phase transition in the chiral limit. The method using nonperturbative
flow equations within the linear quark meson model may also help to shed
some light on the remaining pressing questions at high temperature, like
the nature of the phase transition for realistic values of the strange
quark mass. 

\pagebreak

\section{High baryon number density}

Over the past years, considerable progress has been achieved
in our understanding of high temperature QCD, where simulations on the
lattice and universality arguments played an essential role.
The results of the renormalization group approach to the 
effective quark meson model provides the link between
the low temperature chiral perturbation theory
domain of validity and the high temperature domain of critical
phenomena~\cite{BJW}. On the other hand, our knowledge of the high density 
properties of strongly interacting matter
is rudimentary so far. There are severe problems to use standard simulation
algorithms at nonzero chemical potential on the lattice because of a
complex fermion determinant \cite{barbour}.     
In this section we will try to get some insight into matter at 
high density using nonperaturbative
flow equations \cite{BJW2}.

\subsection{Cold dense matter}

We will consider here a few general aspects of QCD at nonzero
baryon number density and vanishing temperature. For a more detailed
recent review see ref.\ \cite{SB}.  
It is instructive to consider for a moment a free theory of fermions 
with mass $m$ carrying one unit of baryon charge 
where the associated chemical potential is $\mu$. When $\mu > 0$
the ground state is the Fermi sphere with radius 
$p_F=\sqrt{\mu^2-m^2}$ and therefore $n(\mu)=(\mu^2-m^2)^{3/2}/(3 \pi^2)$.
For $\mu < m$ the density vanishes identically and $n$ represents an
order parameter which distinguishes the two phases. One clearly
observes the nonanalytic behavior at $\mu=m$ which denotes
the ``onset'' value for nonzero density. 

{\em Nuclear liquid gas transition.} ---
What is the ``onset'' value for nonzero density in QCD?  The low density 
properties of QCD may be inferred from what is known empirically about
bulk nuclear matter. Here nuclear matter denotes a
uniform, isospin symmetric large sample of matter. Uniform infinite
symmetric matter is an idealization whose properties, however,
are related to those of finite nuclei in the liquid droplet
model of the nucleus. Based on 
information about finite nuclei one finds an energy per baryon of 
$m_N-16 \MeV$ where
$m_N \simeq m_n \simeq m_p$ is the nucleon mass and $16 \MeV$
is the binding energy per nucleon. In QCD one expects the density to jump 
at $\mu_{\rm nuc} \simeq m_N-16 \MeV$ from zero to saturation density $n_0$ at 
which the pressure is zero. A schematic plot is shown in figure
\ref{densitypsa}. 
\begin{figure}
\centerline{\epsfysize 2in\epsfbox{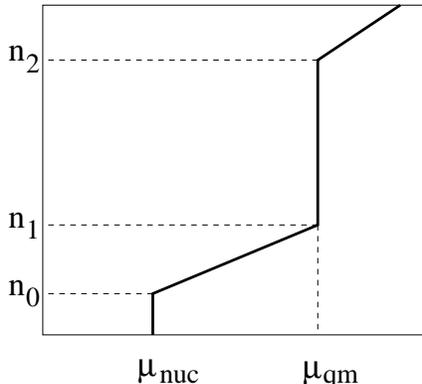}}
\caption{Schematic plot of the baryon number density in QCD as a function
of the chemical potential for zero temperature.}
\label{densitypsa}
\end{figure}
For small temperature this corresponds to 
the transition between a gas of nucleons and nuclear matter.   
The situation changes once other interactions, in 
particular electromagnetic 
interactions, are taken into account. 
Including electromagnetism and adding electrons for charge neutrality
the ``onset'' density is that of the lowest energy state of hadronic 
matter -- ordinary iron with a density about $10^{-14}$ times nuclear 
matter density.

{\em Chiral phase transition} ---
We have reasonable expectations, but much less evidence, for the
behavior of QCD at densities much larger than nuclear matter density.
One may suppose as a starting
point that at very high density quarks behave nearly freely and 
form large Fermi surfaces. The leading interactions 
for particle--hole excitations near the Fermi surface are elastic
scatterings among these particles and holes, since other 
possibilities are blocked by the Pauli exclusion principle. 
Because the momenta are large, scattering processes 
typically involve large momentum transfers and are therefore indeed
characterized by small couplings due to asymptotic freedom. (This
self--consistent argument can be worked out to include also
the neglected ``small angle'' scatterings.)

What can we expect at a chemical potential 
$\mu \simeq \mu_{\rm qm}$ (cf.\ figure \ref{densitypsa}) 
associated with the quark hadron transition,
or at densities of about two to ten times nuclear matter density?
In vacuum chiral symmetry breaking is caused by a condensate
of quark--antiquark pairs with zero net momentum.
Since at nonzero density all particle states
up to the Fermi momentum $p_F$ are occupied, 
there is no possibility for correlated pairs
at low momenta, and at high momentum the quark--antiquark pairs have
large energy (cf.\ also figure \ref{fermis} in section \ref{colorsc}). 
At sufficiently high density we therefore expect the chiral condensate
$\langle \ovl{\psi} \psi \rangle$ to drop significantly.
On the other hand there are attractive quark--quark channels. 
A particularly interesting 
possibility is the formation of Cooper pairs of quarks at high density
which condense. We will discuss this striking feature in section 
\ref{colorsc}. 
First we will
concentrate on the fundamental question for QCD about the nature
of the transition that restores chiral symmetry.

\subsection{The quark meson model at nonzero density}

We extend our discussion of the linear quark meson model to nonzero
baryon number density \cite{BJW2}. 
The model can be viewed as a generalization
of the Nambu--Jona-Lasinio model for QCD~\cite{klevansky}.   
Calculations at nonzero baryon
density are typically based on a mean field approximation and it was
claimed~\cite{klevansky,SB} that the order of the chiral 
phase transition is
ambiguous. Recently, the order of this high density transition has raised
considerable interest. One can argue on general grounds~\cite{BR,SB} that
if two--flavor QCD has a second order transition (crossover) at high
temperatures and a first order transition at high densities, then there
exists a tricritical point (critical endpoint) with long--range
correlations in the phase diagram. As pointed out in section
\ref{intro}, the physics around this point is
governed by universality and may allow for distinctive signatures in heavy
ion collisions~\cite{BR,SB,SRS}. We will discuss this important
issue in section \ref{tricriticalpoint}.

We will observe that in a proper treatment
beyond mean field theory the order of the chiral phase
transition can be fixed within the models under investigation. For this
purpose, a crucial observation is the strong attraction of the flow to
partial infrared fixed points discussed in section 
\ref{FlowEquationsAndInfraredStability}~\cite{JW,BJW}. 
The two remaining relevant or
marginal parameters can be fixed by the phenomenological values of
$f_{\pi}$ and the constituent quark mass. For two massless quark flavors we
find that chiral symmetry restoration occurs via a first order transition
between a phase with low baryon density and a high density phase.
We emphasize, nevertheless, that the linear quark meson model captures the
low density properties of QCD only incompletely since the effects of
confinement are not included.  In particular, for a discussion of the
liquid--gas nuclear transition the inclusion of nucleon degrees of freedom
seems mandatory~\cite{NucMat}. On the other hand this model is expected to
provide a reasonable description of the high density properties 
of QCD.

In quantum field theory the effects of a non--vanishing baryon density in
thermal equilibrium or the vacuum are described by adding to the classical
action a term proportional to the chemical potential $\mu$,
\begin{equation}
  \label{AAA001}
  \Delta_\mu S=i\mu\sum_j 
  \int_0^{1/T}d x_0
  \int d^3 \vec{x}\ \ovl{\psi}_j\gm^0\psi_j\equiv
  -3\frac{\mu}{T}B\; .
\end{equation}
For quarks the sum is over $N_{\rm c}$ colors and $N_{\rm F}$ flavors.
With our conventions
$\mu$ corresponds to the chemical potential for the quark number density.
The baryon number density $n$ can be obtained from the $\mu$--dependence of
the Euclidean effective action $\Gamma$, evaluated at its minimum for fixed
temperature $T$ and volume\footnote{More precisely, $B$ counts the number
  of baryons minus antibaryons. For $T\ra0$ the factor $T/V$ is simply the
  inverse volume of four dimensional Euclidean space.} $V$
\begin{equation}
  \label{AAA002}
  n\equiv\frac{\VEV{B}}{V}=
  -\frac{1}{3}\frac{\partial}{\partial\mu}
  \left.\frac{\Gamma_{\rm min}T}{V}\right|_{\rm T,V}\; .
\end{equation}

We will restrict our discussion here to two massless quarks. A
more realistic treatment would have to include the small up and down quark
masses and the finite, but much heavier, strange quark mass. Though their
inclusion will change certain quantitative estimates, they most likely do
not change the qualitative outcome of the investigation as is
discussed below. The linear quark meson model is defined in section 
\ref{TheQuarkMesonModelAtT=0}. 
At nonzero
temperature $T$ and chemical potential $\mu$ our ansatz for $\Gamma_k$ reads
\begin{eqnarray}
  \label{truncation} 
  \Gm_k &=& \ds{
    \int^{1/T}_0 dx^0\int d^3x \Bigg\{  
    i \ovl{\psi}^a (\gamma^{\mu}\pa_{\mu} + \mu \gamma^0) \psi_a
    +\ovl{h}_k {\ovl{\psi}}^a \left[ \frac{1+\gamma^5}{2} {\Phi_a}^b
    - \frac{1-\gamma^5}{2} {(\Phi^{\dagger})_a}^b\right] \psi_b
    }\nnn 
  && \ds{
    \qquad\qquad\qquad \quad +Z_{\Phi,k} 
    \pa_{\mu}\Phi^*_{ab}\pa^{\mu}\Phi^{ab}
    +U_k(\ovl{\rho};\mu,T)\Bigg\} 
    }\, .
\end{eqnarray}
A nonzero chemical
potential $\mu$ to lowest order results in the term $\sim i\mu
\ovl{\psi}^a\gamma^0\psi_a$ appearing on the right hand side
of~(\ref{truncation}).  Our approximation neglects the dependence of
$Z_{\Phi,k}$ and $\ovl{h}_k$ on $\mu$ and $T$. We also neglect a possible
difference in normalization of the quark kinetic term and the baryon number
current.  The form of the effective action at the compositeness scale,
$\Gamma_{k_\Phi}[\psi,\Phi]$, serves as an initial value for the
renormalization group flow of $\Gamma_{k}[\psi,\Phi]$ for $k<k_\Phi$.  We
will consider here the case that $Z_{\Phi,k_\Phi}\ll1$.  
The limiting case $Z_{\Phi,k_{\Phi}}=0$ can be considered as a
solution of the corresponding Nambu--Jona-Lasinio model where the effective
four--fermion interaction has been eliminated by the introduction of
auxiliary meson fields (cf.\ section \ref{TheQuarkMesonModelAtT=0}).

\subsection{Renormalization group flow}

\centerline{\em 1. Flow equation for the effective potential}\vspace*{0.5cm}

The dependence on the infrared cutoff scale $k$ of the effective action
$\Gamma_k$ is given by the exact flow equation 
(\ref{bferge}). We employ the same infrared
cutoff function for the bosonic fields $R_{kB}$ (\ref{pk}) 
as in the previous sections. A few comments about the fermionic 
cutoff function are in place. 
The infrared
cutoff function for the fermions $R_{kF}$ should be consistent with 
chiral symmetries. This can be
achieved if $R_{kF}$ has the same Lorentz structure as the
kinetic term for free fermions~\cite{Wet90-1}.  In presence of a chemical
potential $\mu$ we use
\begin{equation}
  \label{fermir}
  R_{kF}=(\gamma^\mu q_\mu + i \mu \gamma^0) r_{kF} \, . 
\end{equation}
The effective squared inverse fermionic propagator is then of
the form
\begin{eqnarray}
  \label{fermprop}
  P_{kF} &=& \ds{
    [(q_0+i\mu)^2+\vec{q}^{\,2}](1+r_{kF})^2
    } \nnn 
  &=& \ds{
    (q_0+i \mu)^2+\vec{q}^{\,2}+k^2 \Theta 
    (k_{\Phi}^2-(q_0+i\mu)^2-\vec{q}^{\,2})
    }\; ,
\end{eqnarray}
where the second line defines $r_{kF}$ and one observes that
the fermionic infrared cutoff acts as an additional mass--like
term $\sim k^2$.\footnote{The exponential form (\ref{pk}) of the cutoff
  function $R_{kB}$ renders the first term on the right hand side of
  (\ref{bferge}) both infrared and ultraviolet finite. No need for an
  additional ultraviolet regularization arises in this case. This is
  replaced by the necessary specification of an initial value
  $\Gamma_{k_{\Phi}}$ at the scale $k_{\Phi}$.  Here $k_{\Phi}$ is
  associated with a physical ultraviolet cutoff in the sense that
  effectively all fluctuations with $q^2>k_{\Phi}^2$ are already included
  in $\Gamma_{k_{\Phi}}$. A similar property for the fermionic contribution
  is achieved by the $\Theta$--function in~(\ref{fermprop}).
  We note that in the previous sections 
  an exponential form of the fermionic infrared cutoff function $R_{kF}$ 
  was used. At nonzero density the mass--like IR cutoff simplifies
  the computations considerably because of the trivial momentum
  dependence.}  
Here the $\Theta$--function can be
thought of as the limit of some suitably regularized function, e.g.\ 
$\Theta^{\eps}=[{\rm exp}\{(q_0+i \mu)^2+\vec{q}^{\,2}-
k_{\Phi}^2\}/\eps+1]^{-1}$.

We compute the flow equation for the effective potential $U_k$ from
equation (\ref{frame}) using the ansatz (\ref{truncation}) for $\Gamma_k$
and we introduce a renormalized field $\rho=Z_{\Phi,k}\ovl{\rho}$ and
Yukawa coupling $h_k=Z_{\Phi,k}^{-1/2}\ovl{h}_k$. In complete
analogy to section \ref{FlowEquationsAndInfraredStability} we find that
the flow equation for $U_k$
obtains contributions from  bosonic and fermionic fluctuations,
respectively,
\begin{eqnarray}
  \ds{\frac{\pa }{\pa k}U_{kB}(\rho;T,\mu)} &=& \ds{
    \frac{1}{2} T \sum\limits_n
    \int\limits_{-\infty}^{\infty}
    \frac{d^3\vec{q}}{(2 \pi)^3} \frac{1}{Z_{\Phi,k}}
    \frac{\pa R_{kB}(q^2)}{\pa k} \left\{
    \frac{3}{P_{kB}\left(q^2\right) + U_k'(\rho;T,\mu)}\right. 
    }\nnn 
  && \ds{
    +\left. \frac{1}{P_{kB}\left(q^2\right) + U_k'(\rho;T,\mu)
      + 2 \rho U_k''(\rho;T,\mu)}\right\}  
    \label{dtub} 
    }\, ,\\[2mm]
  \ds{\frac{\pa }{\pa k}U_{kF}(\rho;T,\mu)} &=& \ds{
    -8 N_c T \sum\limits_n
    \int\limits_{-\infty}^{\infty}
    \frac{d^3\vec{q}}{(2\pi)^3} 
    \frac{ k \, \Theta (k_{\Phi}^2-(q_0+i \mu)^2-\vec{q}^{\,2})}
    {P_{kF}\left((q_0+i\mu)^2+\vec{q}^{\,2}\right)
      +h_k^2 \rho/2} 
    \label{dtuf} 
    }\, .
\end{eqnarray}
Here $q^2=q_0^2+\vec{q}^{\,2}$
with $q_0(n)=2n\pi T$ for bosons, $q_0(n)=(2n+1) \pi T$ for fermions $(n
\in \ZZ)$ and $N_c=3$ denotes the number of colors.  The scale dependent
propagators on the right hand side contain the momentum dependent pieces
$P_{kB}=q^2+Z_{\Phi,k}^{-1}R_k(q^2)$ and $P_{kF}$ given by~(\ref{fermprop})
as well as mass terms.  The only explicit
dependence on the chemical potential $\mu$ appears in the fermionic
contribution (\ref{dtuf}) to the flow equation for $U_k$. It is instructive
to perform the summation of the Matsubara modes explicitly for the
fermionic part. Since the flow equations only involve one momentum
integration, this can be easily done with standard techniques 
using contour integrals. (See e.g.\ \cite{Kap} chapter 3.) One finds
\begin{eqnarray}
  \lefteqn{
    \frac{\pa }{\pa k}U_{kF}(\rho;T,\mu) = -8 N_c 
    \int\limits_{-\infty}^{\infty}\,\,\,
    \frac{d^4q}{(2 \pi)^4} \frac{k \, \Theta (k_{\Phi}^2-q^2)}
    {q^2+k^2  + h_k^2 \rho/2}  
    +4 N_c \int\limits_{-\infty}^{\infty}
    \frac{d^3\vec{q}}{(2 \pi)^3} 
    \frac{k}{\sqrt{\vec{q}^{\,2}+k^2+h_k^2 \rho/2}} 
    }\nnn && \ds{
    \quad \times
    \left\{ \frac{1}
      {\ds \exp\left[(\sqrt{\vec{q}^{\,2}+k^2+h_k^2 \rho/2}-\mu)/T\right]+1}
      +\frac{1}
      {\ds \exp\left[(\sqrt{\vec{q}^{\,2}+k^2+h_k^2 \rho/2}+\mu)/T\right]+1}
      \label{uapprox}
    \right\} }
\end{eqnarray}
where, for simplicity, we sent $k_{\Phi}\to\infty$ in the $\mu,T$ dependent
second integral. This is justified by the fact that in the
$\mu,T$ dependent part the high momentum modes are exponentially
suppressed.

For comparison, we note that within the present approach one obtains
standard mean field theory results for the free energy if the
meson fluctuations are neglected, $\partial U_{kB}/\partial k \equiv 0$,
and the Yukawa coupling is kept constant, $h_k=h$ in (\ref{uapprox}). The
remaining flow equation for the fermionic contribution could then easily be
integrated with the (mean field) initial condition 
$U_{k_{\Phi}}(\rho)=\ovl{m}_{k_\Phi}^2\rho$. In the following we will
concentrate on the case of vanishing temperature.  
We find (see below) that a mean field
treatment yields relatively good estimates only for the
$\mu$--dependent part of the free energy $U(\rho;\mu)-U(\rho;0)$. On
the other hand, mean field theory does not give a very reliable
description of the vacuum properties which are important for a
determination of the order of the phase transition at $\mu \not = 0$.\\ 

\centerline{\em 2. Zero temperature physics}\vspace*{0.5cm}

In the limit of vanishing temperature one expects and observes a
non--analytic behavior of the $\mu$--dependent integrand of the fermionic
contribution (\ref{uapprox}) to the flow equation for $U_k$ because of the
formation of Fermi surfaces. Indeed, the explicit $\mu$--dependence of the
flow equation reduces to a step function
\begin{eqnarray}
  \label{dtuf0} 
  \ds{\frac{\pa }{\pa k}U_{kF}(\rho;\mu) =} &-& \ds{
    8 N_c 
    \int\limits_{-\infty}^{\infty}\,\,\,
    \frac{d^4q}{(2 \pi)^4} \frac{k \Theta (k_{\Phi}^2-q^2)}
    {q^2+k^2  + h_k^2 \rho/2}  
    }\nnn 
  &+& \ds{
    4 N_c \int\limits_{-\infty}^{\infty}
    \frac{d^3\vec{q}}{(2 \pi)^3} 
    \frac{k}{\sqrt{\vec{q}^{\,2}+k^2+h_k^2 \rho/2}} \,\,\,\,
    \Theta\! \left( \mu-\sqrt{\vec{q}^{\,2}+k^2+h_k^2 \rho/2} \right) 
    }\, .
\end{eqnarray}
The quark chemical potential $\mu$ enters the bosonic part (\ref{dtub}) of
the flow equation only implicitly through the meson mass terms
$U_k'(\rho;\mu)$ and $U_k'(\rho;\mu) + 2 \rho U_k''(\rho;\mu)$ for the
pions and the $\si$--meson, respectively.  For scales $k > \mu$ the
$\Theta$--function in (\ref{dtuf0}) vanishes identically and there is no
distinction between the vacuum evolution and the $\mu\not = 0$ evolution.
This is due to the fact that our infrared cutoff adds to the effective
quark mass $(k^2+h_k^2 \rho/2)^{1/2}$. For a chemical potential smaller
than this effective mass the ``density'' $-\partial U_k/\partial \mu$
vanishes whereas for larger $\mu$ one can view
$\mu=[\vec{q}_F^{\,2}(\mu,k,\rho)+k^2+h_k^2 \rho/2]^{1/2}$ as an effective
Fermi energy for given $k$ and $\rho$. A small infrared cutoff $k$ removes
the fluctuations with momenta in a shell close to the physical Fermi
surface\footnote{If one neglects the mesonic fluctuations one can
  perform the $k$--integration of the flow equation~(\ref{dtuf0}) in the
  limit of a $k$--independent Yukawa coupling. One recovers (for
  $k_\Phi^2\gg k^2+h^2\rho/2,\mu^2$) mean field theory results except for a
  shift in the mass, $h^2\rho/2\ra h^2\rho/2+k^2$, and the fact that modes
  within a shell of three--momenta
  $\mu^2-h^2\rho/2-k^2\le\vec{q}^{\,2}\le\mu^2-h^2\rho/2$ are not yet
  included. Because of the mass shift the cutoff $k$ also suppresses the
  modes with $q^2<k^2$.}
$\mu^2-h_k^2\rho/2-k^2<q^2<\mu^2-h_{k=0}^2\rho/2$. Our flow equation
realizes the general idea~\cite{Pol92-1} that for $\mu\neq0$ the lowering
of the infrared cutoff $k\ra0$ should correspond to an approach to the
physical Fermi surface. For a computation of the meson effective potential
the approach to the Fermi surface in (\ref{dtuf0}) proceeds from below and
for large $k$ the effects of the Fermi surface are absent.  By lowering $k$
one ``fills the Fermi sea''.

As discussed in section \ref{FlowEquationsAndInfraredStability}
the observed fixed point behavior in the symmetric regime
allows us to fix the model by
only two phenomenological input parameters and we use 
$f_{\pi}=92.4\MeV$ and $300\MeV\ltap M_q\ltap 350\MeV$.
The results for the evolution in vacuum described in section 
\ref{TheQuarkMesonModelAtT=0}~\cite{JW,BJW} show that for
scales not much smaller than $k_{\Phi}\simeq 600 \MeV$ chiral symmetry
remains unbroken. This holds down to a scale of about
$k_{\chi SB} \simeq 400\MeV$
at which the meson potential $U_k(\rho)$ develops a minimum at
$\rho_{0,k}>0$ thus breaking chiral symmetry spontaneously.  Below the
chiral symmetry breaking scale running couplings are no longer governed by
the partial fixed point. In particular, for $k \ltap k_{\chi SB}$ the
Yukawa coupling $h_k$ and the meson wave function renormalization
$Z_{\Phi,k}$ depend only weakly on $k$ and approach their infrared values.
At $\mu \not = 0$ we will follow the evolution from $k=k_{\chi SB}$ to
$k=0$ and neglect the $k$--dependence of $h_k$ and $Z_{\Phi,k}$ in this
range.  According to the above discussion the initial value 
$U_{k_{\chi SB}}$ is $\mu$--independent for $\mu < k_{\chi SB}$. 
We solve the flow equation
(\ref{dtu}) with (\ref{dtub}), (\ref{dtuf0}) numerically as a 
partial differential equation for the potential
depending on the two variables $\rho$ and $k$ for given $\mu$.
Nonzero current quark masses, which are neglected in our approximation, 
result in
a pion mass threshold and would effectively stop the renormalization
group flow of renormalized couplings at a scale around $m_{\pi}$. To
mimic this effect one may stop the evolution by hand at $k_f \simeq
m_{\pi}$.  We observe that our results are very insensitive to such a
procedure which can be understood from the fact that a small infrared
cutoff -- induced by $k_f$ or by the nonzero current quark masses --
plays only a minor role for a sufficiently strong first order 
transition.\footnote{For the results presented in the next section 
we use $k_f=100 \MeV$. The
  flow of the potential $U_k$ around its minima and for its outer convex
  part stabilizes already for $k$ somewhat larger than $k_f$. Fluctuations
  on larger length scales lead to a flattening of the barrier between the
  minima. The approach to convexity is not relevant for the present
  discussion.}

In the fermionic part (\ref{dtuf0}) of the flow equation the
vacuum and the $\mu$--dependent term contribute with opposite signs. This
cancelation of quark fluctuations with momenta below the Fermi
surface is crucial for the restoration of chiral symmetry at high
density\footnote{We note that the renormalization group 
investigation in \cite{HS} of a linear sigma model in $4-\epsilon$ 
dimensions misses this property.}.
In vacuum, spontaneous chiral symmetry breaking is induced in
our model by quark fluctuations which drive the scalar mass
term $U_k'(\rho=0)$ from positive to negative values at the scale $k =
k_{\chi SB}$.  (Meson fluctuations have the tendency to restore chiral
symmetry because of the opposite relative sign,
cf.~(\ref{frame}).) As the chemical potential becomes larger than the
effective mass $(k^2+h_k^2 \rho/2)^{1/2}$ quark fluctuations with momenta
smaller than $\vec{q}_F^{\,2}(\mu,k,\rho)=\mu^2-k^2-h_k^2 \rho/2$
are suppressed.  Since $\vec{q}_F^{\,2}$ is monotonically
decreasing with $\rho$ for given $\mu$ and $k$ the origin of the effective
potential is particularly affected.  We will see in the next section that
for large enough $\mu$ this leads to a second minimum of
$U_{k=0}(\rho;\mu)$ at $\rho=0$ and a chiral symmetry restoring first order
transition.

\subsection{High density chiral phase transition}
\label{hdcpt}

In vacuum or at zero density the effective potential $U(\si)$,
$\si\equiv\sqrt{\rho/2}$, has its minimum at a nonvanishing value
$\si_0=f_{\pi}/2$ corresponding to spontaneously broken chiral symmetry.
As the quark chemical potential $\mu$ increases, $U$ can develop different
local minima. The lowest minimum corresponds to the state of lowest free
energy and is favored.  In figure~\ref{potps}
\begin{figure}[t]
  \unitlength1.0cm
  \begin{center}
  \begin{picture}(13.0,8.0)
  \put(0.0,0.5){
  \epsfysize=8.cm
  \epsfbox[140 525 525 760]{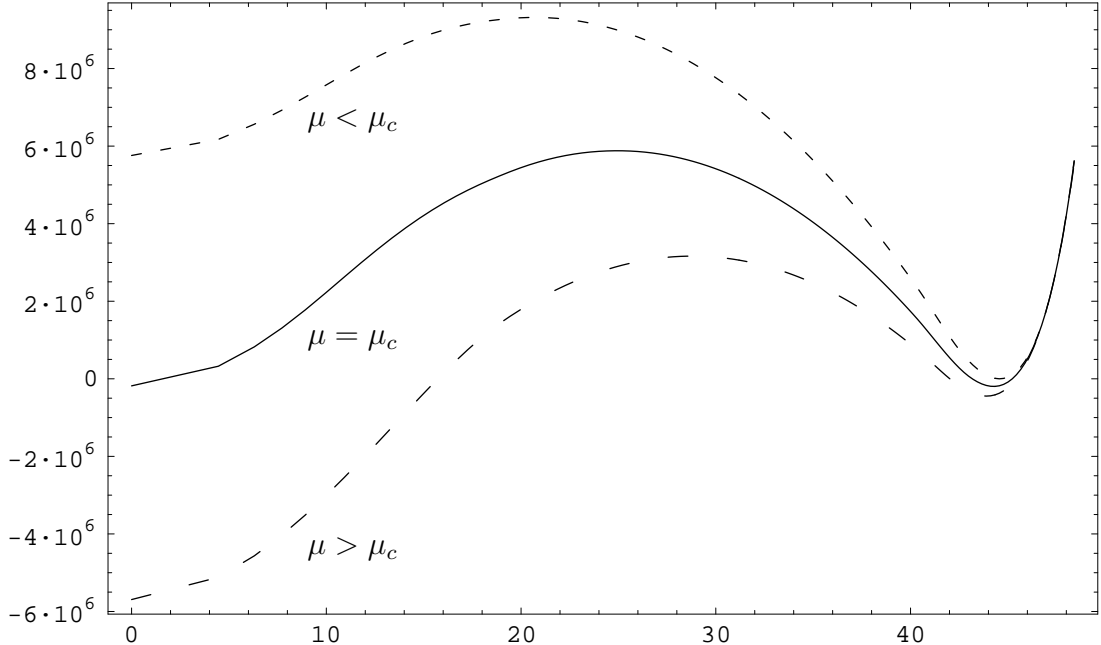}}
  \put(2.8,6.9){$\ds{\mu<\mu_c}$}
  \put(2.8,4){$\ds{\mu=\mu_c}$}
  \put(2.8,1.2){$\ds{\mu>\mu_c}$}
  \end{picture}
  \end{center}
\caption{The zero temperature effective potential $U$ (in ${\rm MeV}^4$)
  as a function of $\si\equiv(\rho/2)^{1/2}$ for different chemical
  potentials. One observes two degenerate minima for a critical chemical
  potential $\mu_c/M_q=1.025$ corresponding to a first order phase
  transition at which two phases have equal pressure and can coexist
  ($M_q=316.2 \MeV$). \label{potps}}
\end{figure}
we plot the free energy as a function of $\si$ for different values
of the chemical potential $\mu=322.6, 324.0,325.2$ MeV.  Here the effective
constituent quark mass is $M_q=316.2 \MeV$.
We observe that for $\mu < M_q$ the potential at its minimum does not
change with $\mu$. The quark number density is 
\begin{equation}
  n_q=-\frac{\pa U}{\pa \mu}_{|{\rm min}}
\end{equation}
and we conclude that the corresponding phase has zero density. In contrast,
for a chemical potential larger than $M_q$ we find a low density 
phase where chiral symmetry is still 
broken. The quark number density as a function of $\mu$ is shown in
figure~\ref{densityps}. One clearly observes the non--analytic behavior
at $\mu=M_q$ which denotes the ``onset'' value for nonzero density. 
\begin{figure}[t]
  \unitlength1.0cm
  \begin{center}
  \begin{picture}(13.0,8.0)
  \put(0.3,0.){
  \epsfysize=8.cm
  \epsfbox[125 415 560 700]{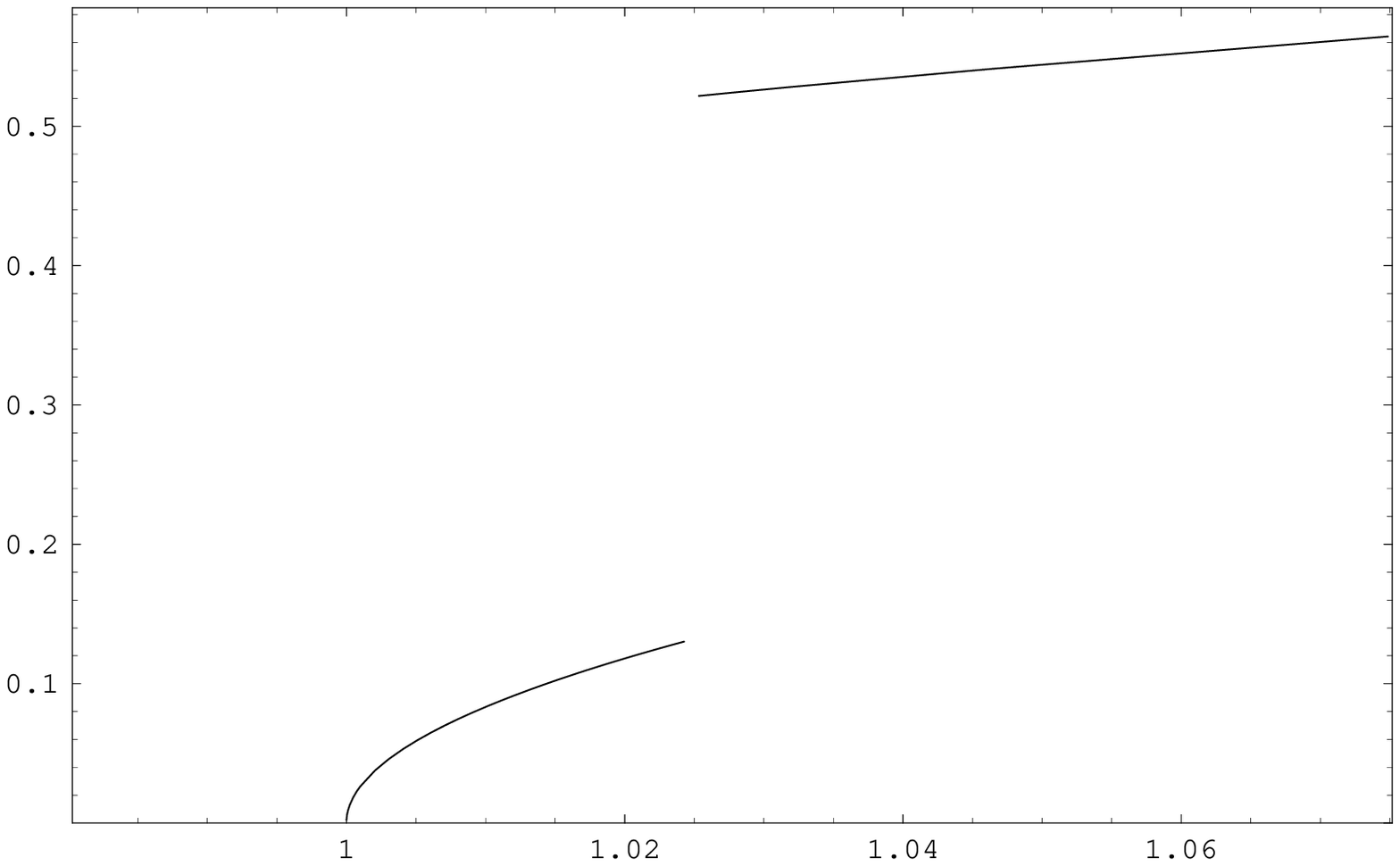}}
  \put(-1.2,5.2){$\ds{\frac{n_q^{1/3}}{M_q}}$}
  \put(13.0,0.5){$\ds{\mu/M_q}$}
  \end{picture}
  \end{center}
\caption{The plot shows $n_q^{1/3}$, where $n_q$ denotes
  the quark number density as a function of $\mu$ in units of the effective
  constituent quark mass ($M_q=316.2\MeV$).} 
  \label{densityps}
\end{figure}
From figure \ref{potps} one also notices the appearance of an additional
local minimum at the origin of $U$.  As the pressure $p=-U$ increases in
the low density phase with increasing $\mu$, a critical value $\mu_c$ is
reached at which there are two degenerate potential minima. Before $\mu$
can increase any further the system undergoes a first order phase
transition at which two phases have equal pressure and can coexist.  In the
high density phase chiral symmetry is restored as can be seen from the
vanishing order parameter for $\mu>\mu_c$. We note that the
relevant scale for the first order transition is $M_q$. 
For this reason we
have scaled our results for dimensionful quantities in units of $M_q$. 
For the class of quark meson models considered here (with $M_q/f_{\pi}$ in
a realistic range around $3$ -- $4$) the first order nature of the high
density transition has been clearly established. In particular, these
models comprise the corresponding Nambu--Jona-Lasinio models where the
effective fermion interaction has been eliminated by the introduction of
auxiliary bosonic fields.

The extent to which the transition from nuclear matter to quark matter in
QCD differs from the transition of a quark gas to quark matter in the
quark meson model or NJL--type models has to be clarified by further
investigations. At the phase transition, the quark number density in the
symmetric phase $n_{q,c}^{1/3}=0.52M_q$ turns out in this model to be not
much larger than nuclear matter density, $n_{q,nuc}^{1/3}=152\MeV$.
Here the open problems are related to the description of the low density
phase rather than the high density phase.  Quite generally, the inclusion
of nucleon degrees of freedom for the description of the low density phase
shifts the transition to a larger chemical potential and larger baryon
number density for both the nuclear and the quark matter
phases~\cite{NucMat}. An increase in $\mu_c$ results also from
the inclusion of vector mesons. This would not change the
topology of the phase diagram inferred from our model.  
Nevertheless, the topology of the
QCD phase diagram is also closely connected with the still unsettled
question of the order of the high temperature ($\mu=0$) transition for
three flavors with realistic quark masses. If the strange quark mass is too
small, or if the axial $U(1)$ symmetry is effectively restored in the
vicinity of the transition, then one may have a first order transition at
high $T$ which is driven by fluctuations~\cite{PW84-1}
(see also the discussion in section \ref{tricriticalpoint}). 

Finally, the low--density first order transition 
from a gas of nucleons or the
vacuum to nuclear matter (nuclear gas--liquid transition) can only be
understood if the low--momentum fermionic degrees of freedom are described
by nucleons rather that quarks~\cite{NucMat}. The low--density branch of
figure~\ref{densityps} cannot be carried over to QCD. We emphasize that
nucleon degrees of freedom can be included in the framework of
nonperturbative flow equations.

\subsection{The tricritical point}
\label{tricriticalpoint}

Already from the qualitative information about the order of the
chiral symmetry restoring transition (a) as a function of 
temperature $T$ for vanishing chemical potential $\mu$, 
and (b) as a function of $\mu$ for $T=0$, one can argue that there 
exists a {\em tricritical point} 
with long--range correlations in the phase diagram. The physics around 
this point is governed by universality and may allow for distinctive 
signatures in heavy ion collisions \cite{BR,SB,BTW}. We will concentrate
first on the ``robust'' universal properties of this point. 

In section \ref{TheQuarkMesonModelAtTNeq0} 
we find that at $\mu=0$, the chiral phase
transition as a function of increasing temperature is
second order for two massless quarks. In the chiral limit 
$\langle \ovl{\psi}{\psi}\rangle$ vanishes identically
in the high temperature phase whereas it is nonzero 
for $T<T_c$. A singularity separates the two phases which can also
be seen from the fact that the function 
$\langle \ovl{\psi}{\psi}\rangle \equiv 0$
cannot be continued to $\langle \ovl{\psi}{\psi}\rangle \not = 0$
analytically. In the vicinity of this singularity the physics
is governed by universality. There are only two independent
critical exponents near this ordinary (bi)critical point.
There is no reason to expect a small chemical potential to 
change this result, since this introduces no new massless
degrees of freedom in the effective three dimensional theory which
describes the long wavelength modes near $T=T_c$. Since the order
parameter $\langle \ovl{\psi}{\psi}\rangle$ is identically zero
on the one side of the transition, the line of  
phase transitions emanating from the
point $T=T_c$, $\mu=0$ cannot end in the $T\mu$--plane. 
On the other hand, in the previous section we find 
that chiral symmetry restoration
at $T=0$ proceeds via a first order transition within the considered models.
Therefore, the minimal possibility is that the line of second order
transitions coming from $T=T_c$, $\mu=0$ turns into a first
order transition at a specific value $T=T_{\rm tc}$, $\mu=\mu_{\rm tc}$.
The point in the phase diagram 
where this occurs is a tricritical point. 

Let us first consider the physics around this point in terms of the 
effective Landau--Ginsburg theory for the long wavelength modes
$\phi \sim \langle \bar{\psi}{\psi} \rangle$. In the vicinity of 
$T=T_{\rm tc}$, $\mu=\mu_{\rm tc}$ this requires a $\phi^6$
potential which has the form
\begin{equation}
U(\phi,0;\mu,T)= U(0,0;\mu_{\rm tc},T_{\rm tc})+
\frac{a(\mu,T)}{2} \phi^2 + \frac{b(\mu,T)}{4} \phi^4
+ \frac{c(\mu,T)}{6}\phi^6 - h\phi \ , 
\label{omegatc}
\end{equation}
where the
coefficient $h$ of the linear term is proportional to the current quark mass.
The coefficients $a$ and $b$ are both zero at the tricritical
point, and barring accidental cancellations both will be
linear in both $(T-T_{\rm tc})$ and $(\mu-\mu_{\rm tc})$.
The $\mu$ and $T$ dependence of $c$ is not important,
as $c$ does not vanish at the tricritical point, and it
is convenient to set $c=1$.
It is easy to verify that, near the
tricritical point, the line of second order transitions
is given by $a=0$, $b>0$ and the line of first order 
transitions is given by $a=3b^2/16$, $b<0$.
Minima of $\Omega$ are described by the scaling form
\begin{equation}
h=\phi_0^5\,\left( \frac{a}{\phi_0^4} + \frac{b}{\phi_0^2} + 1 \right)
\label{eqofstate} \, .
\end{equation}
{}From this, we read off the exponents $\delta=5$, $1/\beta=4$,
and $\phi_t/\beta=2$ where $\phi_t$ is called the crossover 
exponent because tricritical (as opposed to first or second
order) scaling is observed for $b<a^{\phi_t}$. For more
details see ref.\ \cite{tricrit}. 
The critical fluctuations are described by an effectively
three dimensional theory (cf.\ sect.\ \ref{FiniteTemperatureFormalism})
and the $\phi^6$ coupling in (\ref{omegatc}) is dimensionless,
becoming a marginal operator. This explains why 
the upper critical dimension for the $\phi^6$ theory is $d=3$,
and the behavior in the vicinity of the 
tricritical point is correctly described by mean field exponents
up to logarithmic corrections \cite{tricrit}. So if we trust the 
qualitative feature that the
transition is second order at high temperatures and first order
at low temperatures, then QCD with two flavors will
have a tricritical point in the same universality 
class as that in our model, with the critical exponents given
above.  
    
{\em Physics away from the chiral limit.} What, then, happens in this
region of temperature and chemical potential in the
presence of a small quark mass?
We have seen that a nonzero quark mass does have 
a qualitative effect on the second order transition 
which occurs at temperatures above $T_{\rm tc}$:  The $O(4)$ transition
becomes a smooth crossover.
A small quark mass cannot eliminate the first order
transition below $T_{\rm tc}$.  Therefore, whereas we previously had a line
of first order transitions and a line of second order
transitions meeting at a tricritical point, 
with $m\neq 0$ we now
have a line of first order transitions ending at an
ordinary critical point. The situation is precisely
analogous to critical opalescence in a liquid-gas system.
At this critical point, one degree of freedom (that associated
with the magnitude of $\phi$) becomes massless, while
the pion degrees of freedom are massive
since chiral symmetry is explicitly broken.  Therefore,
this transition is in the same universality class as
the three-dimensional Ising model. 

With the nonperturbative methods presented in this 
lecture we can obtain the equation of state
near the critical endpoint, in the same way as we obtained the universal 
properties of the high temperature ($\mu=0$) transition in section
\ref{CriticalBehavior}. If we are just interested in the universal properties
we can directly study a three dimensional model with a single
component field.
The flow equation for the effective potential in the
vicinity of the critical endpoint corresponds to (\ref{udl})
if the fermionic contribution $\sim N_c$ and the contribution
from the three pions are neglected. The only contributions 
comes then from the $\sigma$--resonance. This flow equation
for $d=3$ has been solved numerically in ref.\ \cite{BTW}
and detailed results and comparison with other methods 
can be found there.

From our studies of the chiral phase transition at nonzero 
temperature in section \ref{FiniteTemperatureFormalism} we have
observed the possibility of a second order transition,
with infinite correlation lengths in an unphysical world
in which there are two massless quarks.  
It is exciting
to realize that if the finite density transition is first
order at zero temperature, as in the
models we have considered, then there is a tricritical
point in the chiral limit which becomes an Ising second
order phase transition in a world with chiral symmetry
explicitly broken.  In a sufficiently energetic heavy ion collision, one 
may create conditions in approximate local thermal
equilibrium in the phase in which spontaneous chiral symmetry breaking
is lost.
Depending on the initial density and temperature, when
this plasma expands and cools it will traverse the 
phase
transition at different points in the ($\mu,T)$ plane.
Our results suggest that
in
heavy ion collisions in which the chiral symmetry breaking
transition is traversed at baryon densities which
are not too high and not too low, 
a very long correlation length in the $\sigma$ channel
and critical slowing
down may be manifest even though the pion is massive.

We observe the tricritical point (critical endpoint) to emerge 
from the qualitative result of 
a second order transition (crossover) in one region 
of phase space (high $T$, small $\mu$) and a first order transition 
for the same order parameter in another region (high $\mu$, low $T$).  
The universal properties of this point can be determined
quantitatively. The location of the tricritical point in the phase diagram
is a nonuniversal property, and the crude knowledge of the 
critical chemical potential for the high density, zero temperature
transition makes a precise estimate difficult (cf.\ section \ref{hdcpt}). 
The values for $T_{tc}$ and $\mu_{tc}$  
have been estimated within a NJL--type model in a mean field approximation
to be about $\mu_{\rm tc}\simeq 200$ MeV and $T=T_{\rm tc}\simeq100$ MeV
\cite{BR}, which agrees rather well with an estimate using
a random matrix model \cite{SB}. 

Finally, the question of whether the tricritical point is 
realized in QCD is closely connected with the question
of the order of the high temperature ($\mu=0$) transition for
realistic strange quark masses. If the strange quark 
mass turns out to be too small then we may have a
first order transition at high $T$ which is driven by fluctuations 
\cite{PW84-1}\footnote{The same situation arises if the 
axial $U(1)$ symmetry is 
effectively restored about the transition temperature.}. 
As a consequence, for a particular value of the strange quark mass,
$m_s=m_s^*$, one expects the presence of a tricritical point even for $\mu=0$
\cite{PW84-1,RaWi93-1,Raj95-1,GGP}. 
In this case and also for $m_s < m_s^*$, 
a line of first order transitions connects the 
$T$ and $\mu$ axes. Endpoints would then only occur if the high
temperature, low density region and the low temperature, high density  
region are disconnected.

\pagebreak

\subsection{Color superconductivity}
\label{colorsc}

Chiral symmetry breaking in vacuum is due to a quark--antiquark pairing
with zero net momentum.
As the density grows, more and
more low momentum states are excluded from pairing and the 
chiral condensate is suppressed. We have verified this behavior 
in section \ref{hdcpt} within the effective quark meson model. On the
other hand, there are attractive quark--quark interactions.
In contrast to quark--antiquark pairing there is little cost in free
energy for correlated pairs of quarks (or antiquarks) 
near the Fermi surface. This situation 
is visualized in figure \ref{fermis}. (Figure taken from \cite{mark}.) 
\begin{figure}[h]
\begin{center}
\epsfxsize=4.in
\hspace*{0in}
\epsffile{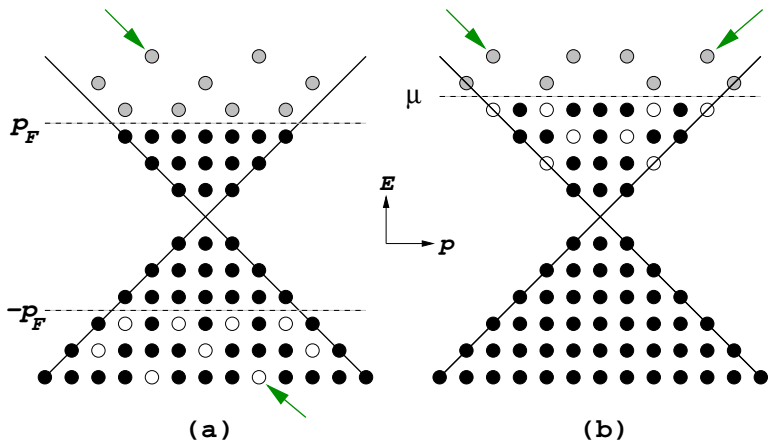} 
\end{center}
\caption{The figure shows a comparison of quark--antiquark (a)
and quark--quark pairing (b). Arrows indicate typical pairs,
having net zero momentum.}
\label{fermis}
\end{figure}
As in the case of ordinary
BCS superconductivity in electron systems, an arbitrarily weak attractive 
interaction between quarks renders the quark Fermi surface unstable
and leads to the formation of a condensate. Since pairs of quarks
cannot be color singlets, a diquark condensate breaks color symmetry.   
It was originally suggested
by Bailin and Love \cite{bailin} (see also \cite{Barrois}) that QCD at
very high density behaves as a ``color superconductor'':
Cooper pairs of quarks condense in an attractive channel opening
up a gap at the Fermi surface. Recent mean field analyses of
NJL--type models indicate that quark pair condensation does
indeed occur and that the gaps are phenomenologically significant
(of order $100$ MeV) at densities of about a few times nuclear matter
density \cite{CSC}. Different 
attractive channels may lead to a possible (simultaneous) formation of  
condensates. In particular, the vacuum of QCD already
has a condensate of a quark--anti-quark pair. An important ingredient
for the understanding of the high density phase structure is the
notion of {\it competing condensates}. 
One expects, that the breaking of color symmetry 
due to a $\langle \psi \psi \rangle$ condensate is suppressed 
by the presence
of a chiral condensate. Likewise, chiral symmetry restoration 
may be induced at lower densities by the presence of a color superconductor
condensate. This behavior has been demonstrated explicitly within
an NJL--type model and led
to a first model calculation of the phase diagram \cite{BR}, 
similar to the one sketched in the figure of section 
\ref{intro}. 
At low temperatures, chiral symmetry restoration
occurs via a first order transition between a phase
with low (or zero) baryon density and a high density
color superconducting phase\cite{BR,JBiele,CarDia}. 
Color superconductivity is found
in the high density phase for temperatures less than of order
tens to 100 MeV.
The strong competition between chiral and superconducting 
condensates also simplifies the discussion: The behavior of the
chiral condensate within the low density phase is well described
without taking into account diquark condensation once the phase
boundary is known. We implicitly exploited 
this fact in section \ref{hdcpt}. 

A systematic calculation in perturbation theory is only possible
at asymptotically high densities, where the strong gauge coupling
at the Fermi energy is small. The corresponding renomalization of quark
operators in the vicinity of the Fermi surface has been studied
in \cite{RG}. The method using nonperturbative flow equations,
which were discussed in these lectures, may help to give some insight
about these questions at lower densities. Since this remains to be done
I will content myself here with some general aspects and  
an outlook.

The explicit form of the favored diquark condensate crucially determines
the symmetry properties of the ground state at high density. 
Diquark operators have the structure
\begin{equation}
\psi^T \Sigma_{\rm Dirac}\Sigma_{\rm Color}\Sigma_{\rm Flavor} \psi
\end{equation}
where the combination of Dirac, color and flavor matrices have to
be antisymmetric according to the Pauli principle. 
It is instructive to do some classification. Let us divide
matrices in symmetric and anti--symmetric ones, where we consider
the case of vanishing angular momentum 
\ben
\Sigma_{\rm Dirac}^{\rm A}&=&
\bigg\{C \gamma_5, C, C \gamma_{\mu} \gamma_5\bigg\} \nnn 
\Sigma_{\rm Dirac}^{\rm S}&=&
\bigg\{C \gamma_\mu, C \sigma_{\mu \nu}\bigg\} \nnn
\Sigma_{\rm color}^{\rm A}&=&
\bigg\{\lambda_7,-\lambda_5,\lambda_2\bigg\}
\equiv\bigg\{\lambda^{\rm A}_{i=1,2,3}\bigg\} \; ;\qquad 
(\lambda^{\rm A}_{i})_{jk}=-i\epsilon_{ijk}  \nnn
\Sigma_{\rm color}^{\rm S}&=&
\bigg\{\lambda_0\equiv \sqrt{\frac{2}{3}} {\bf 1},\lambda_3,
\lambda_8,\lambda_1,\lambda_4,\lambda_6\bigg\} 
\equiv\bigg\{\lambda^{\rm S}_{i=1,\dots,6}\bigg\} \nnn
\Sigma_{\rm flavor}^{\rm A}&=&\bigg\{\tau_2\bigg\}\; ;\qquad 
(\tau_{2})_{ij}=-i\epsilon_{ij} \nnn
\Sigma_{\rm flavor}^{\rm S}&=&\bigg\{\tau_0\equiv \frac{1}{2} \, {\bf 1},
\tau_1,\tau_3\bigg\}=\bigg\{\tau_{i=1,2,3}^{\rm S}\bigg\}
\een
Here ``A'' and ``S'' denotes antisymmetric and symmetric,
respectively; the $\tau$'s are the Pauli matrices; 
$C$ is the charge conjugation matrix, 
$\Sigma_{\rm color}^{\rm A}$ are the three antisymmetric 
and $\Sigma_{\rm color}^{\rm S}$ the six symmetric Gell--Mann matrices.
This split corresponds to the fact that the two quarks can be
comined to form a color antitriplet and a symmetric color ${\bf 6}$: 
$\, {\bf 3}_c \otimes {\bf 3}_c = \bar{\bf 3}_c + {\bf 6}_c$. 
Consider first the color ${\bf \ovl{3}}$ representation.
Overall antisymmetry
then requires that combined Dirac--flavor has to be symmetric.
If we consider the antisymmetric flavor singlet $\tau_2$
then the Dirac matrices must be either $C\gamma_{5}$, 
corresponding to a Lorentz scalar condensate, or $C$ which is a
pseudoscalar. The other possible combination involving 
$C\gamma_{\mu}\gamma_5$ does not pair states near
the Fermi surface and is suppressed. Instead it realizes pairing of a state 
near the Fermi surface and one much below with net zero momentum.

Let us turn to the question if there is indeed an attractive channel between
quarks of different color. We may content ourselves here by looking
at the signs for the scalar channels in the effective four quark interaction
${\cal M}$ (\ref{QCDFourFermi})
\be
 {\cal M}(p_1,p_2,p_3,p_4) = \ds{
-\left\{\bar\psi^i_a(-p_1)\gamma^\mu\left(\frac{\lambda_z}{2}
\right)_i^{\ j}\psi_j^a(-p_3)\right\}\left\{\bar\psi^k_b(p_4)
\gamma_\mu\left(\frac{\lambda_z}{2}\right)_k^{\ \ell}
\psi_\ell^b(p_2)\right\}}\; .
 \label{onegluon}
\ee
The curled brackets indicate  contractions over spinor indices, 
$i,j,k,l=1,2,3$ are the colour
indices and $a,b=1,2$ the flavour indices of the quarks.
Let us neglect for a moment the  
color and flavor structure of the interaction. Using the identity
\ben
(\gamma^{\mu})_{\alpha \beta}(\gamma_{\mu})_{\gamma \delta}
&=&-(i C)_{\alpha \gamma}(i C)_{\delta \beta}
-(\gamma^{5} C)_{\alpha \gamma} (C \gamma^{5})_{\delta \beta}\nnn
&& +\frac{1}{2}(i \gamma^{\mu} C)_{\alpha \gamma} 
(i C \gamma_{\mu})_{\delta \beta}
-\frac{1}{2}(\gamma^{5}\gamma^{\mu} C)_{\alpha \gamma} 
(C \gamma^{5}\gamma_{\mu})_{\delta \beta}
\een
we can rewrite the interaction in terms of quark and antiquark
bilinears\footnote{In our 
Euclidean conventions \cite{Wet90-1} 
$(\bar\psi i C \bar\psi^T)^\dagger=(\psi^T i C \psi)$,
$(\bar\psi \gamma^{5} C  \bar\psi^T)^\dagger=(\psi^T C \gamma^{5}  \psi)$. 
Here $\dagger$ denotes the operation of Euclidean reflection
in analogy to hermitean conjugation in Minkowskian 
space time.}, respectively
\ben
{\cal M}(p_1,p_2,p_3,p_4) &=&
+\left\{\bar\psi(-p_1) i C \bar\psi^T(p_4) \right\}
\left\{\psi^T(p_2) i C \psi(-p_3) \right\} \nnn
&&+\left\{\bar\psi(-p_1) \gamma^{5} C \bar\psi^T(p_4) \right\}
\left\{\psi^T(p_2) C \gamma^{5} \psi(-p_3) \right\} \nnn
&&-\left\{\bar\psi(-p_1) i \gamma^{\mu} C \bar\psi^T(p_4) \right\}
\left\{\psi^T(p_2) i C \gamma_{\mu} \psi(-p_3) \right\} \nnn
&&+\left\{\bar\psi(-p_1) \gamma^{5}\gamma^{\mu} C \bar\psi^T(p_4) \right\}
\left\{\psi^T(p_2) C \gamma^{5}\gamma_{\mu} \psi(-p_3) \right\} \label{qbi}
\; .
\een
From the signs in (\ref{qbi}) one may expect
repulsion in the pseudoscalar and scalar diquark channels
(compare also with the attractive scalar and pseudoscalar
meson channels in (\ref{sigmac})).
However, we neglected color and flavor so far. The appropriate
Fierz identity for color manifests the separation in antitriplet
and color $\bf{6}$ diquark channels
\be
\sum\limits_{z=1}^8 \left(\frac{\lambda_z}{2}\right)_i^{\ j}
\left(\frac{\lambda_z}{2}\right)_k^{\ \ell}
= -\frac{4}{3} \sum\limits_{z=1}^3 
\left(\frac{\lambda_z^{\rm A}}{2}\right)_i^{\ k}
\left(\frac{\lambda_z^{\rm A}}{2}\right)_\ell^{\ j} 
+\frac{2}{3} \sum\limits_{z=1}^6 
\left(\frac{\lambda_z^{\rm S}}{2}\right)_i^{\ k}
\left(\frac{\lambda_z^{\rm S}}{2}\right)_\ell^{\ j} \; .
\label{colorf}
\ee
For flavor we use  
\be
\delta_{a b}\delta_{c d}= \frac{1}{2} (\tau_2)_{a c} (\tau_2)_{d b}
+ \sum\limits_{z=1}^3 (\tau_z^{\rm S})_{a c} (\tau_z^{\rm S})_{d b} \; .
\ee
From the different relative sign for the $\bf{3}_c$ and
the $\bf{6}_c$ channels in (\ref{colorf}) we conclude that
one may expect condensation in the color antitriplet 
Lorentz scalar and pseudoscalar channels. The corresponding color symmetric  
channels may not be expected to condense.

We note, however, that the four quark interaction (\ref{momint})
(``dressed one gluon exchange'') does not break the axial $U(1)_A$ of QCD.
It has been shown \cite{Ho} that the instanton induced interaction 
between light quarks properly reflects the
chiral symmetry of QCD: axial baryon number is broken, while
chiral $SU(2)_L\times SU(2)_R$ is respected.
(The corresponding $U(1)_A$--breaking term can be written as a
determinant (\ref{Invariants}) which has been taken into 
account in the effective quark meson model in section 
\ref{TheQuarkMesonModelAtT=0}.)
Here we only note that the instanton induced interaction leads
to an attractive scalar diquark channel 
but exhibits repulsion
in the pseudoscalar channel for $\bf{3}_c$ \cite{CSC,BR}. 
For two flavors triplet pairing of the form
\begin{equation}
\langle \psi^i_a (p) C\gamma_5 \psi^j_b (-p) \rangle ~=~
\kappa (p^2)  \epsilon^{ij\ell} \epsilon_{ab}
\label{twoflavcondensate}
\end{equation}
has been demonstrated to lead to significant gaps of the order
$100$ MeV for instanton induced interactions \cite{CSC,BR}.
For a nonzero expectation value of 
(\ref{twoflavcondensate}) the diquark bilinear has a color index 
$\ell$ which chooses
a direction in color space, say $\ell=3$, thus breaking 
color symmetry $SU(3) \rightarrow SU(2)$. 
It leaves all flavor
symmetries and, in particular, the chiral 
$SU(2)_L \times SU(2)_R$ intact.
One may expect it to be especially favorable because it maintains
a large symmetry, so that the interaction of a given pair can obtain
contributions from other pairs with different particle combinations.
The above pairing does not involve quarks of the third color. 
It is possible that the remaining ungapped Fermi surface becomes
unstable too, which is of course color symmetric. 
An additional condensation of the form 
\begin{equation}
\langle \psi^3_a(p) C\sigma_{0i} \psi^3_b(-p) \rangle = \eta (p^2) {\hat
p}_i \delta_{ab}
\end{equation}
has been investigated \cite{CSC}.
This condensate violates rotation
symmetry and the gaps were found to be of order several keV at best.

Breaking of the color $SU(3)$ symmetry
generates Goldstone bosons, formally. However, since color is a
gauge symmetry and the Higgs mechanism operates, the spectrum does 
not contain massless scalars but massive vectors. The number 
of broken generators is $8-3=5$ for condensation of the form 
(\ref{twoflavcondensate}) and, therefore, five gluons become massive.
We have decribed here the color superconducting phase as a Higgs
phase. One expects, however, that there is a complementary description
in which this is a confining phase. Indeed, the color superconducting
phase for two flavors may be considered as a realization of confinement
without chiral symmetry breaking.  

The $U(1)_Q$ of electromagnetism is also spontaneously broken
by the condensate. However, the photon mixes with the eighth
component of the gluon to generate an unbroken $U(1)_{Q'}$.
Let $T_3$ denotes the diagonal generator of isospin $SU(2)$
and $B$ baryon number then $Q=T_3+B/2$. For a nonvanishing
color antitriplet condensate (\ref{twoflavcondensate})
$\Delta \sim \epsilon^{i j 3}$ one finds
\begin{equation}
B\, \Delta = \frac{2}{3} \Delta\, \; , \quad
Q\, \Delta = \frac{1}{3} \Delta\, \; , \quad
\lambda_8\, \Delta = \frac{2}{\sqrt{3}} \Delta
\end{equation} 
and one observes a linear combination of electric charge and color
hypercharge $Y\equiv \lambda_8/\sqrt{3}$ under which the
condensate is neutral: $Q'=Q-Y/2$, so a 
modified $U(1)$ persists. (Instead, one may also want to
refer to baryon number and finds the  
combination $B'=B-Y$ which leaves the condensate invariant.)\\

{\em Three quark flavors.} The situation is further complicated by 
the question about the number of light quark flavors participating 
in the condensate. If the strange quark
were heavy relative to fundamental QCD scales, the idealization
of assuming two light flavors would be obviously sufficient.
The situation becomes more involved since we are interested in densities
above the transition for which the chiral condensate vanishes.
Without spontaneous chiral symmetry breaking the relevant strange 
quark is of the order $100$ MeV. Since we are interested in 
chemical potentials larger than $100$ MeV it is a quantitative 
question if an approximation with two light flavors is realistic.     

In the limit of three massless quark flavors a very interesting
and compelling possibility for condensation has been proposed
by Alford, Rajagopal and Wilczek \cite{CFL}. Consider a 
generalization of the two flavor condensate (\ref{twoflavcondensate})
where the antisymmetric flavor singlet $\sim \epsilon_{ab}$ is replaced
by the antisymmetric triplet for three flavors $\sim \epsilon_{abc}$
Since there are an equal number of flavors and colors there is the
possibility for a color--flavor structure 
$\sim~\epsilon^{ijz} \epsilon_{abz}=\delta^i_{a}\delta^j_{b}
-\delta^i_{b}\delta^j_{a}$ by summing over $z$. 
In \cite{CFL} a slightly more general variant
\begin{equation}
\langle \psi^i_a(p) C\gamma_5 \psi^j_b(-p) \rangle ~=~
\kappa_1 (p^2) \delta^i_a \delta^j_b +
\kappa_2 (p^2) \delta^i_b \delta^j_a 
\end{equation}
has been investigated. The mixed Kronecker matrices are invariant
only under matched vectorial color/flavor rotations. 
Ignoring electromagnetism the symmetry of QCD
with three massless flavors is 
$SU(3)_{\rm color}\times SU(3)_L\times SU(3)_R\times U_B(1)$
(cf.\ section \ref{shorttolong}). 
The $SU(3)_{\rm color}$ is a local gauge symmetry
while the chiral flavor symmetry is global, as is the $U(1)_B$ for 
baryon number. As a consequence of the above matching between 
color and flavor symmetries for a nonvanishing condensate 
the only remaining symmetry is the global diagonal $SU(3)_{{\rm color}+L+R}$ 
subgroup. In particular, the gauged color symmetries and the global
axial flavor symmetries are spontaneously broken. The latter means
that chiral symmetry is broken by a new mechanism. The breaking of
the $U(1)_B$ of baryon number leads to superfluidity as in liquid
helium. All fermions aquire a gap and there are Goldstone modes associated
with the spontaneous chiral symmetry breaking. The high density phase 
may therefore be in many ways quite similar to low density QCD \cite{SW}!

It is not clear what phase will be
realized with physical mass $u$, $d$ and $s$ quarks. As pointed out 
in \cite{S},
one may think of having the two flavor condensate first and color--flavor
locking at higher density. A first investigation may be performed
along the lines presented in \cite{BR} where simultaneous condensation
phenomena have been discussed, but finally more sophisticated
methods than the mean field analysis will be needed
to settle these questions. An interesting possibility will be the use 
of truncated nonperturbative flow equations along the lines presented 
in these lectures.    
 
\acknowledgments

I thank M.\ Alford, D.--U.\ Jungnickel, K.\ Rajagopal and C.\ Wetterich
for collaboration on work presented in these lectures. I would like
to express my gratitude to the organizers of the Nuclear Physics
Summer School and Symposium (NuSS'98) for the 
invitation to give these lectures and for providing a most stimulating 
environment.

\newpage
\thispagestyle{empty}

\end{document}